%% file: Optimized_Configurable_Architectures.tex
\documentclass[journal,draftclsnofoot,onecolumn]{IEEEtran}

\usepackage{ifpdf}
\ifCLASSINFOpdf
  \usepackage[pdftex]{graphicx}
  \graphicspath{{}}
  \DeclareGraphicsExtensions{.pdf,.jpeg,.png}
\else
  \usepackage[dvips]{graphicx}
  \graphicspath{{}}
  \DeclareGraphicsExtensions{.eps}
\fi

\ifCLASSOPTIONcaptionsoff
  \usepackage[nomarkers]{endfloat}
 \let\MYoriglatexcaption\caption
 \renewcommand{\caption}[2][\relax]{\MYoriglatexcaption[#2]{#2}}
\fi

\usepackage{cite}
\usepackage[cmex10]{amsmath}
\usepackage{amsfonts}
\usepackage{mathtools}
\usepackage{ctable}
\usepackage[para]{threeparttable}
\mathtoolsset{showonlyrefs}
\usepackage{amssymb}
\usepackage{bbm}
\usepackage{dsfont}
\usepackage{amscd}
\usepackage{dsfont}
\usepackage{leftidx,tensor}
\usepackage[algo2e]{algorithm2e}
\SetAlgoCaptionSeparator{.}
\SetAlFnt{\small}
\SetAlCapFnt{\small}
\SetAlCapNameFnt{\small}
\usepackage{xcolor}
\definecolor{darkgreen}{rgb}{0,0.75,0}
\usepackage{multirow,array}
\usepackage{thinsp}
\usepackage{stfloats}
\usepackage{units}
\usepackage[ddmmyyyy]{datetime}
\usepackage{tikz}

\newcommand{\abs}[1]{\left|{#1}\right|}
\newcommand{\norm}[1]{\left\|{#1}\right\|}
\newcommand{\slice}[2]{{\left\lfloor{#1}\right\rceil}_{#2}}
\newcommand{\mbf}[1]{\mathbf{#1}}
\newcommand{\mbb}[1]{\mathbb{#1}}

\newcommand{\E}[1]{{\mbb{E}\!\left[{#1}\right]}}

\newcommand{\bigO}[1]{\ensuremath{\mathop{}\mathopen{}O\mathopen{}\left(#1\right)}}
\newcommand{\bs}[1]{\boldsymbol{#1}}

\newcommand{\MLL}[0]{\mathrm{ML}}
\newcommand{\MAP}[0]{\mathrm{MAP}}
\newcommand{\WL}[0]{\mathrm{WL}}

\newcommand{\re}[0]{\mathrm{R}}
\newcommand{\im}[0]{\mathrm{I}}

\newcommand{\mxH}[0]{\mbf{H}}

\newcommand{\nth}[1]{{#1}{\text{th}}}

\newcommand{\blue}[1]{{\color{blue}{#1}}} 
\newcommand{\red}[1]{{\color{red}{#1}}} 
\newcommand{\magenta}[1]{{\color{magenta}{#1}}} 
\newcommand{\orange}[1]{{\color{orange}{#1}}} 

\renewcommand{\a}[0]{\alpha}
\renewcommand{\b}[0]{\beta}
\renewcommand{\c}[0]{\gamma}

\renewcommand{\IEEEQED}{\IEEEQEDclosed}

\begin{document}

\title{Optimized Configurable Architectures for Scalable Soft-Input Soft-Output MIMO Detectors with 256-QAM}

\author{Mohammad~M.~Mansour,~\IEEEmembership{Senior~Member,~IEEE}, and Louay M.A. Jalloul,~\IEEEmembership{Senior~Member,~IEEE}
\thanks{M. M. Mansour is with the Department of Electrical and Computer Engineering at the American University of Beirut, Lebanon, e-mail: mmansour@ieee.org. L. Jalloul is with Qualcomm Inc., San Jose, CA, e-mail: jalloul@ieee.org.}
}


\markboth{IEEE~Transactions~on~Signal~Processing, draft (\ddmmyyyydate\today)}%
{M. M. Mansour: Optimized Configurable Architectures for Scalable Soft-Input Soft-Output MIMO Detectors with 256-QAM}

%
\maketitle

%
\begin{abstract}
This paper presents an optimized low-complexity and high-throughput multiple-input multiple-output (MIMO) signal detector core for detecting spatially-multiplexed data streams. The core architecture supports various layer configurations up to 4, while achieving near-optimal performance, as well as configurable modulation constellations up to 256-QAM on each layer. The core is capable of operating as a soft-input soft-output log-likelihood ratio (LLR) MIMO detector which can be used in the context of iterative detection and decoding. High area-efficiency is achieved via algorithmic and architectural optimizations performed at two levels. First, distance computations and slicing operations for an optimal 2-layer maximum a posteriori (MAP) MIMO detector are optimized to eliminate the use of multipliers and reduce the overhead of slicing in the presence of soft-input LLRs. We show that distances can be easily computed using elementary addition operations, while optimal slicing is done via efficient comparisons with soft decision boundaries, resulting in a simple feed-forward pipelined architecture. Second, to support more layers, an efficient channel decomposition scheme is presented that reduces the detection of multiple layers into multiple 2-layer detection subproblems, which map onto the 2-layer core with a slight modification using a distance accumulation stage and a post-LLR processing stage. Various architectures are accordingly developed to achieve a desired detection throughput and run-time reconfigurability by time-multiplexing of one or more component cores. The proposed core is applied as well to design an optimal multi-user MIMO detector for LTE. The core occupies an area of \unit[1.58]{MGE} and achieves a throughput of \unit [733]{Mbps} for 256-QAM when synthesized in \unit[90]{nm} CMOS.
\end{abstract}

%

\IEEEpeerreviewmaketitle

\vspace{-0.15in}
\section{Introduction}\label{s:intro}\vspace{-0.05in}
Multiple-input multiple-output (MIMO) systems have become mainstream technology for achieving high spectral efficiencies in wireless communications standards such as IEEE 802.11ac~\cite{802.11ac} and the 3GPP Long-Term Evolution (LTE)~\cite{LTE_36.211}. Detection of spatially-multiplexed MIMO streams plays a key role in receiver design, both in terms of performance and complexity, and has remained to be an active area of research~\cite{2003_Paulraj,2006_Giannakis,2007_Biglieri,2010_Oestges,2014_Chockalingam}. The focus has been on developing area-/energy-efficient VLSI implementations of MIMO detectors that are capable of achieving close to optimal performance.


A plethora of MIMO detectors have appeared in the literature on this subject, offering various performance-complexity tradeoffs. Suboptimal zero-forcing (ZF) and minimum mean-squared error (MMSE) detectors~\cite{2007_Biglieri}, as well as nonlinear parallel and successive interference cancellation schemes~\cite{2000_Hassibi,1999_Golden,2003_Wubben,2011_Studer_JSSC}, require relatively low complexity but sacrifice performance. On the other hand, tree-search or list-based detectors require substantially higher complexity but can offer (near-)ML performance, such as the well-known sphere decoding algorithm~\cite{1993_Viterbo_Biglieri,1999_Viterbo_Boutros,2000_Damen,2002_Agrell,2003_Hochwald,2005_Hassibi_Vikalo,2005_Jalden,2011_Seethaler}. Other tree-search schemes, such as the K-Best algorithm~\cite{2002_Wong,2006_Wenk_ISCAS,2010_Mondal,2010_Liu,2010_Shen_Eltawil,2012_Shabany,2013_Mahdavi}, address the non-deterministic throughput aspects of sphere decoders. Practical implementation aspects have been investigated in~\cite{2004_Garret,2004_Guo,2005_Jalden,2005_Burg,2008_Studer,2009_Yang_Markovic_TCASI,2009_Yang_Markovic_ESSCIRC,2010_Liu,2011_Borlenghi,2012_Shabany,2012_Liu_Lofgren,2012_Sun_Cavallaro,2013_Mahdavi,2013_Chen,2014_sphereP1_mansour,2014_sphereP2_mansour,2014_Huang}.

Subspace detection based on channel decomposition offers a good compromise between performance and complexity (e.g. see~\cite{2005a_Jiang,2005b_Jiang,2008_Ariyavisitakul,2011_Chen}). In these schemes, the effective MIMO channel matrix is decomposed into parallel subchannels that can be used to detect subsets of streams in parallel. By allowing subspaces to overlap, additional diversity can be gathered by putting a low reliable data stream into several detection sets. The LORD algorithm proposed in~\cite{2006_Fitz,2007a_Siti} can be viewed as a special class of subspace MIMO detectors. It achieves ML performance (in the max-log-MAP~\cite{2005_Yee_ICASSP} sense) on 2 transmit antennas, but its performance degrades when the number of antennas increases. In~\cite{2009_Ojard}, the LORD algorithm was generalized to 4-transmit antennas by using matrix inversion to decompose the channel into single streams.

Support for ever increasing data rates has come through an increase in the number of supported spatial streams, or through the use of more bandwidth via carrier aggregation~\cite{2012_Zhang}. LTE-Advanced uses up to 8 spatial streams, or the aggregation of five component carriers for a bandwidth of \unit[100]{MHz}, which lead to staggering speeds of over \unit[1]{Gbps}. While the receiver complexity to detect 8 spatial layers remains to be very challenging especially for dense constellations, the use of carrier aggregation with distinct or separate physical layers and convergence at higher layers seems more tractable. Since each physical layer of a component carrier is required to support 2 or 4 spatial layers, the need for the hardware optimization of these MIMO detector cores becomes paramount, especially if near-ML performance is desired, higher-order modulations such as 256-QAM are to be supported, and high-throughput processing is a must.

\emph{Contributions}: We propose in this work an optimized and configurable $2\!\times\!2$ soft-input soft-output maximum a posteriori (MAP) MIMO detector, and use it as a basic building block for constructing high-throughput detectors for higher-order layers. The key features and advantages of the proposed detector core are: 1) scalability in supporting multiple layers, 2) flexibility in accommodating multiple layer-configurations and detection of subsets of layers, 3) configurability of supported constellations per layer, 4) support for soft-input log-likelihood ratios (LLRs) from channel decoder, 5) near-ML performance, and 6) reduced-complexity and high-throughput operation. We develop extensive optimizations at both the algorithmic and architectural levels targeted for a $2\!\times\!2$ soft-input soft-output MAP MIMO detector, as well as its extension to support more spatial layers. In particular, optimizations of distance computations (to eliminate multipliers and simplify slicing) are shown to result in substantial reduction in computational complexity when supporting constellations up to 256-QAM. Furthermore, the complexity of a 1D slicer is shown to play a key role in the overall complexity of the detector, when soft-input LLRs are supported. To this end, an efficient slicing scheme based on \emph{soft} decision boundaries is presented. Moreover, a low-complexity scheme that decomposes a MIMO channel into multiple subsets of decoupled streams is proposed. It is shown that decoupled streams can be detected efficiently and in parallel using the optimized $2\!\times\!2$ core. Moreover, the $2\!\times\!2$ core is applied in the context of multi-user (MU-MIMO) for joint modulation classification and data detection. The core has been implemented on an FPGA, and synthesized as well using a generic \unit[90]{nm} ASIC CMOS library.

The rest of the paper is organized as follows. After introducing the system model in Section~\ref{s:system_model}, Section~\ref{s:2x2MAP} presents the optimizations targeted for a 2-layer MAP MIMO detector in terms of distance computations and slicing. Key equations for distances and soft decision boundaries are derived assuming both zero and non-zero input LLRs. Section~\ref{s:WLD} proposes a matrix decomposition scheme to support detection of more spatial streams. We show that the key distance equations scale in a straightforward fashion from the 2-layer case, where only a new distance-accumulation and a post-LLR processing phases are needed. In Section~\ref{s:complexity_par_arch}, single and multi-core detector architectures are developed. The core is applied in Section~\ref{s:MU_MIMO} part of MU-MIMO detection for constellation estimation and data detection. Synthesis and simulations results are reported in Section~\ref{s:sim}. Finally, Section~\ref{s:conclusion} ends with concluding remarks.

%
\vspace{-0.15in}
\section{System Model}\label{s:system_model}\vspace{-0.05in}
Consider a MIMO system with $N$ transmit and $M\!\geq\!N$ receive antennas. The equivalent complex baseband input-output system relation can be modeled as $\mbf{\tilde{y}}\!=\!\mbf{Hx}\!+\!\mbf{n}$, where $\mbf{\tilde{y}}\!\in\!\mathcal{C}^{M\times 1}$ is the received complex signal vector, $\mbf{H}\!\in\!\mathcal{C}^{M\times N}$ is the complex channel matrix, $\mbf{x}\!=\!\!\left[ x_1\ x_2\cdots x_N \right]^T\in\mathcal{X}\!=\!\mathcal{X}_1\!\times\cdots\times\!\mathcal{X}_N$ is the $N\!\times\! 1$ transmitted complex symbol vector, and $\mbf{n}\!\in\!\mathcal{C}^{M\times 1}$ is a zero-mean complex Gaussian circularly symmetric random noise vector with covariance $\sigma_{}^2\mbf{I}_M$. Each symbol $x_n$ belongs to a complex constellation $\mathcal{X}_n$ of size $Q_n\!=\!2^{q_n}$, and is associated via the map $\mbf{b}(\cdot)$ with a coded bit-interleaved vector $\mbf{b}(x_n)\!=\!\mbf{b}_n\!=\!\left[ b_{n,1}\ b_{n,2}\ \cdots b_{n,q_n} \right]^T$ of length $q_n$ over the set $\{-1,+1\}$, where binary 0 maps to $+1$. Let $\abs{\mathcal{X}}\!=\!Q\!=\!2^q$, and denote the binary vector associated with the overall symbol vector $\mbf{x}$ as $\mbf{b}(\mbf{x})\!=\![\mbf{b}_1;\cdots;\mbf{b}_{N}]\!=\![b_{n,j}]$, for $n\!=\!1,\cdots,N$, and $j\!=\!1,\cdots,q_n$. Motivated by recent standards, we assume rectangular QAM constellations, where $\mathcal{X}_n\!=\!\mathcal{P}_n\!\times\!\mathcal{P}_n$, and $\mathcal{P}_n$ is a 1D $P_n$-PAM constellation with $P_n\!=\!\sqrt{Q_n}$.

We assume $\mbf{H}$ is known to the receiver, has full column rank and is decomposed as $\mbf{H}\!=\!\mbf{QL}$, where $\mbf{Q}\!\in\!\mathcal{C}^{M\times N}$ is a unitary matrix and  $\mbf{L}\!\in\!\mathcal{C}^{N\times N}$ is a lower triangular matrix (LTM) with positive and real diagonal elements. Since $\mbf{Q}$ is unitary, it preserves Euclidean norm as well as noise statistics. Hence we use the transformed relation $\mbf{y}\!\triangleq\! \mbf{Q}^*\mbf{\tilde{y}}\!=\!\mbf{Lx}\!+\!\mbf{Q}^*\mbf{n}\!\in\!\mathcal{C}^{N\times 1}$ to model the MIMO system.

A hard-decision (HD) maximum \emph{a posteriori} (MAP) MIMO detector achieves log-max~\cite{2005_Yee_ICASSP} optimal performance by finding the symbol vector $\mbf{x}$ in $\mathcal{X}$ that is closest to the received vector $\mbf{y}$ under the unscaled ``distance" metric~\cite{2003_Hochwald}:
\begin{IEEEeqnarray}{rCl}
    d(\mbf{x}) &\triangleq& \norm{\mbf{y}-\mbf{Lx}}^2 - \mbf{b}^T\!(\mbf{x})\bs{\lambda}, \label{eq:dx_def}
\end{IEEEeqnarray}
where $\bs{\lambda}\!=\![\bs{\lambda}_1;\cdots;\bs{\lambda}_{N}]\!=\![\lambda_{n,j}]$ is a column vector of \emph{a priori} LLR values $\lambda_{n,j}\!\in\!\mathcal{R}$ associated with the bits in   $\mbf{b}(\mbf{x})$, assuming these bits are statistically independent:
\begin{equation}\label{eq:LLRin_def}
  \lambda_{n,j} = \frac{1}{\sigma_{}^2}\ln\frac{\textrm{Prob}\!\left(b_{n,j}=+1\right)}{\textrm{Prob}\!\left(b_{n,j}=-1\right)}.
\end{equation}
The subvector $\bs{\lambda}_n\!=\![\lambda_{n,1},\cdots,\lambda_{n,q_n}]^T$ is associated with the bits $\mbf{b}(x_n)$ of the $\nth{n}$ symbol $x_n$. The hard-decision MAP solution of the MIMO detection problem is given by\footnote{The quantities $d^{\MAP}$ in~\eqref{eq:dMAP_def} and $\Lambda_{n,j}^{\MAP}$ in~\eqref{eq:LLRMAP_def} need to be scaled by $\sigma_{}^2/2$.}
\begin{align}
    d^{\MAP} = \underset{\mbf{x}\in \mathcal{X}}{\mathop{\min }}\ {d(\mbf{x})}  ~~~\text{and}~~~
    \mbf{x}^{\MAP} =\underset{\mbf{x}\in \mathcal{X}}{\mathop{\arg\min }}\ {d(\mbf{x})}.\label{eq:dMAP_def}
\end{align}

For joint iterative MIMO detection and decoding however, soft-input soft-output MIMO detectors are required. A log-max optimal soft-input soft-output MAP MIMO detector computes $2q$ other minimum distance metrics as follows:
\begin{IEEEeqnarray}{rCl}
    \Lambda_{n,j}^{\MAP} \!&=&\!   \! \underset{\mbf{x}\in \mathcal{X}_{n,j}^{(+1)}}{\mathop{\min}}\!{d(\mbf{x})} - \! \underset{\mbf{x}\in \mathcal{X}_{n,j}^{(-1)}}{\mathop{\min }}\!{d(\mbf{x})}, \label{eq:LLRMAP_def}
\end{IEEEeqnarray}
for $n\!=\!1,\cdots,N$ and $j\!=\!1,\cdots,q_n$, where $\mathcal{X}_{n,j}^{(+1)}\!=\!\left\{\mbf{x}\!\in\! \mathcal{X}:b_{n,j}\!=\!+1 \right\}$ and $\mathcal{X}_{n,j}^{(-1)}\!=\!\left\{ \mbf{x}\!\in\! \mathcal{X}:b_{n,j}\!=\!-1 \right\}$ are the subsets of symbol vectors in $\mathcal{X}$ that have their corresponding $\nth{j}$ bit in the $\nth{n}$ symbol $+1$ and $-1$, respectively.

%
\section{Optimized MIMO MAP Detection for 2 Layers}\label{s:2x2MAP}
Finding the MAP solutions in~\eqref{eq:dMAP_def} and~\eqref{eq:LLRMAP_def} require computing $\prod_{n=1}^N\! Q_n$ distance metrics. When $N\!=\!2$, a simplification~\cite{2006_Fitz} can be applied to reduce the number of computations from $Q_1\!\cdot\!Q_2$ to $Q_1\!+\!Q_2$. Triangularizing the channel matrix as $\mbf{H}\!=\!\mbf{Q}\mbf{L}$ with $\mbf{Q}$ being unitary, we obtain:
\begin{align}
     \mbf{y}\!-\!\mbf{L}\mbf{x} \!=\!
    \begin{bmatrix}
      y_1 \\
      y_2
    \end{bmatrix}
    \!\!-\!\!
    \begin{bmatrix}
      \a & 0 \\
      \c & \b
    \end{bmatrix}\!\!
    \begin{bmatrix}
      x_1 \\
      x_2
    \end{bmatrix},\label{eq:QR1}
  \end{align}
where $\mbf{y}\!=\!\mbf{Q}^*\mbf{\tilde{y}}$, with $\a,\b\in\mathcal{R}^+$ and $\c\!\in\!\mathcal{C}$. Then~\eqref{eq:dx_def} becomes
\begin{align}
  d(\mbf{x}) &= f_1(x_1) + f_2(x_2\,|\,x_1),~\text{where}\label{eq:dx_f1_f2}\\
  f_1(x_1) &= \abs{y_1\!-\!\a x_1}^2 \!-\!\mbf{b}_{}^T\!(x_1)\bs{\lambda}_1^{},~\text{and}\label{eq:f1_x1}\\
  f_2(x_2\,|\,x_1) &= \abs{y_2\!-\!\c x_1\!-\!\b x_2}^2 \!-\!\mbf{b}_{}^T\!(x_2)\bs{\lambda}_2^{}.\label{eq:f2_x2_given_x1}
\end{align}
The minimum distance in~\eqref{eq:dMAP_def} can then be computed as
\begin{IEEEeqnarray}{rCl}\hspace{-0.15in}
\underset{\mbf{x}\in \mathcal{X}}{\mathop{\min}}{\,d(\mbf{x})} \!&=&\!
\underset{\substack{x_1\in \mathcal{X}_1\\ x_2\in \mathcal{X}_2}}{\mathop{\min}}\!\!\left\{
        f_1(x_1) + f_2(x_2\,|\,x_1)\right\}\label{eq:min_dx_f1_f2}\\[-0.45em]
        \!&=&\!
\underset{x_1\in \mathcal{X}_1}{\mathop{\min}}\!\!\left\{
        f_1(x_1) +\underset{x_2\in \mathcal{X}_2}{\mathop{\min}}\! f_2(x_2\,|\,x_1)\right\}\notag\\
        \!&=&\!
\underset{x_1\in \mathcal{X}_1}{\mathop{\min}}\!\!\left\{
        f_1(x_1) + f_2(\hat{x}_2(x_1)\,|\,x_1)\right\}\notag\\[-0.35em]
        \!&=&\!
\underset{x_1\in \mathcal{X}_1}{\mathop{\min}}
        d(x_1,\hat{x}_2(x_1))\label{eq:min_dx1}
\end{IEEEeqnarray}
where\vspace{-0.05in}
\begin{equation}\label{eq:x2hat_prob}
  \hat{x}_2(x_1) = \underset{x_2\in \mathcal{X}_2}{\mathop{\arg\min}}\! \left\{\abs{y_2\!-\!\c x_1\!-\!\b x_2}^2 \!-\!\mbf{b}_{}^T\!(x_2)\bs{\lambda}_2^{}\right\}.
\end{equation}
Denote the set of sliced symbol vectors for all $x_1$ in~\eqref{eq:min_dx1} by
\begin{equation}\label{eq:O_1}
  \mathcal{O}_1 = \left\{ \left[x_1\ \hat{x}_2(x_1)\right]^T\,:\, x_1\!\in\!\mathcal{X}_1 \right\}.
\end{equation}
The bit LLRs of symbol $x_1$, for $j\!=\!1,\cdots,q_1$, are given by
\begin{IEEEeqnarray}{rCl}\label{eq:LLR_x_1_j}\hspace{-0.1in}
    \Lambda_{1,j}^{\MAP} \!&=&\!
        \!\!\!\underset{x_1^{}\in \mathcal{X}_{1,j}^{(+1)}}{\mathop{\min}}{\!d(x_1,\hat{x}_2(x_1))} -\!\!\!\!\!
        \underset{x_1^{}\in \mathcal{X}_{1,j}^{(-1)}}{\mathop{\min}}{\!d(x_1,\hat{x}_2(x_1))}.
\end{IEEEeqnarray}

To obtain the bit LLRs of $x_2$ however, we triangularize $\mbf{H}$ as $\mbf{Q}'\, \mbf{L}'$ so that a zero appears in the upper left corner:
\begin{align}
     \mbf{y}'\!-\!\mbf{L}'\mbf{x} =
    \begin{bmatrix}
      y_{1}' \\
      y_{2}'
    \end{bmatrix}
    \!\!-\!\!
    \begin{bmatrix}
      0     & \a' \\
      \b' & \c'
    \end{bmatrix}
    \!\!
    \begin{bmatrix}
      x_{1} \\
      x_{2}
    \end{bmatrix},\label{eq:QR2}
  \end{align}
where $\mbf{y}'\!=\!{\mbf{Q}'}^*\mbf{\tilde{y}}$; $\a',\b'\in\mathcal{R}^+$ and $\c'\in\mathcal{C}$. Then~\eqref{eq:dx_def} becomes
\begin{align*}
  d(\mbf{x}) &= f_{2}'(x_2) + f_{1}'(x_1\,|\,x_2),~\text{where}\\
  f_{2}'(x_2) &= \abs{y_{1}'\!-\!\a'x_2}^2 \!-\!\mbf{b}_{}^T\!(x_2)\bs{\lambda}_2^{},~\text{and}\\
  f_{1}'(x_1\,|\,x_2) &= \abs{y_{2}'\!-\!\c'x_2\!-\!\b'x_1}^2 \!-\!\mbf{b}_{}^T\!(x_1)\bs{\lambda}_1^{},
\end{align*}
and the minimum distance in~\eqref{eq:dMAP_def} can be computed as
\begin{IEEEeqnarray}{rCl}\hspace{-0.15in}
\underset{\mbf{x}\in \mathcal{X}}{\mathop{\min}}{\,d(\mbf{x})}
        \!&=&\!
\underset{x_2\in \mathcal{X}_2}{\mathop{\min}}\!\!\left\{
        f_{2}'(x_2) +\underset{x_1\in \mathcal{X}_1}{\mathop{\min}} f_{1}'(x_1\,|\,x_2)\right\}\notag\\
        \!&=&\!
\underset{x_2\in \mathcal{X}_2}{\mathop{\min}}\!\!\left\{
        f_{2}'(x_2) + f_{1}'(\hat{x}_1\,|\,x_2)\right\}\notag\\
        \!&=&\!
\underset{x_2\in \mathcal{X}_2}{\mathop{\min}}
        d(\hat{x}_1(x_2),x_2),\label{eq:min_dx2}
\end{IEEEeqnarray}
where $\hat{x}_1(x_2) \!=\! \underset{x_1\in \mathcal{X}_1}{\mathop{\arg\min}}\, f_{1}'(x_1\,|\,x_2)$. Denote the set of sliced symbol vectors for all $x_2$ in~\eqref{eq:min_dx2} by
\begin{equation}\label{eq:O_2}
  \mathcal{O}_2 = \left\{ \left[\hat{x}_1(x_2)\ x_2\right]^T\,:\, x_2\!\in\!\mathcal{X}_2 \right\}.
\end{equation}
The bit LLRs of symbol $x_2$, for $j\!=\!1,\cdots,q_2$, are given by
\begin{IEEEeqnarray}{rCl}\label{eq:LLR_x_2_j}\hspace{-0.15in}
    \Lambda_{2,j}^{\MAP} \!&=&\!
        \!\!\!\underset{x_2^{}\in \mathcal{X}_{2,j}^{(+1)}}{\mathop{\min}}{\!d(\hat{x}_1(x_2),x_2)} -\!\!\!\!
        \underset{x_2^{}\in \mathcal{X}_{2,j}^{(-1)}}{\mathop{\min}}{\!d(\hat{x}_1(x_2),x_2)}.
\end{IEEEeqnarray}

Since $\mbf{Q}$ and $\mbf{Q}'$ are unitary, the MAP solutions in~\eqref{eq:min_dx1} and~\eqref{eq:min_dx2} are identical. To find the hard-decision (HD)-MAP solution, only 1-sided QLD is needed on either layer 1 or 2. If $Q_1\!\leq\!Q_2$, a list of $Q_1$ distances $\mathcal{D}_1\!=\!\left\{d(x_1,\hat{x}_2(x_1))\,:\,  x_1\!\in\!\mathcal{X}_1\right\}$ is generated by enumerating all symbols $x_1\!\in\!\mathcal{X}_1$ and the minimum is selected. If $Q_2\!<\!Q_1$, a list of $Q_2$ distances $\mathcal{D}_2\!=\!\left\{d(\hat{x}_1(x_2),x_2)\,:\,  x_2\!\in\!\mathcal{X}_2\right\}$ is generated and the minimum is selected. However, to generate soft LLRs, 2-sided QLDs are needed, and both lists of distances must be generated to select the appropriate minima according to~\eqref{eq:LLR_x_1_j} and~\eqref{eq:LLR_x_2_j}.
\begin{figure}[t]
\centering
\includegraphics[scale=1]{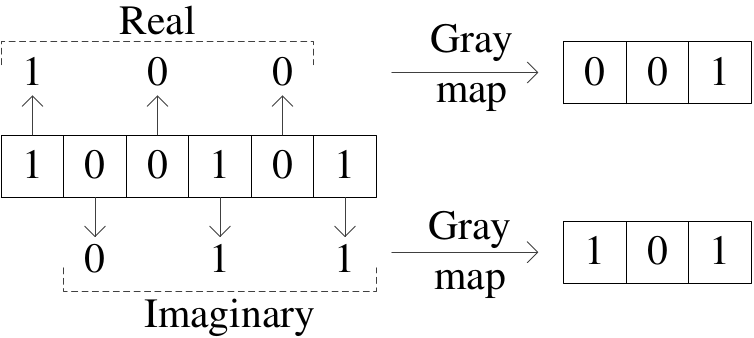}\vspace{-0.05in}
\caption{Gray-coded mapping for 64-QAM in LTE~\cite{LTE_36.211}.}
\label{f:gray_mappingLTE}
\end{figure}

\vspace{-0.15in}
%
\subsection{Distance Metric Optimizations}\label{s:dist_opt_2x2}
For efficient distance computations, we separate the real and imaginary parts of all complex variables, and exploit the fact that the real and imaginary parts of each QAM symbol are mapped independently into 1D PAM symbols, i.e., some bits are used only for mapping of the real part and some only for the imaginary part (see Fig.~\ref{f:gray_mappingLTE}). Note that this mapping is used in, e.g., the IEEE 802.11ac~\cite{802.11ac} and LTE~\cite{LTE_36.211} standards. Under this assumption, we can split the bias term $\mbf{b}_{}^T\!(x_n)\bs{\lambda}_n^{}$ into a part $\mbf{b}_{n\re}^T\!\bs{\lambda}_{n\re}^{}\!\triangleq\!\mbf{b}_{}^T\!(x_{n\re})\bs{\lambda}_{n\re}^{}$ associated with the bits of the real part of the QAM symbol, and a part $\mbf{b}_{n\im}^T\!\bs{\lambda}_{n\im}^{}\!\triangleq\!\mbf{b}_{}^T\!(x_{n\im})\bs{\lambda}_{n\im}^{}$ associated with the bits of the imaginary part. Let $\c\!=\!\c_{\re}\!+\!j\c_{\im}$, $x_n\!=\!x_{n\re}\!+\!jx_{n\im}$, $y_n\!=\!y_{n\re}\!+\!jy_{n\im}$ for $n\!=\!1,2$. Then the distance in~\eqref{eq:dx_f1_f2} becomes
\begin{align}\label{eq:dx_f1R_f1I_f2R_f2I}\hspace{-0.05in}
  d(\mbf{x}) &\!=\!f_{1\re}(x_{1\re}) \!+\! f_{1\im}(x_{1\im})
        \!+\! f_{2\re}(x_{2\re}|x_1) \!+\! f_{2\im}(x_{2\im}|x_1),
\end{align}
where the terms on the righthand side are given by
\begin{align*}
  f_{1\re}(x_{1\re}) &\!=\!( y_{1\re}\!-\!\a x_{1\re})^2\!-\!\mathbf{b}_{1\re}^{T}\mathbf{\lambda}_{1\re}^{} \notag\\
  f_{1\im}(x_{1\im}) &\!=\!(y_{1\im}\!-\!\a x_{1\im})^2 \!-\!\mbf{b}_{1\im}^{T}\bs{\lambda}_{1\im}^{}\notag\\
  f_{2\re}(x_{2\re}|x_1) &\!=\! ( y_{2\re}\!-\! \c_{\re} x_{1\re}\!+\! \c_{\im} x_{1\im} \!-\! \b x_{2\re})^2 \!-\!\mbf{b}_{2\re}^{T}\bs{\lambda}_{2\re}^{}\\
  f_{2\im}(x_{2\im}|x_1) &\!=\!( y_{2\im} \!-\!\c_{\re} x_{1\im} \!-\! \c_{\im} x_{1\re} \!-\!\b x_{2\im})^2\!-\!\mbf{b}_{2\im}^{T}\bs{\lambda}_{2\im}^{}.
\end{align*}
Expanding~\eqref{eq:dx_f1R_f1I_f2R_f2I}, minimizing with respect to $x_{2\re}$ and $x_{2\im}$, and removing irrelevant terms, we obtain the following key equation:
\begin{align} \hspace{-0.1in}
  \bar{d}(\mbf{x}) &\!=\!\bar{f}_{1\re}(x_{1\re}) \!+\! \bar{f}_{1\im}(x_{1\im}) \notag\\
        &\!+\! \underset{x_{2\re}\in \mathcal{P}_2}{\mathop{\min}}\! \bar{f}_{2\re}(x_{2\re}\,|\,x_1) \!+\! \underset{x_{2\im}\in \mathcal{P}_2}{\mathop{\min}}\!\bar{f}_{2\im}(x_{2\im}\,|\,x_1),\label{eq:dbar_x}
\end{align}
where $\mathcal{P}_2$ is the 1D PAM constellation in $\mathcal{X}_2$ of layer 2, and
\begin{align}
  \bar{f}_{1\re}(x_{1\re}) &= A x_{1\re}^2 \!+\! Cx_{1\re} \!-\!\mbf{b}_{1\re}^T \bs{\lambda}_{1\re}^{}\label{eq:f1Rbar_2x2}\\
  \bar{f}_{1\im}(x_{1\im}) &= A x_{1\im}^2 \!+\! Dx_{1\im} \!-\!\mbf{b}_{1\im}^T \bs{\lambda}_{1\im}^{}\label{eq:f1Ibar_2x2}\\
  \bar{f}_{2\re}(x_{2\re}\,|\,x_1) &= (Ex_{1\re} \!+\! Fx_{1\im})x_{2\re} \notag\\
                             & + (Bx_{2\re}^2 \!+\! Gx_{2\re} \!-\! \mbf{b}_{2\re}^T\bs{\lambda}_{2\re}^{})\label{eq:f2Rbar_2x2}\\
  \bar{f}_{2\im}(x_{2\im}\,|\,x_1) &= (Ex_{1\im} \!-\! Fx_{1\re})x_{2\im} \notag\\
                              &+ (Bx_{2\im}^2\!+\! Hx_{2\im} \!-\! \mbf{b}_{2\im}^T\bs{\lambda}_{2\im}^{}).\label{eq:f2Ibar_2x2}
\end{align}
The constant coefficients in~\eqref{eq:f1Rbar_2x2}-\eqref{eq:f2Ibar_2x2} are given by
\begin{align}
  A &= \a^2 + \abs{\c}^2, \quad
  B = \b^2 \label{eq:constant_A_B_2x2},\\[-0.3em]
  C &= -2\left( \a y_{1\re} + \c_{\re}y_{2\re} + \c_{\im}y_{2\im} \right) \label{eq:constant_C_2x2},\\[-0.3em]
  D &= -2\left( \a y_{1\im} - \c_{\im}y_{2\re} + \c_{\re}y_{2\im} \right) \label{eq:constant_D_2x2},\\[-0.3em]
  E &= +2\b \c_{\re},~
  F = -2\b\c_{\im},~
  G = -2\b y_{2\re},~
  H = -2\b y_{2\im},~~ \label{eq:constant_E_F_G_H_2x2}
\end{align}
and can be precomputed off-line from $\mbf{H}$ and $\mbf{y}$. The HD-MAP solution is obtained by populating all $Q_1$ distances in~\eqref{eq:dbar_x} and selecting the minimum. The same applies for the LLRs.

%
\subsection{Slicing Assuming \textbf{Zero} Prior LLRs}\label{s:2x2ML}
Assuming the input LLRs $\bs{\lambda}$ are zero, the rightmost term in~\eqref{eq:dx_def} vanishes and the MAP detection problem reduces to a least-squares integer ML problem. Then
$\hat{x}_2$ in~\eqref{eq:x2hat_prob} can be obtained by slicing $(y_2-\c x_1)/\b\!\in\! \mathcal{C}$ to the nearest constellation point in $\mathcal{X}_2$ using the operator $\slice{u}{\mathcal{X}_n}\triangleq \underset{x\in\mathcal{X}_n}{\arg\min} \abs{u-x}$:
\begin{IEEEeqnarray}{rCl}
\label{eq:x2hat_determ}
    \hat{x}_2 &=& \slice{(y_2\!-\!\c x_1)/\b}{\mathcal{X}_2} \in \mathcal{X}_2.
\end{IEEEeqnarray}
By separating the real and imaginary parts as $\hat{x}_2\!=\!\hat{x}_{2\re}\!+\!j\hat{x}_{2\im}$, the slicing operation in~\eqref{eq:x2hat_determ} splits into:
\begin{align}
  \hat{x}_{2\re} &= \slice{(y_{2\re}\!-\! \c_{\re} x_{1\re}\!+\! \c_{\im} x_{1\im})/\b}{\mathcal{P}_2} \in \mathcal{P}_2 \label{eq:x2hatR_determ},\\
  \hat{x}_{2\im} &= \slice{(y_{2\im} \!-\!\c_{\re} x_{1\im} \!-\! \c_{\im} x_{1\re})/\b}{\mathcal{P}_2} \in \mathcal{P}_2 \label{eq:x2hatI_determ},
\end{align}
where $\mathcal{P}_2\!=\!\{p_1,p_2,\cdots,p_{P_2}\}$ is the $P_2$-PAM constellation, and $P_2\!=\!\sqrt{Q_2}$. The operations in~\eqref{eq:x2hatR_determ}-\eqref{eq:x2hatI_determ} reduce to simple comparisons with the (deterministic) decision boundaries of $\mathcal{P}_2$ as follows. Let $z_2 \!=\!y_2\!-\!\c x_1\!=\!z_{2\re}\!+\!jz_{2\im}$ where
\begin{align}
  z_{2\re} &\!=\! y_{2\re}\!-\! \c_{\re} x_{1\re}\!+\! \c_{\im} x_{1\im},\label{eq:z2R}\\
  z_{2\im} &\!=\! y_{2\im} \!-\!\c_{\re} x_{1\im} \!-\! \c_{\im} x_{1\re}.\label{eq:z2I}
\end{align}
Assume the constellation points are ordered such that $p_i\!<\!p_k$ if $i\!<\!k$. Then $\hat{x}_{2\re}$ maps to the point $p_i$ that satisfies
\begin{equation}\label{eq:decbound_determ_R}
   \b\,\frac{p_{i-1}+p_i}{2}  \leq z_{2\re} < \b\,\frac{p_{i}+p_{i+1}}{2}
\end{equation}
for $i\!=\!1,\cdots,P_2$, where $p_0\!=\!-\infty$ and $p_{P_2+1}\!=\!+\infty$. Similarly for $\hat{x}_{2\im}$. Hence the actual distances $f_2(x_2|x_1)$ themselves \emph{need not be computed} for all $x_2$ and a given $x_1$ in order to find the symbol $x_2$ that minimizes $f_2(x_2|x_1)$ in~\eqref{eq:x2hat_prob}. Therefore,~\eqref{eq:min_dx1} requires only $\abs{\mathcal{X}_1}\!=\!Q_1$ distance computations. By the same argument,~\eqref{eq:min_dx2} requires only $\abs{\mathcal{X}_2}\!=\!Q_2$ distance computations.

\vspace{-0.1in}

%
\subsection{Slicing Assuming \textbf{Non-Zero} Prior LLRs}\label{s:nonzero_priors}
When the prior terms are included in the distance computations, slicing cannot be directly applied in~\eqref{eq:x2hat_prob} since the decision boundaries now depend on the bias term $\mbf{b}_{}^T\!(x_2)\bs{\lambda}_2^{}$. We develop next an optimal scheme that enables efficient slicing similar to~\eqref{eq:decbound_determ_R} based on~\cite{2014_b_Gomma}. In~\cite{2012_Yeung}, a scheme that computes suboptimal slicing boundaries was presented. Compared to our approach,~\cite{2012_Yeung} incurs a performance loss with equivalent complexity.

The real part of $\hat{x}_2$ in~\eqref{eq:x2hat_prob} is given by
\begin{align}\label{eq:x2hatR_prob}
  \hat{x}_{2\re} &\!=\! \underset{x_{2\re}\in \mathcal{P}_2}{\mathop{\arg\min}}\left\{ (z_{2\re}\!-\!\b x_{2\re})^2
       \!-\!\mbf{b}_{}^T\!(x_{2\re})\bs{\lambda}_{2\re}^{}\right\}.
\end{align}
To decide in favor of $p_i\!\in\!\mathcal{P}_2$, then $\forall\,k\!\neq\! i$, we must have
\begin{equation}\label{eq:dec_bound_2R}
  (z_{2\re}\!-\!\b p_i)^2
       \!-\!\mbf{b}_{}^T\!(p_i)\bs{\lambda}_{2\re}^{} <
       (z_{2\re}\!-\!\b p_k)^2
       \!-\!\mbf{b}_{}^T\!(p_k)\bs{\lambda}_{2\re}^{}.
\end{equation}
This condition can be formulated in terms of \emph{decision boundaries} $R(p_i,p_k)\!=\!R(p_k,p_i)$:
\begin{equation}\label{eq:decision_bound_R}
  R(p_i,p_k) = B\!\cdot\!(p_i+p_k) - \frac{\mbf{b}_{}^T\!(p_i)\!-\!\mbf{b}_{}^T\!(p_k)}{p_i-p_k}\bs{\lambda}_{2\re}^{}, ~\forall\,k\!\neq\! i,
\end{equation}
between $p_i$ and all other $p_k$'s in $\mathcal{P}_2$. Assuming the points in $\mathcal{P}_2$ are ordered such that $p_i\!<\!p_k$ if $i\!<\!k$, then for $p_1$ to satisfy~\eqref{eq:dec_bound_2R},
we must have $2\b z_{2\re}\!<\!R(p_1,p_k)$ for all $p_k\!>\!p_1$. For $p_{P_2}$ to satisfy~\eqref{eq:dec_bound_2R}, we must have $2\b z_{2\re}\!>\!R(p_{P_2},p_k)$ for all $p_k\!<\!p_{P_2}$. For any other internal point $p_i$, $i\!\neq\!1,i\!\neq\!P_2$, we must have $2\b z_{2\re}\!<\!R(p_i,p_k)$ for all $p_k\!>\!p_i$, \emph{and} $2\b z_{2\re}\!>\!R(p_i,p_k)$ for all $p_k\!<\!p_i$. These conditions can be combined into a single condition for $i\!=\!1,\cdots,P_2$, as follows:
\begin{equation}\label{eq:minmax_decbound_R}
   \underset{k=0,\cdots,i-1}{\mathop{\max}} R(p_i,p_k) \leq 2\b z_{2\re} < \underset{k=i+1,\cdots,P_2+1}{\mathop{\min}} R(p_i,p_k),
\end{equation}
where $p_0\!=\!-\infty$, $p_{P_2+1}\!=\!+\infty$, $\mbf{b}_{}(p_0)\!=\!\mbf{b}_{}(p_{P_2+1})\!=\!\mbf{0}_{1 \times q_2/2}$. Note that~\eqref{eq:decision_bound_R} and~\eqref{eq:minmax_decbound_R} reduce to~\eqref{eq:decbound_determ_R} when $\bs{\lambda}_{2\re}^{}\!=\!\mbf{0}_{1 \times q_2/2}$.

Substituting~\eqref{eq:z2R} for $z_{2\re}$ in~\eqref{eq:minmax_decbound_R}, using the constants~\eqref{eq:constant_E_F_G_H_2x2}-\eqref{eq:constant_E_F_G_H_2x2}, and accounting for sign change, we obtain the following slicing condition that is suitable for hardware implementation:
\begin{align}
   \underset{k=i+1,\cdots,P_2+1}{\mathop{\max}}  R(p_i,p_k) - G
            \leq Ex_{1\re} + Fx_{1\im}
            < \underset{k=0,\cdots,i-1}{\mathop{\min}}  R(p_i,p_k) - G.\label{eq:minmax_decbound_simplified_R}
\end{align}
Note that in~\eqref{eq:minmax_decbound_simplified_R}, the maximum on the lefthand side is now taken over all points $p_k\!\in\!\mathcal{P}$ that are \emph{greater} than $p_i$ as opposed to \emph{smaller} than $p_i$ as was done in~\eqref{eq:minmax_decbound_R} due to the change in sign. Similarly for the minimum on the righthand side in~\eqref{eq:minmax_decbound_simplified_R}.

A similar analysis applied to compute $\hat{x}_{2\im}\!=\!\underset{x_{2\im}\in \mathcal{P}_2}{\mathop{\min}} f_{2\im}(x_{2\im}|x_1)$ leads to the decision regions $I(p_i,p_k)$:
\begin{equation}\label{eq:decision_bound_I}
  I(p_i,p_k) = B\!\cdot\!(p_i+p_k) - \frac{\mbf{b}_{}^T\!(p_i)\!-\!\mbf{b}_{}^T\!(p_k)}{p_i-p_k}\bs{\lambda}_{2\im}^{},
\end{equation}
using now $\bs{\lambda}_{2\im}^{}$, and the associated slicing condition:
\begin{align}
   \underset{k=i+1,\cdots,P_2+1}{\mathop{\max}} I(p_i,p_k) - H
            \leq  Ex_{1\im} - Fx_{1\re}
            <  \underset{k=0,\cdots,i-1}{\mathop{\min}} I(p_i,p_k) - H. \label{eq:minmax_decbound_simplified_I}
\end{align}

Note that by construction of the decision boundaries in~\eqref{eq:minmax_decbound_R} (and their imaginary counterparts), the proposed approach is \emph{optimal}. The approach in~\cite{2012_Yeung} however is suboptimal because it employs heuristics to compute simplified but suboptimal decision boundaries.

\vspace{-0.1in}
%
\section{Extension to Higher-Order Layers}
\begin{figure}[t]
\centering
\includegraphics[scale=1.5]{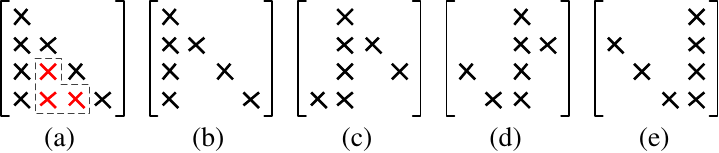}\vspace{-0.15in}
\caption{$4\!\times\!4$ structures: (a) Full; (b)-(e) punctured structures for every layer.}
\label{f:matrices}
\end{figure}

\vspace{-0.00in}\label{s:WLD}
The previous optimizations cannot be directly extended to $N\!\geq\!3$ layers because the structure of the lower triangular matrix $\mbf{L}$ includes off-diagonal terms that prevent searching for the MAP solution by enumerating symbols in one layer and finding the minima through slicing individually on all other layers in parallel. More specifically, in Fig.~\ref{f:matrices}(a), the presence of the demarked entries in the LTM implies that determining the MAP solution requires enumerating symbols on the first $N\!-\!1$ layers and slicing only on the last layer, as is typically done in tree-search detectors (e.g.~\cite{2008_Studer}), and hence still requiring $\bigO{\prod_{n}{Q_n}}$ complexity rather than $\bigO{\sum_n Q_n}$.

One desirable structure of $\mbf{H}$ for a 4-layer MIMO system would be as shown in Fig.~\ref{f:matrices}(b), in which the demarked entries are zeroed out. Here, by enumerating symbols on layer 1, the minimum distances and associated symbols on layers 2 to 4 can be searched for in parallel through slicing only on the corresponding layers, similar to the 2-layer system. This suffices to compute the LLRs associated with the bits on layer-1 symbol. A similar process is repeated by decomposing $\mxH$ according to the structures shown in Figs.~\ref{f:matrices}(c)-(e)~\cite{2009_Ojard} to compute the LLRs for bits associated with layers 2 to 4.

\begin{figure}[hbtp]
\centering
\includegraphics[scale=1.5]{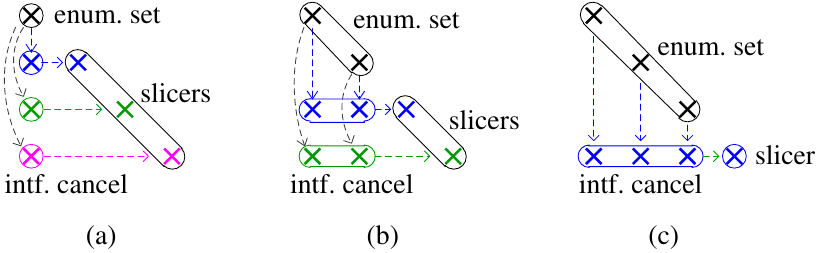}\vspace{-0.15in}
\caption{(a) $(1,3\!\times\!1,3)$, (b) $(2,2\!\times\!2,2)$, (c) $(3,1\!\times\!3,1)$ punctured structures}
\label{f:dataflow_matrices}
\end{figure}

Other ``punctured" structures are also possible for a $4\!\times\!4$ system as shown in Fig.~\ref{f:dataflow_matrices}. They differ in 1) the number of layers over which symbols are enumerated (\emph{enumeration or detection set}), 2) the submatrix structure used to propagate these enumerated symbols and cancel their interference effect from the remaining layers (\emph{interference cancellation set}), and 3) the number of layers in which the minimum distance and associated symbol can be obtained by slicing after interference cancelation (\emph{slicer set}). Let $U$ denote the size of the enumeration set, $S$ the size of the slicer set, and $S \!\times\! U$ the size of the interference cancellation set. We refer to this structure using the triplet $\left(U,S\!\times\!U,S\right)$. For example, in Fig.~\ref{f:dataflow_matrices}(a), we enumerate over $U\!=\!1$ layer only, cancel interference from this layer to the 3 other layers using a $3\!\times\! 1$ structure, and slice over $S\!=\!3$ layers. In the structure in Fig.~\ref{f:dataflow_matrices}(b), we enumerate over $U\!=\!2$ layers, cancel interference using a $2\!\times\! 2$ structure, and slice over $S\!=\!2$ layers.

LLR values are generated for bits in symbols included in the detection set only. Complementary structures that enumerate symbols on other decoupled layers are required to generate their respective LLRs. For example, the $(1,3\!\times\!1,3)$ structure requires 3 similar structures to generate LLRs for layers 2 to 4 (Fig.~\ref{f:matrices}(c)-(e)). When $U\!>\!1$, decoupled layers can overlap by placing a stream with low reliability in multiple detection sets.

%
\vspace{-0.1in}
\subsection{WL Decomposition (WLD) Scheme}\vspace{-0.05in}\label{s:proposed_comp_decomp}
In~\cite{2014_mansour_SPL_WLD}, a decomposition scheme was introduced to transform $\mbf{H}$ into a punctured LTM $\mbf{L}$ with a desired structure via a projection matrix $\mbf{W}$. In this section, we extend the scheme to handle soft-input MIMO detection using prior LLRs fed from a soft-input-soft-output channel decoder. We assume $N\!=\!M$.

We seek a matrix $\mbf{W}\!=\!\left[\mbf{w}_1~\mbf{w}_2~\cdots~\mbf{w}_N \right]\!\in\!\mathcal{C}^{N\times N}$ such that $\mbf{W}^{*}\mbf{H} \!=\! \mbf{L}$ is a punctured LTM and $\mbf{L}\!=\![l_{ij}]\!\in\!\mathcal{C}^{N\times N}$ with $l_{ii}\!\in\!\mathcal{R}^+$. In general, if $\mbf{L}$ is punctured, then $\mbf{W}$ is non-unitary and hence does not preserve Euclidean norm:
\begin{align}
  \mbf{y}\triangleq\mbf{W}^*\widetilde{\mbf{y}} &= \mbf{L}\mbf{x} + \mbf{W}^*\mbf{n} \label{eq:transformed_IO_relation}\\
  g(\mbf{x})\!\triangleq\!\norm{\mbf{y}-\mbf{Lx}}^2 &\neq \norm{\widetilde{\mbf{y}}-\mbf{Hx}}^2\!=\!d(\mbf{x})\label{eq:norms_inequality}
\end{align}
However, if we impose the condition
\begin{align}\label{eq:unit_length_cond}
  \textrm{diag}{\left(\mbf{W}_{}^*\mbf{W}\right)} = [1~1~\cdots~1]_{1\times N}^T,
\end{align}
then the transformed noise vector $\mbf{W}^{*}\mbf{n}$ has an unaltered covariance matrix $\E{\mbf{W}^{*}\mbf{n}\mbf{n}^*\mbf{W}}\!=\!\sigma_{}^2\mbf{I}_N$.

To induce a specific pattern of zeros below the main diagonal in $\mbf{L}$, we choose the columns of $\mbf{W}$ to be orthogonal to the columns of $\mbf{H}\!=\!\left[\mbf{h}_1~\mbf{h}_2~\cdots~\mbf{h}_N \right]$ where these zeros are to be introduced. More specifically, let $\mathcal{I}_n, n\!=\!1,\cdots,N$, be the column index sets where puncturing is desired in each row $n$ of $\mbf{H}$. Denote $\mbf{H}_{\mathcal{I}_n}$ the submatrix formed by the columns of $\mbf{H}$ whose index belongs to set $\mathcal{I}_n$. Define the column vector $\widetilde{\mbf{w}}_n \!=\! \mbf{P}_{\mathcal{I}_n}^{\perp} \mbf{h}_n^{}$, where
\begin{align}\label{eq:proj_Pn}
    \mbf{P}_{\mathcal{I}_n}^{\perp} &=  \mbf{I}_N-\mbf{H}_{\mathcal{I}_n}^{}\!\! \left(\mbf{H}_{\mathcal{I}_n}^{*}\mbf{H}_{\mathcal{I}_n}^{}\!\right)^{-1} \!\mbf{H}_{\mathcal{I}_n}^{*},
\end{align}
and $\mbf{H}_{\mathcal{I}_n} \!=\! \left\{ \mbf{h}_m \left. \right| m \in \mathcal{I}_n \right\}$. Then the column vectors of $\mbf{W}$ are given by
\begin{equation*}
  \mbf{w}_n\!=\!\frac{\widetilde{\mbf{w}}_n}{\norm{\widetilde{\mbf{w}}_n}}\!=\!
  \frac{\mbf{P}_{\mathcal{I}_n}^{\perp} \mbf{h}_n^{}}{\sqrt{\mbf{h}_n^{*} \mbf{P}_{\mathcal{I}_n}^{\perp} \mbf{h}_n^{}}}.
\end{equation*}

Furthermore, it was shown in~\cite{2014_mansour_SPL_WLD} that $\mbf{L}$ and $\mbf{W}^*\widetilde{\mbf{y}}$ can be derived using a simple modification to the standard QL decomposition procedure~\cite{1996_Golub}. This avoids the need for expensive matrix inversion operations in~\eqref{eq:proj_Pn}. On modern vector digital signal processors (DSPs), matrix QLD operations are natively supported and optimized part of the instruction set. For example, on a CEVA XC-4210 processor~\cite{CEVA}, QL decomposition of a $4\!\times\!4$ complex matrix requires only 12 clock cycles. Hence, we assume that the channel matrix $\mbf{H}$ has been preprocessed by a similar DSP, and detection is performed based on the transformed system in~\eqref{eq:transformed_IO_relation}. Note that because of~\eqref{eq:norms_inequality}, the solution to the detection problem is no longer optimal (but still achieves near-optimal performance as demonstrated in Section~\ref{s:sim}).

%
\vspace{-0.1in}
\subsection{Optimized Detection Algorithm Using WLD}\vspace{-0.05in}\label{s:opt_det_algo}
We next present an optimized detection algorithm based on the WLD scheme, by extending the $N\!=\!2$ case of Section~\ref{s:2x2MAP}. For simplicity, we only consider decompositions of the form $(1,S\!\times\! 1,S)$, similar to Fig.~\ref{f:matrices}. The $N$ layers are decoupled by first circularly shifting the columns of $\mbf{H}$, and then performing WLD on the permuted $\mbf{H}$. We refer to the decomposition whose detection set is the $\nth{m}$ layer as the $\nth{m}$ WLD of $\mbf{H}$. To simplify notation, we describe the detection steps for $m\!=\!1$. The same steps apply to detect the other layers with an appropriate adjustment to the layer indices. Let
\begin{align}\label{eq:partitioned_L}\hspace{-0.15in}
\arraycolsep=2.55pt
\mbf{x}\!=\! \left[
               \begin{array}{c}
                 x_1 \\
                 x_2 \\
                 x_3 \\
                 \vdots \\
                 x_N \\
               \end{array}
             \right]\!,\,
\mbf{y}\!=\! \left[
               \begin{array}{c}
                 y_1 \\
                 y_2 \\
                 y_3 \\
                 \vdots \\
                 y_N \\
               \end{array}
             \right]\!,\,
\mbf{L}\!=\! \left[
               \begin{array}{ccccc}
                 \a & 0 & 0 & 0 & 0 \\
                 \c_2 & \b_2 & 0 & 0 & 0 \\
                 \c_3 & 0 & \b_3 & 0 & 0 \\
                 \vdots & \vdots & \vdots & \ddots & \vdots\\
                 \c_N & 0 & \cdots & 0 & \b_N
               \end{array}
             \right],
\end{align}
be the transmitted symbol vector, received signal vector, and the WL-decomposed channel matrix in normal order, respectively, where: $y_n\!\in\!\mathcal{C}$, $x_n\!\in\!\mathcal{X}_n$ for $n\!=\!1,\cdots,N$; $\a,\b_n\!\in\!\mathcal{R}^+$ and $\c_n\!\in\!\mathcal{C}$ for $n\!=\!2,\cdots,N$. Then the distance metric $g(\mbf{x})$ of $\mbf{x}$ from $\mbf{y}$ based on $\mbf{L}$ in~\eqref{eq:norms_inequality} can be written as\vspace{-0.05in}
\begin{align}
  g(\mbf{x}) &\!=\! f_1(x_1) \!+\! \sum_{n=2}^N f_n(x_n\,|\,x_1),\label{eq:gx_f1_fn}
\end{align}\\[-2em]
where\vspace{-0.05in}
\begin{align*}
  f_1(x_1) &\!=\! \abs{y_1\!-\!\a x_1}^2 \!-\!\mbf{b}_{}^T\!(x_1)\bs{\lambda}_1^{},~\text{and}\\
  f_n(x_n|x_1) &\!=\! \abs{y_n\!-\!\c_n x_1\!-\!\b_n x_n}^2 \!-\!\mbf{b}_{}^T\!(x_n)\bs{\lambda}_n^{},
\end{align*}
for $n\!=\!2,\cdots,N$. We next minimize $g(\mbf{x})$ similar to~\eqref{eq:min_dx_f1_f2}:
\begin{align}\hspace{-0.05in}
  g_{}^{\WL}
    & \!\triangleq\!
        \underset{\mbf{x} \in \mathcal{X}}{\mathop{\min}}{\left\{f_1(x_1) \!+\! \sum_{n=2}^N f_n(x_n\,|\,x_1) \right\}}\label{eq:gWL}\\[-0.25em]
    & \!=\!
        \underset{x_1\in \mathcal{X}_1}{\mathop{\min}}\left\{
        f_1(x_1) + \sum_{n=2}^N\underset{x_n\in \mathcal{X}_n}{\mathop{\min}}\! f_n(x_n\,|\,x_1)\right\}\label{eq:gWL_step1}\\
    & \!=\!
        \underset{x_1\in \mathcal{X}_1}{\mathop{\min}}\left\{
        f_1(x_1) + \sum_{n=2}^N f_n(\hat{x}_n(x_1)\,|\,x_1)\right\}\notag\\
    & \!=\!
        \underset{x_1\in \mathcal{X}_1}{\mathop{\min}}\,
        g\!\left(x_1,\hat{x}_2(x_1),\cdots,\hat{x}_N(x_1)\right)\label{eq:gWL_x1_xnhat}
\end{align}\\[-1.5em]
where
\begin{equation}\label{eq:xnhat_prob}
  \hat{x}_n(x_1) = \underset{x_n\in \mathcal{X}_n}{\mathop{\arg\min}}\! \left\{\abs{y_n\!-\!\c_n x_1\!-\!\b_n x_n}^2 \!-\!\mbf{b}_{}^T\!(x_n)\bs{\lambda}_n^{}\right\}.
\end{equation}\\[-1.5em]
Denote the set of sliced symbol vectors for all possible $x_1$ in~\eqref{eq:gWL_x1_xnhat} by (defined similar to~\eqref{eq:O_1} but for any $N\!\geq\!2$)
\begin{equation}\label{eq:O_1_N}
  \mathcal{O}_1 = \left\{ \left[x_1\ \hat{x}_2(x_1)\ \cdots\ \hat{x}_N(x_1)\right]^T\,:\, x_1\!\in\!\mathcal{X}_1 \right\}.
\end{equation}\\[-2em]
The symbol vector that minimizes~\eqref{eq:gx_f1_fn} is denoted as
\begin{align}\hspace{-0.05in}
  \mbf{x}_{}^{\WL}
    & \!\triangleq\!
        \underset{\mbf{x} \in \mathcal{O}_1}{\mathop{\arg\min}}{\,g(\mbf{x})}.\label{eq:xWL}
\end{align}
To efficiently determine $g_{}^{\WL}$, we optimize the distance computations in~\eqref{eq:gWL} by splitting the complex quantities into their real and imaginary components:\small
\begin{align*}
  f_1(x_1) &\!=\! f_{1\re}(x_{1\re})\!+\!f_{1\im}(x_{1\im})\\
  f_{1\re}(x_{1\re}) &\!=\! (y_{1\re}\!-\!\a x_{1\re})^2 \!-\! \mbf{b}_{1\re}^T\bs{\lambda}_{1\re}\\
  f_{1\im}(x_{1\im}) &\!=\! (y_{1\im}\!-\!\a x_{1\im})^2 \!-\! \mbf{b}_{1\im}^T\bs{\lambda}_{1\im}
\end{align*}\normalsize
and\vspace{-0.1in}
\begin{align*}
  f_n(x_n) &\!=\! f_{n\re}(x_{n\re})\!+\!f_{n\im}(x_{n\im})\\
  f_{n\re}(x_{n\re}|x_1) &\!=\! (y_{n\re}\!-\!\c_{n\re} x_{1\re}\!+\!\c_{n\im}x_{1\im}\!-\!\b_nx_{n\re})^2 \!-\! \mbf{b}_{n\re}^T\bs{\lambda}_{n\re}\\
  f_{n\im}(x_{n\im}|x_1) &\!=\! (y_{n\im}\!-\!\c_{n\re} x_{1\im}\!-\!\c_{n\im}x_{1\re}\!-\!\b_nx_{n\im})^2 \!-\! \mbf{b}_{n\im}^T\bs{\lambda}_{n\im}
\end{align*}\normalsize
for $n\!\geq\!2$. Substituting back in~\eqref{eq:gWL_step1}, expanding terms, minimizing w.r.t. $x_{n\re}$ and $x_{n\im}$, and eliminating irrelevant terms, we obtain
\begin{IEEEeqnarray}{rCl}\hspace{-0.2in}
\bar{g}(\mbf{x}) = \bar{f}_{1\re}(x_{1\re}) +  \bar{f}_{1\im}(x_{1\im})
 +\sum_{n=2}^N \left( \underset{x_{n\re}\in \mathcal{P}_n}{\mathop{\min}} \bar{f}_{n\re}(x_{n\re}|x_1) + \underset{x_{n\im}\in \mathcal{P}_n}{\mathop{\min}} \bar{f}_{n\im}(x_{n\im}|x_1)\right)\label{eq:gbar_x}
\end{IEEEeqnarray}
where
\begin{align}
  \bar{f}_{1\re}(x_{1\re}) &\!=\! Ax_{1\re}^2 \!+\!Cx_{1\re} \!-\! \mbf{b}_{1\re}^T\bs{\lambda}_{1\re}\label{eq:f1Rbar_NxN}\\
  \bar{f}_{1\im}(x_{1\im}) &\!=\! Ax_{1\im}^2 \!+\!Dx_{1\re} \!-\! \mbf{b}_{1\im}^T\bs{\lambda}_{1\im}\label{eq:f1Ibar_NxN}\\
  \bar{f}_{n\re}(x_{n\re})
        &\!=\! (E_nx_{1\re}\!+\!F_nx_{1\im})x_{n\re} \notag\\
        &\!+\! (B_nx_{n\re}^2\!+\!G_nx_{n\re}\!-\!\mbf{b}_{n\re}^T\bs{\lambda}_{n\re})\label{eq:f2Rbar_NxN}\\
  \bar{f}_{n\im}(x_{n\im})
        &\!=\! (E_nx_{1\im}\!-\!F_n x_{1\re})x_{n\im} \notag\\
        &\!+\! (B_nx_{n\im}^2\!+\!H_n x_{n\im}\!-\!\mbf{b}_{n\im}^T\bs{\lambda}_{n\im})\label{eq:f2Ibar_NxN}
\end{align}
Similar to~\eqref{eq:constant_A_B_2x2}-\eqref{eq:constant_E_F_G_H_2x2}, the constants above are given by:
\begin{align}
  A &= \a^2 + \sum_{n=2}^N\abs{\c_n}^2, \quad
  B_n = \b_n^2 \label{eq:constant_A_B_NxN}\\[-0.75em]
  C &= -2\a y_{1\re} -2 \sum_{n=2}^N\left(\c_{n\re}y_{n\re} + \c_{n\im}y_{n\im}\right) \label{eq:constant_C_NxN}\\[-0.3em]
  D &= -2\a y_{1\im} +2 \sum_{n=2}^N\left(\c_{n\im}y_{n\re} - \c_{n\re}y_{n\im} \right) \label{eq:constant_D_NxN}\\[-0.3em]
  E_n &\!=\! +2\b_n \c_{n\re},~
  F_n \!=\! -2\b_n\c_{n\im},~
  G_n \!=\! -2\b_n y_{n\re},~
  H_n \!=\! -2\b_n y_{n\im}\label{eq:constant_E_F_G_H_NxN}.
\end{align}

Using $g(\mbf{x})$ (or $\bar{g}(\mbf{x})$), the LLRs of the bits in layer 1 are
\begin{IEEEeqnarray}{rCl}
    \Lambda_{1,j}^{\WL} =
        \underset{x_1^{}\in \mathcal{X}_{1,j}^{(+1)}}{\mathop{\min}}{g\!\left(x_1,\hat{x}_2(x_1),\cdots,\hat{x}_N(x_1)\right)} -\underset{x_1^{}\in \mathcal{X}_{1,j}^{(-1)}}{\mathop{\min}}{g\!\left(x_1,\hat{x}_2(x_1),\cdots,\hat{x}_N(x_1)\right)}.\label{eq:LLR_x_1_j_NxN}
\end{IEEEeqnarray}
The bit-LLRs in the remaining $N\!-\!1$ layers are similarly obtained by using the other $N\!-\!1$ complementary WL structures of $\mbf{H}$ (see Fig.~\ref{f:matrices}). Finally, equations~\eqref{eq:decision_bound_R}-\eqref{eq:minmax_decbound_simplified_R} for $N\!=\!2$ can be used to slice $\hat{x}_{n\re}\!\!=\!\!\underset{x_{n\re}\in \mathcal{P}_n}{\mathop{\min}}f_{n\re}(x_{n\re}|x_1)$, and~\eqref{eq:decision_bound_I}-\eqref{eq:minmax_decbound_simplified_I} to slice $\hat{x}_{n\im}\!=\!\!\underset{x_{n\im}\in \mathcal{P}_n}{\mathop{\min}} f_{n\im}(x_{n\im}|x_1)$, but with the constants $B,E,F,G,H$ replaced by $B_n,E_n,F_n,G_n,H_n$, and $\mathcal{P}_2$, $\bs{\lambda}_{2\re}^{}$, $\bs{\lambda}_{2\im}^{}$ by $\mathcal{P}_n$, $\bs{\lambda}_{n\re}^{}$, $\bs{\lambda}_{n\im}^{}$.

%
\vspace{-0.15in}
\subsection{Post LLR Processing}\label{s:post_processing}\vspace{-0.05in}
Since $g(\mbf{x})\!\neq\!d(\mbf{x})$, there is no guarantee that the $g_{}^{\WL}$ and $\mbf{x}_{}^{\WL}$ obtained in~\eqref{eq:gWL_x1_xnhat} and~\eqref{eq:xWL} using one WLD structure of $\mbf{H}$, are the same ones obtained using the other $N\!-\!1$ WLD structures with the columns of $\mbf{H}$ permuted. To avoid confusion, we refer to the quantities in~\eqref{eq:gx_f1_fn},~\eqref{eq:gWL_x1_xnhat}, and~\eqref{eq:xWL} pertaining to the $\nth{m}$ layer WL decomposition using the subscript $m$: $g_m(\mbf{x})$, $g_{m}^{\WL}$, $\mbf{x}_{m}^{\WL}$.

The ``WL-minimal'' HD solution, denoted as $g_{\mathrm{min}}^{\WL}$ and $\mbf{x}_{\mathrm{min}}^{\WL}$, corresponds to the minimum of the $N$ various $g_{m}^{\WL}$ values:
\begin{align}\label{eq:g_opt_WL}
  g_{\mathrm{min}}^{\WL} = \underset{m}{\mathop{\min}}\,g_{m}^{\WL},\quad
  \mbf{x}_{\mathrm{min}}^{\WL} = \underset{\mbf{x}_{m}^{\WL}}{\mathop{\arg\min}}\,g_{m}^{\WL}.
\end{align}

A similar minimization is required as well to adjust the LLR values $\Lambda_{n,j}^{\WL}$ relative to the global minimum $g_{\mathrm{min}}^{\WL}$ and the bits of its corresponding symbol vector $\mbf{x}_{\mathrm{min}}^{\WL}$. This adjustment cannot be done by comparing the individual $\Lambda_{n,j}^{\WL}$'s alone. One simple way is based on the list of distances $g_m(\mbf{x})$ generated from all decompositions for $m\!=\!1,\cdots,N$, together with their corresponding symbol vectors. Let $\mathcal{O}_m$ denote the set of symbol vectors
\begin{IEEEeqnarray}{rCl}\label{eq:O_m}
  \mathcal{O}_m &\!=\!& \left\{ \left[\hat{x}_1(x_m)\cdots \hat{x}_{m-1}(x_m)\ x_m \ \hat{x}_{m+1}(x_m) \cdots \hat{x}_N(x_m)\right]^T \, : \, x_m\!\in\!\mathcal{X}_m \vphantom{\left[\hat{x}_N(x_m)\right]^T}\right\},~m\!=\!1,\cdots,N,
\end{IEEEeqnarray}
where the $\nth{n}$ sliced symbol in the $\nth{m}$ WLD is
\begin{equation*}
  \hat{x}_n(x_m) = \underset{x_n\in \mathcal{X}_n}{\mathop{\arg\min}}\! \left\{\!\abs{y_{n,m}^{}\!-\!\c_{n,m}^{} x_m\!-\!\b_{n,m}^{} x_n}^2 \!\!-\!\mbf{b}_{}^T\!(x_n)\bs{\lambda}_n^{}\!\right\},
\end{equation*}
for $n\!\neq\!m$. Here $y_{n,m}$, $\c_{n,m}$, and $\b_{n,m}$ are defined as in~\eqref{eq:partitioned_L} but relative to the $\nth{m}$ WLD of $\mbf{H}$. Next, define the partitions on $\mathcal{O}_m$: $\mathcal{O}_{n,j,m}^{(+1)} \!=\! \left\{\mbf{x}\in\mathcal{O}_m \ : b_{n,j}\!=\!+1 \right\}$ and $\mathcal{O}_{n,j,m}^{(-1)} \!=\! \left\{\mbf{x}\in\mathcal{O}_m \ :  b_{n,j}\!=\!-1 \right\}$. Then the ``WL-minimal'' LLRs are given by
\small
\begin{IEEEeqnarray}{rCl}\hspace{-0.25in}
    \Lambda_{n,j,\mathrm{min}}^{\WL} \!&=&\!
    \underset{m}{\mathop{\min}}\! \left\{\! \underset{\mbf{x}\in \mathcal{O}_{n,j,m}^{(+1)}}{\mathop{\min}}{\!g_{m}(\mbf{x})}\! \right\} -
    \underset{m}{\mathop{\min}}\! \left\{\! \underset{\mbf{x}\in \mathcal{O}_{n,j,m}^{(-1)}}{\mathop{\min}}{\!g_{m}(\mbf{x})}\! \right\}
        \label{eq:LLR_x_n_j_NxN}.
\end{IEEEeqnarray}\normalsize\\[-3em]

%
\vspace{-0.1in}
\subsection{Discussion}\label{s:discussion}\vspace{-0.05in}
The key equations for the general $N$-layer case derived above reduce to the optimal equations derived in Section~\ref{s:2x2MAP} for $N\!=\!2$. A comparison between the two shows that the same operations applied to compute $\bar{d}(\mbf{x})$ in~\eqref{eq:dbar_x} are applied to compute $\bar{g}(\mbf{x})$ in~\eqref{eq:gbar_x}, but using the respective constants of layer $n$ instead of layer $2$. Hence, a $2\!\times\!2$ MAP detector can be viewed as a building block for constructing detectors for higher-order layers, with a simple modification to account for the extra accumulated sum terms in~\eqref{eq:gbar_x}, in addition to the LLR processing of~\eqref{eq:LLR_x_n_j_NxN} at the output stage. A parallel architecture will be developed next and its complexity analyzed.

%
\section{Parallel 2-Layer Detector Architecture}\label{s:complexity_par_arch} 
\begin{figure*}[t]
\centering
\includegraphics[scale=0.75]{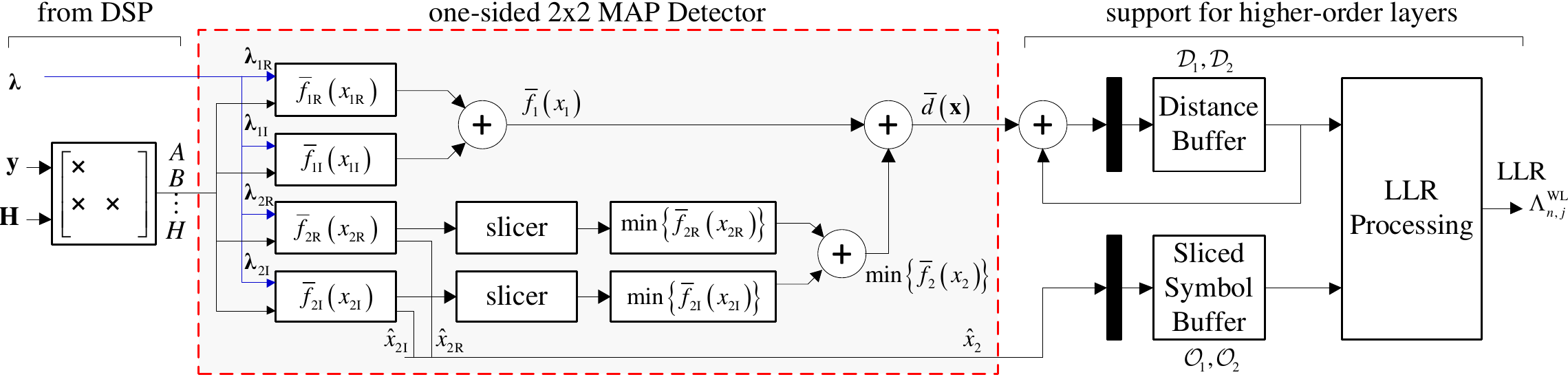}\vspace{-0.1in}
\caption{Block diagram of a parallel one-sided $2\!\times\!2$ MAP detector core, with input and output interfaces}
\label{f:2x2_MAP_detector_block_diagram}
\end{figure*}

\vspace{-0.05in} Figure~\ref{f:2x2_MAP_detector_block_diagram} shows a block diagram of a parallel $2\!\times\!2$ MAP detector core that implements the key equations in~\eqref{eq:dbar_x}-\eqref{eq:f2Ibar_2x2}. For flexibility and scalability to higher-order layers, the constellations supported on each layer are configurable from BPSK up to 256-QAM, and can be distinct on each layer. We assume the input constants~\eqref{eq:constant_A_B_2x2}-\eqref{eq:constant_E_F_G_H_2x2} to the detector are supplied by an external DSP. The outputs are two lists of distances $\mathcal{D}_1,\mathcal{D}_2$ and their associated lists of symbol vectors $\mathcal{O}_1,\mathcal{O}_2$, which are fed to a post LLR processing stage to extract the LLRs values depending on the number of layers.

%
\vspace{-0.1in}
\subsection{Optimized Implementation of Distance Expressions}\label{s:opt_dist_expressions}
A careful inspection of expressions~\eqref{eq:f1Rbar_2x2}-\eqref{eq:f2Ibar_2x2} shows that $\bar{d}(\mbf{x})$ can be evaluated without using multipliers, assuming the constants are pre-processed and fed as inputs to the detector. The reason is that the variables $x_{1\re}$, $x_{1\im}$, $x_{2\re}$, and $x_{2\im}$ are integers that belong to a PAM constellation. More specifically, in LTE~\cite{LTE_36.211}, they are odd integers in the set $\mathcal{P}_2\!=\!\{2m\!+\!1 \,|\,m\!=\!-P/2\!+\!1,\cdots,0,\cdots,P/2\!-\!1\}$ and $P\!=\!\sqrt{Q_2}$. Hence the terms that involve the products of $x_{1\re}$, $x_{1\im}$, $x_{2\re}$, $x_{2\im}$ in~\eqref{eq:f1Rbar_2x2}-\eqref{eq:f2Ibar_2x2} with the constants in~\eqref{eq:constant_A_B_2x2}-\eqref{eq:constant_E_F_G_H_2x2} are simply integer multiples of these constants. These product terms can be computed using basic \emph{addition operations} with appropriate power-of-2 manipulations of the operands without using expensive multipliers. Also from symmetry, only positive multiples need to be computed. Table~\ref{t:constant_multiples} summarizes the number of various distinct product terms that need to be computed for various PAM constellation sizes.

Moreover, the dot products $\mbf{b}_{1\re}^T \bs{\lambda}_{1\re}^{}$ between the input LLR vectors and all the bit vectors are simply all linear binary combinations of the $q_1/2\!=\!(\log_2{Q_1})/2$ individual input LLRs $\lambda_i$ of $\bs{\lambda}_{1\re}^{}$:
\begin{equation}\label{eq:LLRpatterns}
  \pm\lambda_1\pm\lambda_2\pm\cdots\pm\lambda_{q_1/2}.
\end{equation}\\[-2em]
Also from symmetry, only half of these sums actually need to be computed, giving a total of $2^{q_1/2-1}$ different sums. The same applies to other dot product terms in~\eqref{eq:f1Ibar_2x2}-\eqref{eq:f2Ibar_2x2}.

Next, as $x_{1\re}$ runs over the $P_1$ integers in $\mathcal{P}_1$, the expression $\left(A x_{1\re}^2 \!+\! Cx_{1\re}^{} \!-\!\mbf{b}_{}^T\!(x_{1\re}) \bs{\lambda}_{1\re}^{}\right)$ takes $P_1$ different values. However, because of the Gray mapping of the bits, then $\mbf{b}_{}^T\!(-x_{1\re}) \bs{\lambda}_{1\re}^{}\!\neq\!-\mbf{b}_{}^T\!(x_{1\re}) \bs{\lambda}_{1\re}^{}$ and hence there is no symmetry that can be exploited to save in computations here. The same argument applies to the three other expressions $\left(A x_{1\im}^2 \!+\! Dx_{1\im} \!-\!\mbf{b}_{1\im}^T \bs{\lambda}_{1\im}^{}\right)$, $(Bx_{2\re}^2 \!+\! Gx_{2\re} \!-\! \mbf{b}_{2\re}^T\bs{\lambda}_{2\re}^{})$, and $(Bx_{2\im}^2 \!+\! Hx_{2\im} \!-\! \mbf{b}_{2\im}^T\bs{\lambda}_{2\im}^{})$ in~\eqref{eq:f1Ibar_2x2}-\eqref{eq:f2Ibar_2x2}.
\begin{table}[t]
\centering
\caption{Distinct product terms to be computed: $x,y,z\!\in\!\mathcal{P}_2$; $r,s\!\in\!\mathcal{R}$.}\vspace{-0.1in}
\label{t:constant_multiples}
\renewcommand{\arraystretch}{1.25}
\begin{tabular}{|@{\hspace{0.5mm}}c@{\hspace{0.5mm}}|@{\hspace{0.5mm}}c@{\hspace{0.5mm}}|@{\hspace{0.5mm}}c@{\hspace{0.5mm}}|@{\hspace{0.5mm}}c@{\hspace{0.5mm}}|@{\hspace{0.5mm}}c@{\hspace{0.5mm}}|}
  \hline
  $\#$ distinct terms & 2-PAM & 4-PAM & 8-PAM & 16-PAM \\\hline\hline
  $r\!\cdot\!\abs{x}$ & 1 & 2 & 4 & 8 \\\hline
  $r\!\cdot\!\abs{x}\!\cdot\!\abs{y}$ & 1 & 3 & 10 & 33 \\\hline
  $r\!\cdot\!x^2$ & 1 & 2 & 4 & 8 \\\hline
  $(r\!\cdot\!\abs{x}\!\pm\!s\!\cdot\!\abs{y})\!\cdot\!\abs{z}$ & 2 & 14 & 116 & 914\\\hline
  $\abs{\mbf{b}_{1\re}^T \bs{\lambda}_{1\re}^{}}$ & 1 & 2 & 4 & 8\\\hline
\end{tabular}
\end{table}

Finally, for the remaining sum of products of cross terms $(Ex_{1\re} \!+\! Fx_{1\im})x_{2\re}$, as $x_{2\re}$ cycles through the $P_2$ integers in $\mathcal{P}_2$, the expression takes $P_2$ different values for every pair $(x_{1\re},x_{1\im})$. However, for all possible $(x_{1\re},x_{1\im})$, repetitions occur. The number of unique values of $(Ex_{1\re} \!+\! Fx_{1\im})x_{2\re}$ is twice that of $(E\!\abs{x_{1\re}}\!\pm\! F\!\abs{x_{1\im}})\!\abs{x_{2\re}}$ (summarized in Table~\ref{t:constant_multiples}). By symmetry, these are also the same values taken by the other sub-expression $(Ex_{1\im} \!-\! Fx_{1\re})x_{2\im}$ in~\eqref{eq:f2Ibar_2x2}.

Therefore, hardware complexity will be measured in terms of number of adders, in addition to number of (2:1)-multiplexers (muxes) needed to steer operands to these adders. Wider ($n$:1)-muxes can be constructed using $n-1$ (2:1)-muxes.

We next determine the actual number of adders required to compute each of the unique terms in~\eqref{eq:f1Rbar_2x2}-\eqref{eq:f2Ibar_2x2}, assuming 256-QAM and its underlying 1D 16-PAM constellation. The same analysis applies to other constellations. The required multiples $Ax_{1\re}^2$ for 16-PAM are $\{9,25,49,81,121,169,225\}\!\times\!A$, which can be generated using 11 adders as follows:\small
\begin{equation*}
\begin{IEEEeqnarraybox}[
\IEEEeqnarraystrutmode
\IEEEeqnarraystrutsizeadd{1pt}
{1pt}
][c]{rCl}
9A\!&=&\!8A\!+\!A,\\
49A\!&=&\!64A\!-\!15A,\\
121A\!&=&\!128A\!-\!7A,\\
31A\!&=&\!32A\!-\!A
\end{IEEEeqnarraybox}~~
\begin{IEEEeqnarraybox}[
\IEEEeqnarraystrutmode
\IEEEeqnarraystrutsizeadd{1pt}
{1pt}
][c]{rCl}
25A\!&=&\!16A\!+\!9A,\\
81A\!&=&\!32A\!+\!49A,\\
41A\!&=&\!32A\!+\!9A,\\
225A\!&=&\!256A\!-\!31A
\end{IEEEeqnarraybox}~~
\begin{IEEEeqnarraybox}[
\IEEEeqnarraystrutmode
\IEEEeqnarraystrutsizeadd{1pt}
{1pt}
][c]{rCl}
15A\!&=&\!16A\!-\!A\\
7A\!&=&\!8A\!-\!A\\
169A\!&=&\!128A\!+\!41A\\
\end{IEEEeqnarraybox}
\label{eq:example_left_right2}
\end{equation*}\normalsize
Similarly, the 8 multiples $C\abs{x_{1\re}}$ can be generated using 7 adders. For $\mbf{b}_{1\re}^T \bs{\lambda}_{1\re}^{}$, 8 values of can be generated as
\begin{equation*}
\begin{IEEEeqnarraybox}[
\IEEEeqnarraystrutmode
\IEEEeqnarraystrutsizeadd{1pt}
{1pt}
][c]{rCl}
&&(\lambda_1\!+\!\lambda_2)\!\pm\!(\lambda_3\!+\!\lambda_4),\\
&&(\lambda_1\!-\!\lambda_2)\!\pm\!(\lambda_3\!+\!\lambda_4),
\end{IEEEeqnarraybox}~~
\begin{IEEEeqnarraybox}[
\IEEEeqnarraystrutmode
\IEEEeqnarraystrutsizeadd{1pt}
{1pt}
][c]{rCl}
&&(\lambda_1\!+\!\lambda_2)\!\pm\!(\lambda_3\!-\!\lambda_4),\\
&&(\lambda_1\!-\!\lambda_2)\!\pm\!(\lambda_3\!-\!\lambda_4)
\end{IEEEeqnarraybox}
\end{equation*}
with 12 adders. The other 8 are their negatives.

To generate the unique elements of $(E\!\abs{x_{1\re}}\!\pm\! F\!\abs{x_{1\im}})\!\abs{x_{2\re}}$, we first generate all unique sums with $x_{2\re}\!=\!1$, i.e. $(E\!\abs{x_{1\re}}\!\pm\! F\!\abs{x_{1\im}})$, such that $\gcd(\abs{x_{1\re}},\abs{x_{1\im}})\!=\!1$, and then generate all their multiples. The number of unique sums of the form $(E\!\abs{x_{1\re}}\!\pm\! F\!\abs{x_{1\im}})$ with co-prime coefficients $\abs{x_{1\re}}$ and $\abs{x_{1\im}}$ from the set $\{1,3,\cdots,15\}$ is 49. We next enumerate the unique multiples from each of these 49 classes. For $(\abs{x_{1\re}},\abs{x_{1\im}})\!=\!(1,1)$, there are $33\!\times\!2$ distinct multiples of $(E\!\pm\!F)$. For $(\abs{x_{1\re}},\abs{x_{1\im}})\!=\!(1,3)$ or $(3,1)$, there are $18\!\times\!2$ distinct multiples of $(E\!\pm\!3F)$ and $18\!\times\!2$ of $(3E\!\pm\!F)$. For $(\abs{x_{1\re}},\abs{x_{1\im}})\!=\!(1,5)$, $(5,1)$, $(3,5)$, or $(5,3)$, there are $13\!\times\!2$ distinct multiples of each. Finally, for the remaining 42 classes, there are $8\!\times\!2$ distinct multiples of each. Summing all distinct multiples we obtain 914.

Table~\ref{t:integer_multiples} summarizes the various constants that appear in the computation of $(E\!\abs{x_{1\re}}\!\pm\! F\!\abs{x_{1\im}})\!\abs{x_{2\re}}$ for 16-PAM, and how they are generated using addition operations involving powers-of-2 operands and other already computed constants. First, the odd multiples $3E,5E,\cdots,15E$, and $3F,5F,\cdots,15F$, require 14 adders. The term $(E\!\pm\!F)$ and its $33\!\times\!2$ distinct multiples require all the 36 constants in Table~\ref{t:integer_multiples} and hence need $(36\!+\!1)\!\times\!2$ adders. The term $(E\!\pm\!3F)$ requires 18 constants $\{1,3,5,7,9,11,13,15,21,25,27,33,35,39,45,55,65,75\}$, and hence needs $2\!\times\!18$ adders. The same count is needed for $(3E\!\pm\!F)$. On the other hand, the term $(E\!\pm\!5F)$ and its $13\!\times\!2$ distinct multiples require only 13 constants $\{1,3,5,7,9,11,13,15,21,27,33,39,45\}$ and hence need $2\!\times\!13$ adders. The same applies for $(5E\!\pm\!F)$, $(3E\!\pm\!5F)$, and $(5E\!\pm\!3F)$. For the remaining 42 classes, only 8 constants $\{1,3,5,7,9,11,13,15\}$ appear and hence need $2\!\times\!8\!\times\!42$ adders. Summing all counts results in a total of 936 adders. Finally note that $(E\!\abs{x_{1\re}}\!\pm\! F\!\abs{x_{1\im}})\!\abs{x_{2\re}}$ takes the same values as $(E\!\abs{x_{1\re}}\!\mp\! F\!\abs{x_{1\im}})\!\abs{x_{2\im}}$ but in a different order.
\input{table_integer_multiples.tex}

%
\vspace{-0.05in}
\subsection{Minimization by Exhaustive Search}\label{s:min_exh_search_arch}
One approach to implement the minimizations in~\eqref{eq:dbar_x} is by exhaustive search. In~\eqref{eq:f2Rbar_2x2}, for every pair $(x_{1\re},x_{1\im})$, 16 out of 914 distinct values of $(E\!\abs{x_{1\re}}\!\pm\! F\!\abs{x_{1\im}})\!\abs{x_{2\re}}$ pertaining to the 16 different $x_{2\re}$'s are added to $B x_{2\re}^2 \!+\! Gx_{2\re} \!-\!\mbf{b}_{2\re}^T \bs{\lambda}_{2\re}^{}$, and the minimum is selected. Hence a total of $16\!\times\!256$ adders are needed to generate all possible values of $\bar{f}_{2\re}(x_{2\re}\,|\,x_1)$. The same holds for $\bar{f}_{2\im}(x_{2\im}\,|\,x_1)$. To find the minimum among $P_2$ quantities, a binary tree of comparators comprised of $P_2\!-\!1$ adders and $P_2\!-\!1$ (2:1)-multiplexers are needed. A total of $2\!\times\!256$ such comparators are needed. Finally, the 256 minima from each case are added to complete the sum for $\bar{d}(\mbf{x})$ in~\eqref{eq:dbar_x}.

To generate the hard-decision MAP solution, the minimum among the 256 distances $\bar{d}(\mbf{x})$ must be taken and the corresponding constellation symbol be identified. This requires a total of 255 adders and 255 muxes. On the other hand, to compute the output LLRs of the bits in $x_1$ according to~\eqref{eq:LLR_x_1_j}, the 256 distances in~\eqref{eq:dbar_x} must be minimized over two complementary sets for every bit and their difference be taken. The 256-QAM constellation points can be viewed as 16 columns each containing 16 points, or as 16 rows each containing 16 points. In LTE, the 4 bits corresponding to the real part of the constellation points do not change in every column, and the 4 bits corresponding to the imaginary part do not change in every row. Hence it suffices to take the minimum distances among all points in each row and among all points in each column independently. The column minima are used to compute the LLRs of the real bits by partitioning the columns into two groups of 8 columns depending on whether the bit is $+1$ or $-1$ in the column. The minimum distance among each group of columns is taken, and the difference of the two minima generates the LLR of that bit. The same applies to the imaginary bits and the row minima. Hence a total of $2\!\times\!16$ 16-point comparators are needed, amounting to 480 adders and 480 muxes, to extract the minima, followed by 8 adders to take the differences.

Table~\ref{t:num_adders_noslice_arch} summarizes the core complexity using exhaustive search. The core requires 18290 adders and 8160 muxes.
\input{table_num_adders.tex}

%
\vspace{-0.1in}
\subsection{Minimization by Slicing}\label{s:min_slicing_arch}\vspace{-0.05in}
We next analyze the complexity of computing $\underset{x_{2\re}\in \mathcal{P}_2}{\mathop{\min}}\bar{f}_{2\re}(x_{2\re}|x_1)$ in~\eqref{eq:dbar_x} via the slicing approach by first determining $\hat{x}_{2\re}\!=\!\underset{x_{2\re}\in \mathcal{P}_2}{\mathop{\arg\min}}\bar{f}_{2\re}(x_{2\re}|x_1)$ followed by evaluating $\bar{f}_{2\re}(\hat{x}_{2\re}|x_1)$, for all possible $x_1$. To minimize $\bar{f}_{2\re}(x_{2\re}|x_1)$, the decision boundaries $R(x_{2\re},\bar{x}_{2\re})$ in~\eqref{eq:decision_bound_R} must be computed for all $x_{2\re}\!\neq\!\bar{x}_{2\re}\!\in\!\mathcal{P}_2$, and appropriate minima and maxima must be extracted from these boundaries according to~\eqref{eq:minmax_decbound_simplified_R} and compared to $Ex_{1\re}\!+\!Fx_{1\im}$. Similarly, to minimize $\bar{f}_{2\im}(x_{2\im}|x_1)$, the decision boundaries $I(x_{2\im},\bar{x}_{2\im})$ in~\eqref{eq:decision_bound_I} must be computed for all $x_{2\im}\!\neq\!\bar{x}_{2\im}\!\in\!\mathcal{P}_2$, and appropriate minima and maxima must be extracted from these boundaries according to~\eqref{eq:minmax_decbound_simplified_I} and compared to $Ex_{1\im}\!-\!Fx_{1\re}$.
\begin{figure}[t]
\centering
\includegraphics[scale=1.00]{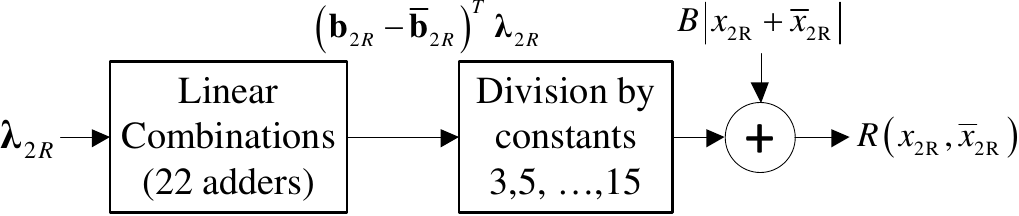}\vspace{-0.1in}
\caption{Computation of decision boundaries}
\label{f:decision_boundaries}
\end{figure}

By analogy, it suffices to analyze the complexity of~\eqref{eq:decision_bound_R} and~\eqref{eq:minmax_decbound_simplified_R}. Since $R(x_{2\re},\bar{x}_{2\re})\!=\!R(\bar{x}_{2\re},x_{2\re})$, only $P_2(P_2\!-\!1)/2\!=\!120$ decision boundaries need to be computed (see Fig.~\ref{f:decision_boundaries}). The sum $\abs{x_{2\re}\!+\!\bar{x}_{2\re}}$ takes $P_2\!-\!2$ distinct non-zero values ($2,4,\cdots,2P_2\!-\!4$), and hence the product $B\!\abs{x_{2\re}\!+\!\bar{x}_{2\re}}$ term in~\eqref{eq:decision_bound_R} requires 6 adders. Similarly, the difference $\abs{x_{2\re}\!-\!\bar{x}_{2\re}}$ takes $P_2\!-\!1$ distinct non-zero values ($2,4,\cdots,2P_2\!-\!2$). For the division of $\left(\mbf{b}_{2\re}\!-\!\overline{\mbf{b}}_{2\re}\right)^{\!T}\!\bs{\lambda}_{2\re}^{}$ by these constants, where $\overline{\mbf{b}}_{2\re}\!=\!\mbf{b}(\bar{x}_{2\re})$, the term $\left(\mbf{b}_{2\re}\!-\!\overline{\mbf{b}}_{2\re}\right)^{\!T}\!\bs{\lambda}_{2\re}^{}$ takes 80 distinct values, 40 of which can be obtained by negation. These 40 values require 22 adders.
The required ratios $\left(\mbf{b}_{2\re}\!-\!\overline{\mbf{b}}_{2\re}\right)^{\!T}\!\bs{\lambda}_{2\re}^{}/(x_{2\re}\!-\!\bar{x}_{2\re})$ take only 40 distinct values, and require divisions by $3,5,7,9,11,13,15$. However, each value of $\left(\mbf{b}_{2\re}\!-\!\overline{\mbf{b}}_{2\re}\right)^{\!T}\!\bs{\lambda}_{2\re}^{}$ need not be divided by all these 7 constants. By going over all various combinations, it is easy to show that the number of divisions by the various values of $\abs{x_{2\re}\!-\!\bar{x}_{2\re}}$ is as follows ($\text{constant}:\text{count}$):\vspace{-0.075in}\small
\begin{equation*}
\begin{IEEEeqnarraybox}[
\IEEEeqnarraystrutmode
\IEEEeqnarraystrutsizeadd{1pt}{1pt}][c]{rCl}
2&:&4,\\
18&:&3,
\end{IEEEeqnarraybox}~~
\begin{IEEEeqnarraybox}[
\IEEEeqnarraystrutmode
\IEEEeqnarraystrutsizeadd{1pt}{1pt}][c]{rCl}
4&:&3,\\
20&:&2,
\end{IEEEeqnarraybox}~~
\begin{IEEEeqnarraybox}[
\IEEEeqnarraystrutmode
\IEEEeqnarraystrutsizeadd{1pt}{1pt}][c]{rCl}
6&:&5,\\
22&:&3,
\end{IEEEeqnarraybox}~~
\begin{IEEEeqnarraybox}[
\IEEEeqnarraystrutmode
\IEEEeqnarraystrutsizeadd{1pt}{1pt}][c]{rCl}
8&:&2,\\
24&:&1,
\end{IEEEeqnarraybox}~~
\begin{IEEEeqnarraybox}[
\IEEEeqnarraystrutmode
\IEEEeqnarraystrutsizeadd{1pt}{1pt}][c]{rCl}
10&:&5,\\
26&:&2,
\end{IEEEeqnarraybox}~~
\begin{IEEEeqnarraybox}[
\IEEEeqnarraystrutmode
\IEEEeqnarraystrutsizeadd{1pt}{1pt}][c]{rCl}
12&:&3,\\
28&:&1,
\end{IEEEeqnarraybox}~~
\begin{IEEEeqnarraybox}[
\IEEEeqnarraystrutmode
\IEEEeqnarraystrutsizeadd{1pt}{1pt}][c]{rCl}
14&:&4,\\
30&:&1
\end{IEEEeqnarraybox}~~
\begin{IEEEeqnarraybox}[
\IEEEeqnarraystrutmode
\IEEEeqnarraystrutsizeadd{1pt}{1pt}][c]{rCl}
16&:&1,\\
&&
\end{IEEEeqnarraybox}
\end{equation*}\normalsize\\[-1em]
Divisions by powers-of-2 are trivial. Division by 3 covers division by $6\!=\!3\!\times\!2$, $12\!=\!3\!\times\!4$, and $24\!=\!3\!\times\!8$, and hence is needed 9 times. Division by 5 covers division by 10 and 20, and hence is needed 7 times. In a similar fashion, division by 7 is needed 5 times, by 9 is needed 3 times, by 11 is needed 3 times, by 13 is needed 2 times, and by 15 is needed once. The total number of such non-trivial divisions is 30. The complexity of a division-by-small-constant circuit is roughly equivalent to a small number of adders for small bit-widths. Specifically, a divide-by-3 is equivalent to 1 adder; by 5, 7, 9, and 11 are equivalent to 2 adders; and by 13 and 15 are equivalent to 3 adders. Hence, the ratios in~\eqref{eq:decision_bound_R} can be computed using 54 adders. Finally, computing all 120 decision boundaries by adding/subtracting the various 14 non-zero values of $B\abs{x_{2\re}\!+\!\bar{x}_{2\re}}$ to the various 40 distinct ratios $\left(\mbf{b}_{2\re}\!-\!\overline{\mbf{b}}_{2\re}\right)^{\!T}\!\bs{\lambda}_{2\re}^{}/(x_{2\re}\!-\!\bar{x}_{2\re})$ requires 112 adders ($B\abs{x_{2\re}\!+\!\bar{x}_{2\re}}\!=\!0$ in 8 cases out of the 120).

\begin{figure}[t]
\centering
\includegraphics[scale=1]{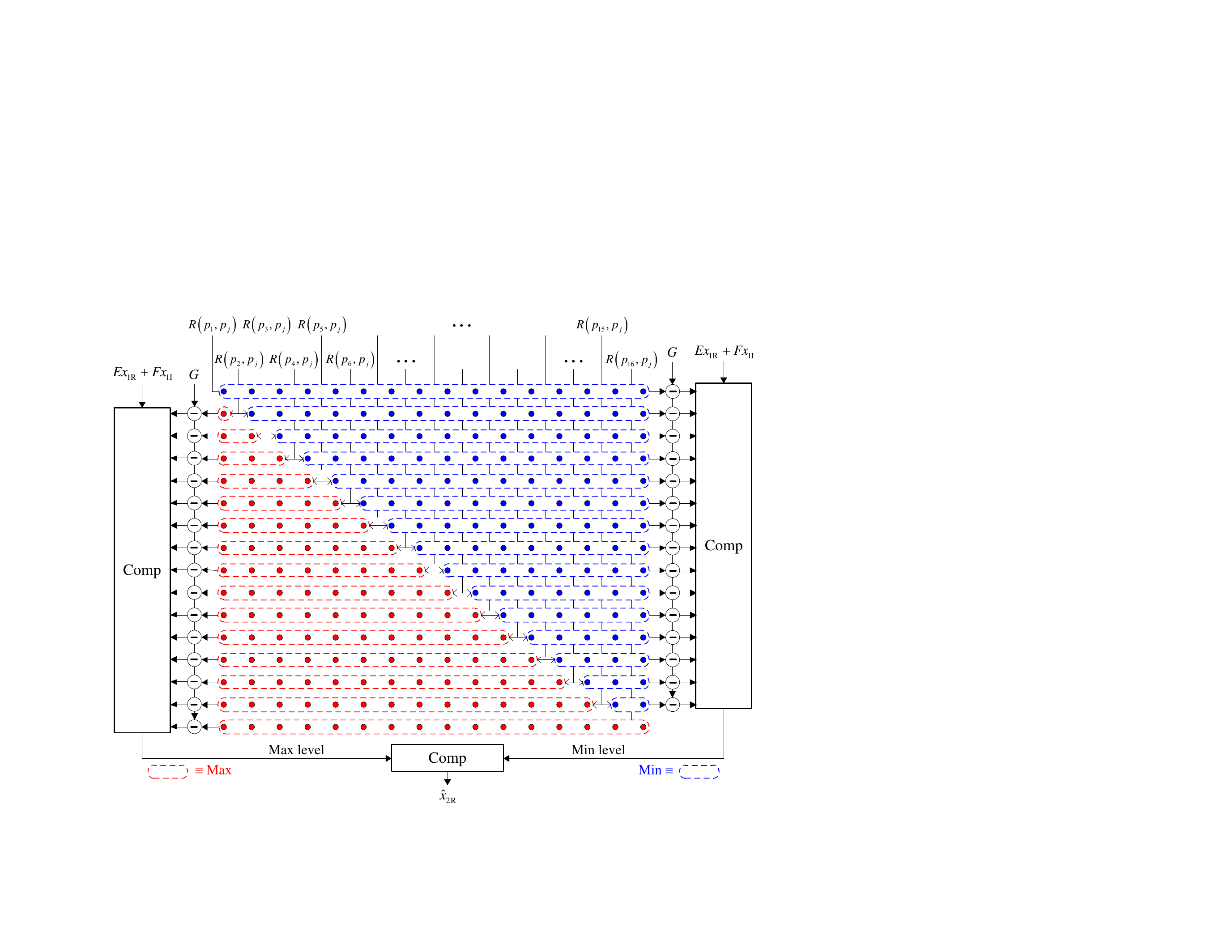}\vspace{-0.15in}
\caption{Block diagram of optimized slicer architecture}
\label{f:slicer}
\end{figure}

Moving to~\eqref{eq:minmax_decbound_simplified_R}, a subset of $P_2\!-\!1$ minimum and $P_2\!-\!1$ maximum regions must be extracted from these boundaries for every hypothesis point $x_{2\re}$ w.r.t. all other $P_2\!-\!1$ points in $\mathcal{P}_2$. These can be obtained using a set of $P_2$ comparator trees, comprising a total of $14\!\times\!15\!=\!210$ adders and 210 (2:1)-MUXs. Next, $G$ is subtracted from each of the $P_2\!-\!1$ min and $P_2\!-\!1$ max boundaries using 30 adders. Finally, comparisons between $Ex_{1\re}\!+\!Fx_{1\im}$ and these min/max boundaries are required to determine $\hat{x}_{2\re}$ according to~\eqref{eq:minmax_decbound_simplified_R}. Each comparison requires 30 adders. Only 128 such comparisons are needed for $\abs{Ex_{1\re}\!\pm\!Fx_{1\im}}$, requiring a total of 3840 adders. The other 128 are derived by symmetry. Figure~\ref{f:slicer} shows the architecture of the slicer block in Fig.~\ref{f:2x2_MAP_detector_block_diagram}.

Based on the results from the slicers, the $\hat{x}_{2\re}$'s are used to evaluate $\bar{f}_{2\re}(\hat{x}_{2\re}|x_1)$. This is done by \emph{selecting} the appropriate multiples $\abs{(Ex_{1\re}\!\pm\!F\!x_{1\im})\hat{x}_{2\re}}$ to be added to $B \hat{x}_{2\re}^2 \!+\! G\hat{x}_{2\re} \!-\!\mbf{b}^T\!(\hat{x}_{2\re}) \bs{\lambda}_{2\re}^{}$. Hence 256 adders are needed, in addition to $128$ (8:1)-MUXES and $256$ (16:1)-MUXES.

Table~\ref{t:num_adders_slicing} summarizes the complexity resources of the slicer-based detector. The architecture requires 11246 adders and 10372 muxes, which amount to a $38.52\%$ savings in adders and an increase of $27.11\%$ in muxes compared with the previous architecture using exhaustive search minimization. The internal pipeline registers, output buffers and accumulators in Fig.~\ref{f:2x2_MAP_detector_block_diagram} are the same between the 2 architectures, and thus are not included in the comparisons.
\input{table_num_adders_slice.tex}

%
\vspace{-0.1in}
\subsection{Multi-Core Detector Architectures}\label{s:arch_config}
\begin{figure}[hbtp]
\centering
\includegraphics[scale=1]{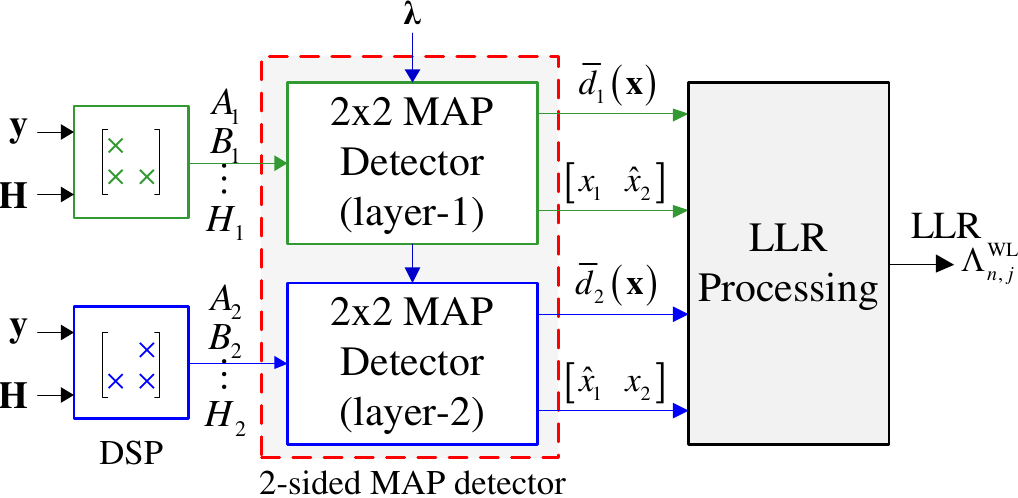}\vspace{-0.05in}
\caption{Block diagram of 2-sided MAP detector}
\label{f:2_sided_MAP_core}
\end{figure}

\vspace{-0.05in}
Depending on the target throughput and the number of antennas $N$ in the MIMO systems, multiple detector cores similar to Fig.~\ref{f:2x2_MAP_detector_block_diagram} can be configured to build a MIMO detector. Figure~\ref{f:2_sided_MAP_core} shows a 2-sided fully parallel $2\!\times\!2$ MIMO detector architecture that uses 2 separate cores to detect the two streams. Since the detection algorithm in this case in optimal, the post LLR processing stage simply implements~\eqref{eq:LLR_x_1_j} and~\eqref{eq:LLR_x_2_j}, without the need for distance buffering and accumulation.
\begin{figure}[hbtp]
\centering
\includegraphics[scale=1]{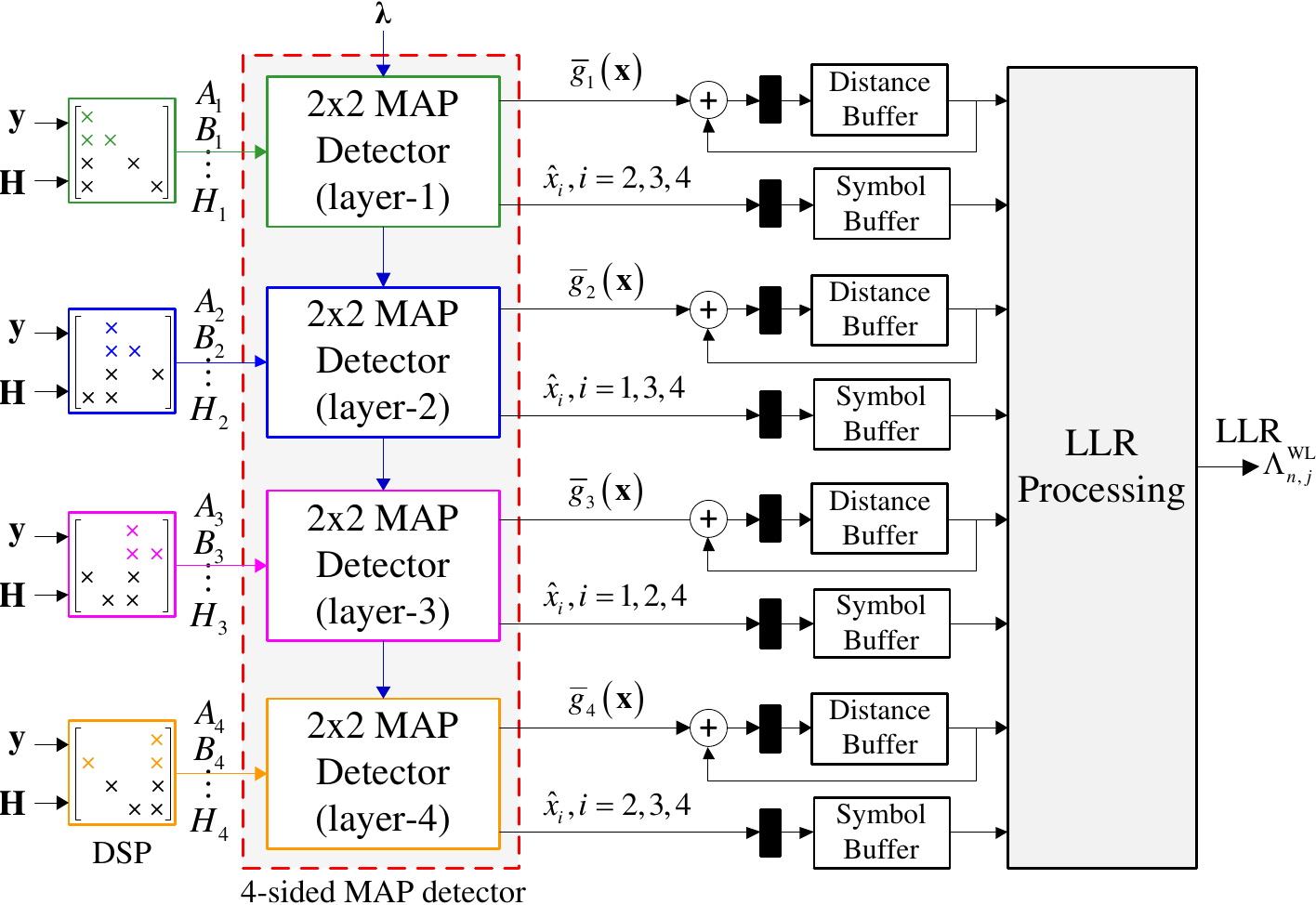}\vspace{-0.05in}
\caption{Block diagram of 4-sided MAP detector}
\label{f:4_sided_MAP_core}
\end{figure}

Figure~\ref{f:4_sided_MAP_core} shows a 4-sided fully parallel $4\!\times\!4$ MIMO detector that uses 4 cores to process the 4 streams. Here distance buffering and accumulation are needed before LLR processing in order to adjust the individual LLRs according to~\eqref{eq:LLR_x_n_j_NxN}. In this case, the WLD matrix inputs for all 4 streams using the decompositions in~\eqref{eq:partitioned_L} are needed. If chip area is the constraining factor, a MIMO detector can be built using a single core that is time-multiplexed among the 4 streams.

%
\vspace{-0.05in}
\section{Application to MU-MIMO Detection}\label{s:MU_MIMO} \vspace{-0.05in}
Multi-user MIMO (MU-MIMO) has been proposed as a method for increasing the capacity of wireless networks~\cite{2009_Lee,2011_Duplicy}. In MU-MIMO, multiple users are scheduled on the same physical resource blocks (PRBs). Several receiver processing methods have been proposed in the literature for MU-MIMO~\cite{2011_Duplicy,2011_Bai,2011_Ghaffar_a,2011_Ghaffar_b}. We consider an optimal MU-MIMO detection method based on the joint constellation estimation of the interfering user and data detection. The optimal MU-MIMO detector can be \emph{efficiently} implemented with a slight modification of the MAP MIMO detector developed in Section~\ref{s:2x2MAP}.

\vspace{-0.05in}
\subsection{MU-MIMO System Model}\label{s:MU_MIMO_system_model}\vspace{-0.05in}
We consider a practical OFDM-based MU-MIMO system where $2$ users are co-scheduled on the same PRBs, and each UE has $2$ receive antennas. Let $K$ be the number of tones in each PRB. Also let user 1 denote the user of interest with \emph{known} constellation $\mathcal{X}_{\mathrm{S}}$, while user 2 denotes the interfering user whose constellation $\mathcal{X}_{\mathrm{I}}$ is \emph{unknown} to user 1's receiver. The received frequency-domain complex signal vector $\mbf{y}[k]\!\in\!\mathcal{C}^{2\times 1}$ at the UE of interest on the $\nth{k}$ resource element (RE) over which the $2$ users are scheduled is given by
\begin{align}
  \mbf{y}[k] &\!=\! \mbf{H}[k] \mbf{x}[k] \!+\! \mbf{n}[k]\notag\\
             &\!=\! \mbf{h}_1[k]x_1[k] \!+\! \mbf{h}_2[k] x_{2}[k] \!+\! \mbf{n}[k],~k\!=\!1,\cdots,K, \label{eq:MUMIMO_sys_model}
\end{align}
where $\mbf{H}[k]\!=\!\left[\mbf{h}_1[k]~\mbf{h}_2[k]\right] \!\in\!\mathcal{C}^{2 \times 2}$ is the complex channel matrix with $\mbf{h}_1[k]$ and $\mbf{h}_2[k]$ representing the cascade of the channel and precoders of user 1 and user 2, respectively; $\mbf{x}[k]\!=\![x_1[k]~x_2[k]]^T$ denotes the transmitted $2\!\times\! 1$ QAM symbol vector where $x_1[k]\!\in\!\mathcal{X}_{\mathrm{S}}$, $x_2[k]\!\in\!\mathcal{X}_{\mathrm{I}}$; and $\mbf{n}[k]\!\in\!\mathcal{C}^{2\times 1}$ is the noise vector at the $\nth{k}$ RE modeled as a zero-mean complex Gaussian random vector with variance $\sigma^2$.

\subsection{ML MU-MIMO Detection}\label{s:ML_MU_MIMO_detection}\vspace{-0.05in}
The maximum likelihood estimate of the constellation of the interfering user based on $\mbf{y}[1],\cdots,\mbf{y}[K]$ is given by
\begin{align}\label{eq:ML_estimate}
    \hat{\mathcal{X}}_{\mathrm{I}} \!=\!  \underset{\mathcal{X}_{\mathrm{I}}\in \mathcal{M}}{\mathop{\arg\max }}\,\mathrm{p}\!\left( \left. \left\{\mbf{y}[k]\right\}_{k=1}^{K} \right|\left\{ \mbf{H}[k] \right\}_{k=1}^{K},\mathcal{X}_{\mathrm{S}},\mathcal{X}_{\mathrm{I}} \right),
\end{align}
where $K$ is the number of REs over which $\mathcal{X}_{\mathrm{I}}$ is constant, and
\begin{equation}\label{eq:set_M}
  \mathcal{M}\triangleq\left\{\text{4-QAM},\text{16-QAM},\text{64-QAM},\text{256-QAM} \right\},
\end{equation}
denotes the set of allowable constellations for the interferer. Assuming that $x_{1}[k],x_{2}[k],\mbf{n}[k]$ are independent for all  $k\!=\!1,\cdots,K$, the ML estimate of the interferer's constellation can then be written as
\begin{align}
  \hat{\mathcal{X}}_{\mathrm{I}}
    =
    \underset{\mathcal{X}_{\mathrm{I}}\in\mathcal{M}}
    {\mathop{\arg\max}}\,\frac{1}{{{\abs{\mathcal{X}_{\mathrm{I}}}}^K}}
    \prod\limits_{k=1}^{K}
    \sum\limits_{x_{1}[k]\in \mathcal{X}_{\mathrm{S}}}
    \sum\limits_{x_2[k]\in \mathcal{X}_{\mathrm{I}}}
    \mathrm{p}\!\left(\! \left. \mbf{y}[k]\, \right|\mbf{H}[k],\mathcal{X}_{\mathrm{S}},\mathcal{X}_{\mathrm{I}},x_1[k],x_2[k] \right),\label{eq:ML_estimate_step3}
\end{align}
where $\abs{\mathcal{X}_{\mathrm{I}}}$ denotes the size of the interfering user's constellation, under the assumption that
\begin{align}\label{eq:prior_prob}
    \mathrm{P}(x_{1}[k]) \!=\! \frac{1}{\abs{\mathcal{X}_{\mathrm{S}}}},~\text{and}~ \mathrm{P}(x_{2}[k]) \!=\! \frac{1}{\abs{\mathcal{X}_{\mathrm{I}}}}, ~k\!=\!1,\cdots,K.
\end{align}
Let $d(\mbf{x}[k])\!=\!\norm{\mbf{y}[k]\!-\!\mbf{H}[k]\mbf{x}[k]}^2/\sigma^2$, we can then write~\eqref{eq:ML_estimate_step3} as
\begin{align}\label{eq:ML_estimate_exp}\hspace{-0.1in}
     \hat{\mathcal{X}}_{\mathrm{I}} \!=\! \underset{ \mathcal{X}_{\mathrm{I}} \in \mathcal{M}}{\mathop{\arg\max }}\,\frac{1}{\abs{\mathcal{X}_{\mathrm{I}}}^{K}}\!\prod\limits_{k=1}^{K} \sum\limits_{\mbf{x}[k]\in \mathcal{X}_{\mathrm{S}}\!\times\! \mathcal{X}_{\mathrm{I}}}{\!\!\!\!\exp{\!(-d(\mbf{x}[k]))}}.
\end{align}
Using the log-max approximation~\cite{1998_Viterbi}, we can approximate the ML estimate $\hat{\mathcal{X}}_{\mathrm{I}}$ by~\cite{2015_Gomaa}
\begin{align}\label{eq:ML_estimate_max_log}\hspace{-0.1in}
    \hat{\mathcal{X}}_{\mathrm{I}} \!\approx\! \underset{\mathcal{X}_{\mathrm{I}}\in \mathcal{M}}{\mathop{\arg\min }}\left(\! K\log\left( \abs{\mathcal{X}_{\mathrm{I}}} \right) + \sum\limits_{k=1}^{K} {\underset{\mbf{x}[k]\in \mathcal{X}_{\mathrm{S}}\times\mathcal{X}_{\mathrm{I}}}{\mathop{\min }}} \!\! d(\mbf{x}[k])  \right),
\end{align}
where $\log(\cdot)$ is the natural logarithmic function.

\begin{figure}[t]
\centering
\includegraphics[scale=1]{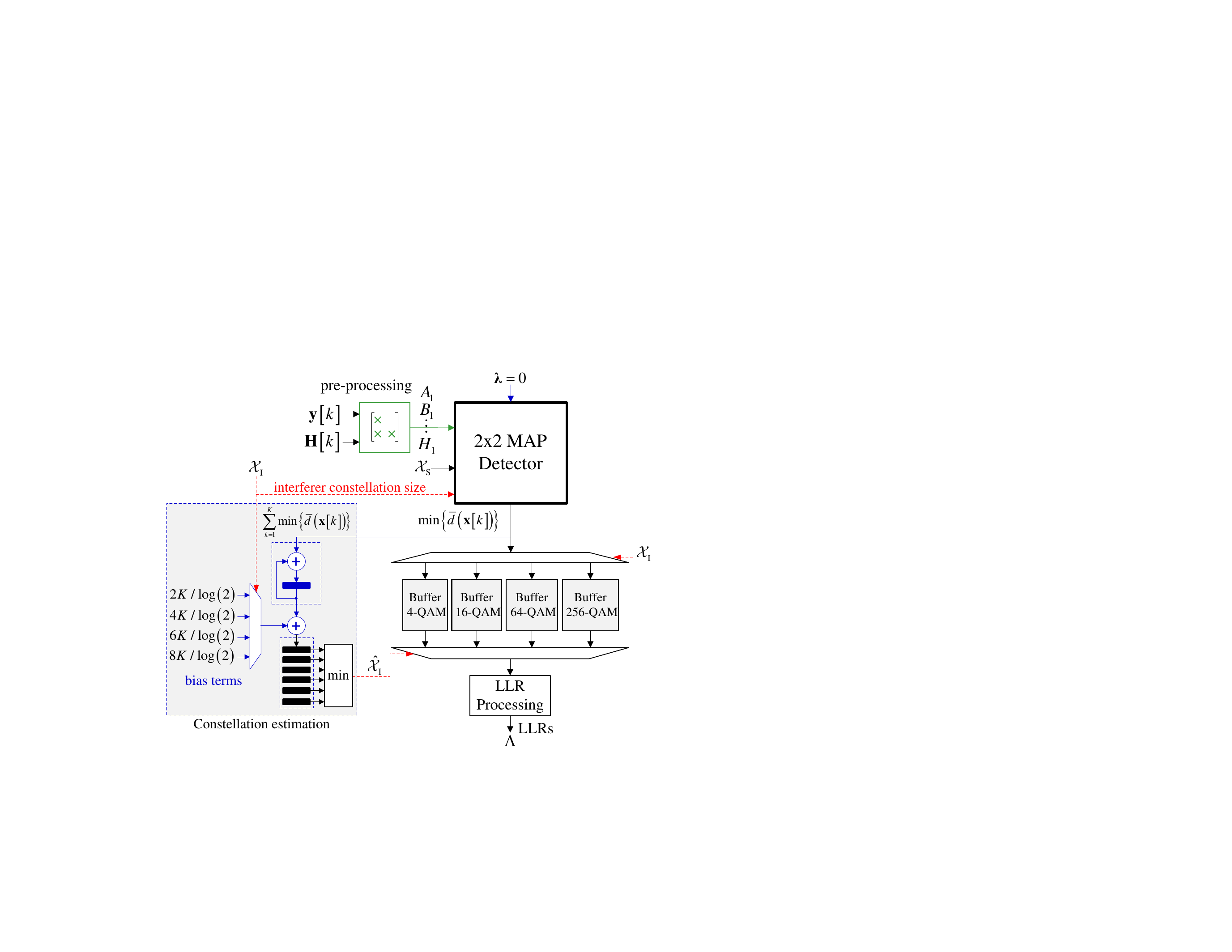}\vspace{-0.15in}
\caption{Block diagram of a MU-MIMO detector}
\label{f:MU_MIMO_core}
\end{figure}

Once the co-scheduled user's constellation, $\hat{\mathcal{X}}_{\mathrm{I}}$, is estimated, then the LLR of the $\nth{j}$ bit of the desired user QAM symbol $x_1[k]$ on the $\nth{k}$ RE is given by \cite{2006_Fitz}
\begin{IEEEeqnarray}{rCl}\label{eq:final_LLR}
    \Lambda_{k,j}^{\MLL} ~\approx
        \underset{\substack{x_1[k]^{}\in \mathcal{X}_{\mathrm{S},j}^{(+1)}\\ x_2[k]^{}\in \hat{\mathcal{X}}_{\mathrm{I}}}}{\mathop{\min}}{\!d\left(\mbf{x}[k]\right)} \,~-
        \underset{\substack{x_1[k]^{}\in \mathcal{X}_{\mathrm{S},j}^{(-1)}\\ x_2[k]^{}\in \hat{\mathcal{X}}_{\mathrm{I}}}}{\mathop{\min}}{\!d\left(\mbf{x}[k]\right)},
\end{IEEEeqnarray}
where $\mathcal{X}_{\mathrm{S},j}^{(+1)}\!=\!\{ x\in \mathcal{X}_{\mathrm{S}}: b_{j}\!=\!+1\}$ and $\mathcal{X}_{\mathrm{S},j}^{(-1)}\!=\!\{ x\in \mathcal{X}_{\mathrm{S}}: b_{j}\!=\!-1\}$. As seen from \eqref{eq:final_LLR}, computing the LLRs involves the same distance computations as those needed for the co-scheduler user's constellation estimation in \eqref{eq:ML_estimate_max_log}. This fact is exploited in the architecture of a joint constellation classifier and data MU-MIMO detector shown in Fig.~\ref{f:MU_MIMO_core}, which uses an optimized one-sided MAP MIMO detector as its core. The MIMO detector processes the received signal $\mbf{y}[k]$ assuming all $4$ possible choices of the interferer's constellation. It generates $4$ corresponding lists of minimum distance metrics $d(\mbf{x}[k])$ and their associated symbol vectors $\mbf{x}[k]$ for all the $\abs{\mathcal{M}}$ possible hypotheses of the interferer's constellation, with $x_1[k]\!\in\!\mathcal{X}_{\mathrm{S}}$. These distances and symbols are stored in $4$ buffers each of size $\abs{\mathcal{X}_{\mathrm{S}}}$ as shown in Fig.~\ref{f:MU_MIMO_core}.

For each tone, the minimum distance from each list is passed to an adder that accumulates the minimum distances over a span of $K$ tones, during which the interferer modulation is assumed to be static. The resulting $4$ minimum accumulated distances for each interferer hypothesis are stored in a buffer. The minimum from this buffer is used to identify the interferer's constellation, and the corresponding stored distances in the buffers are selected and forwarded for LLR processing according to~\eqref{eq:final_LLR}.

Note that since the interferer's modulation constellation remains static over $K$ tones for a duration of $1$ subframe in LTE ($14$ OFDM symbols), the particular choice of $K\!=\!12$ results in substantial savings in computations. The detector only needs to run in the above mode to identify the interferer's constellation for one OFDM symbol in the subframe. It can then switch back to normal ML detection mode (without modulation classification) to generate the LLRs for the remaining 13 OFDM symbols for the user of interest $x_1[k]$.

Taking the LTE scenario for hardware complexity analysis, the total number of possible tones in 1 PRB in a subframe is $12\!\times\!14\!=\!168$. Of these tones, 28 are reserved for pilots (for cell specific reference signals and for UE specific pilots to support the MU-MIMO transmission mode), and 140 for data. In the hardware architecture of Fig.~\ref{f:MU_MIMO_core}, the total number of distance computations needed to generate the LLRs from these 140 data tones is $(140\!+\!12\times 5)\!\times\!\abs{\mathcal{X}_{\mathrm{S}}}$. This corresponds to an increase of only $42.86\%$ compared to the distances computed by an ML detector with perfect knowledge of the interferer.

\begin{figure}[t]
\centering
\includegraphics[scale=1]{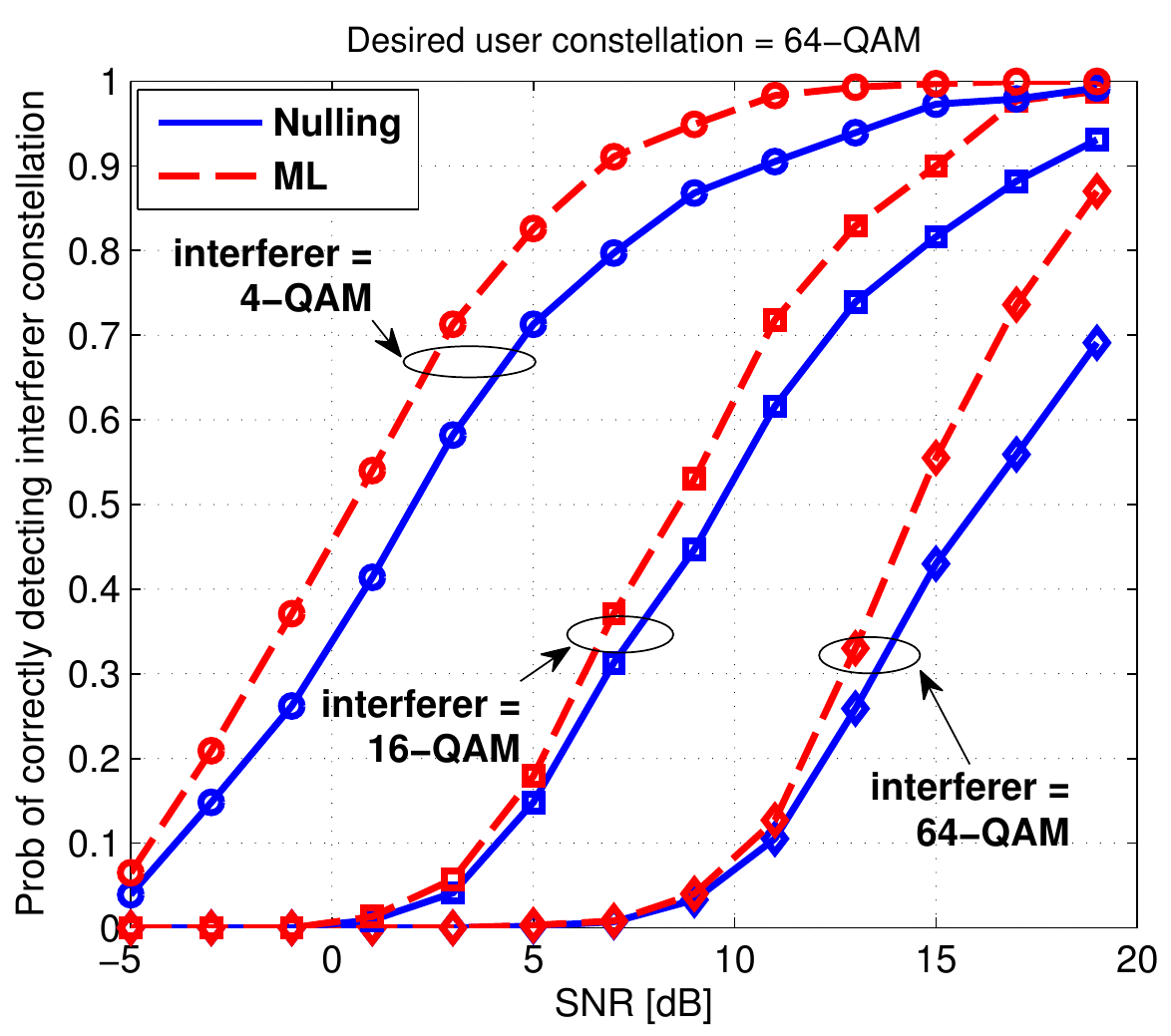}\vspace{-0.1in}
\caption{Probability of correct interferer modulation constellation detection versus SNR~\cite{2015_Gomaa}. Solid lines are for nulling approach, dashed are for the ML approach. Desired user constellation is fixed to 64-QAM and the co-scheduled user constellation is 4-, 16-, and 64-QAM. The channel is i.i.d. block fading.}
\label{f:fading_scenario_sig_64QAM_diffInterfConst}
\end{figure}

Figure~\ref{f:fading_scenario_sig_64QAM_diffInterfConst} shows the results when $\mathcal{X}_{{\mathrm{S}}}$ is 64-QAM, with $\mathcal{X}_{{\mathrm{I}}}$ being 4-, 16-, and 64-QAM using $K\!=\!24$ resource elements. The plots show that the ML classification method has a \unit[5]{dB} gain over the basic nulling approach when $\mathcal{X}_{{\mathrm{S}}}$ is 4-QAM, and \unit[2]{dB} gain in the case of $\mathcal{X}_{{\mathrm{S}}}$ being 64-QAM. Therefore, the gain of the ML classification method is largest for small constellation sizes of the desired signal, i.e., the largest gain is attained when the receiver complexity is minimal.
\begin{figure}[hbtp]
\centering
\includegraphics[scale=1]{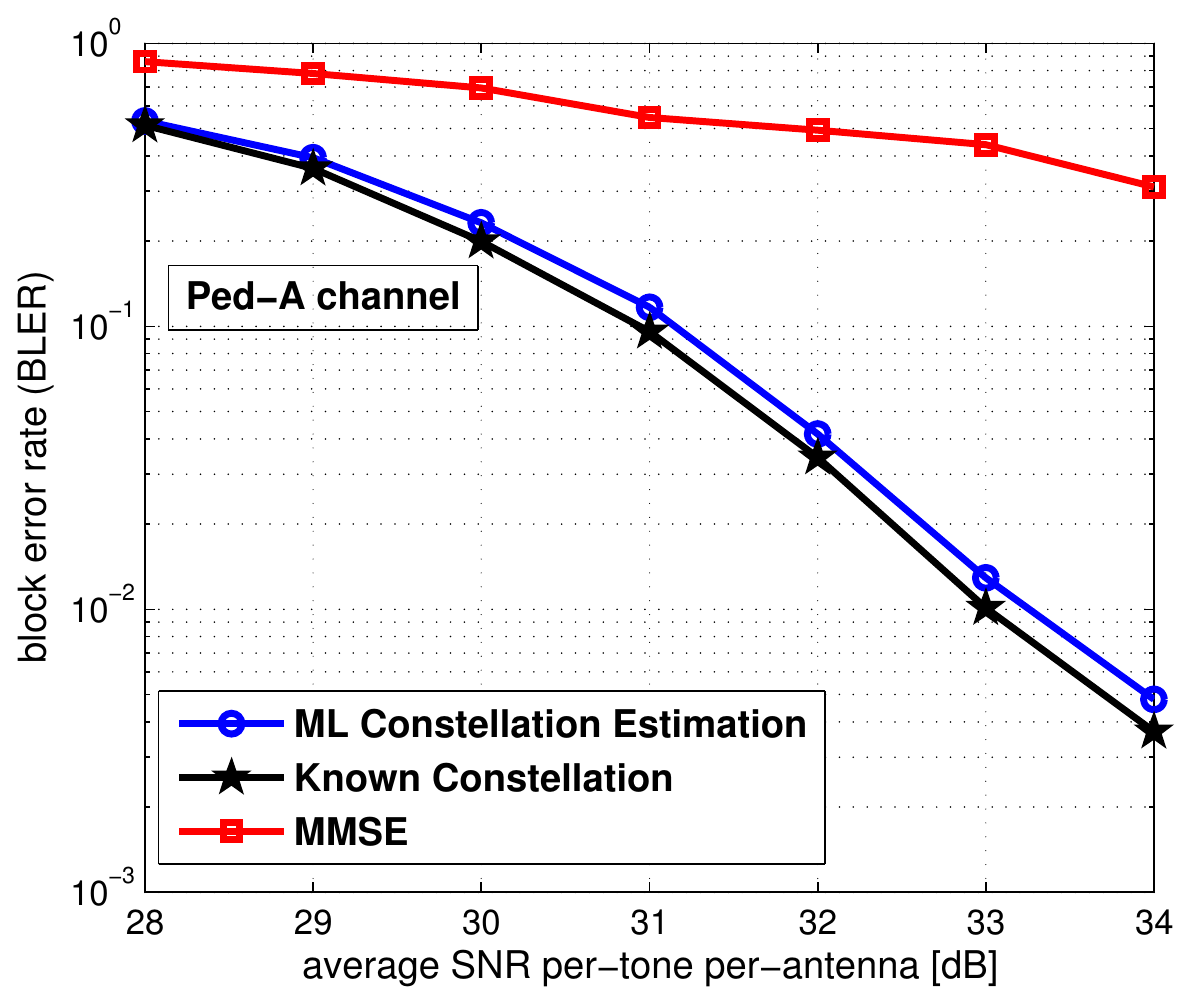}\vspace{-0.1in}
\caption{BLER versus per-tone per-antenna SNR (dB). EPA channel, high correlation (0.9), $64$-QAM for both users and code-rate 1/2.}
\label{f:PedA_BLER}
\end{figure}

Figure~\ref{f:PedA_BLER} shows the performance of the joint ML classification and detection method as compared to an ML receiver that has perfect knowledge of the interfering user's constellation. Also shown in the figure is the performance of the linear MMSE receiver that only uses the knowledge of the interfering user's channel and does not exploit knowledge of the interferer's constellation. Both users use $64$-QAM, with the turbo code of~\cite{LTE_36.212} and encoding rate $1/2$ using block size $6144$ bits. The pedestrian-A (Ped-A)~\cite{PedB_ChannelModel} multi-path fading channel with high antenna correlation was used. The effective channel matrix is given by $\mbf{H}=\mbf{R}^{1/2}_t \mbf{H}_c \mbf{R}^{1/2}_r$, where $\mbf{H}_c$ is channel whose entries are uncorrelated and generated according to the Ped-A model, $\mbf{R}_t$ and $\mbf{R}_r$ are the transmit and receive antenna $2\!\times\!2$ correlation matrices, respectively, which have $1$ on the diagonal entries and $0.9$ on the off-diagonal. As seen from Fig.~\ref{f:PedA_BLER}, the joint ML classification and detection receiver is only \unit[0.1]{dB} away from an ML receiver that has perfect knowledge of the co-scheduled user constellation. The MMSE method has a significant performance degradation as compared to the joint ML classification and detection receiver.

%
\vspace{-0.05in}
\section{Implementation and Simulation Results}\label{s:sim}\vspace{-0.05in}
The proposed $2\times 2$ reconfigurable MIMO detector architecture was modeled in VHDL and synthesized on a Xilinx Virtex$^{\circledR}$-6 FPGA. The core was also synthesized using a \unit[90]{nm} CMOS ASIC library. The experimental simulations below evaluate the coded bit-error rate (BER) performance of the proposed detection algorithm and the implemented core, assuming a MIMO system employing either 2 transmit and 2 receive antennas, or 4 transmit and 4 receive antennas. The channel encoder is based on the LTE turbo encoder specification~\cite{LTE_36.211} with interleaver length 1024, using 16-QAM, 64-QAM, and 256-QAM modulation constellations. The channel entries are assumed to be i.i.d. complex Gaussian random variables with unit variance. At the receiver end, we assume perfect channel knowledge. The turbo decoder implements the true \emph{A Posteriori} Probability algorithm, and performs 4 full decoding iterations. Also, the detector and turbo decoder perform up to 4 outer joint detection and decoding iterations. Channel decomposition is performed externally by a pre-processing stage and the coefficients in~\eqref{eq:constant_A_B_2x2}-\eqref{eq:constant_E_F_G_H_2x2} are fed as input.

\begin{figure}[hbtp]
\centering
\includegraphics[scale=1]{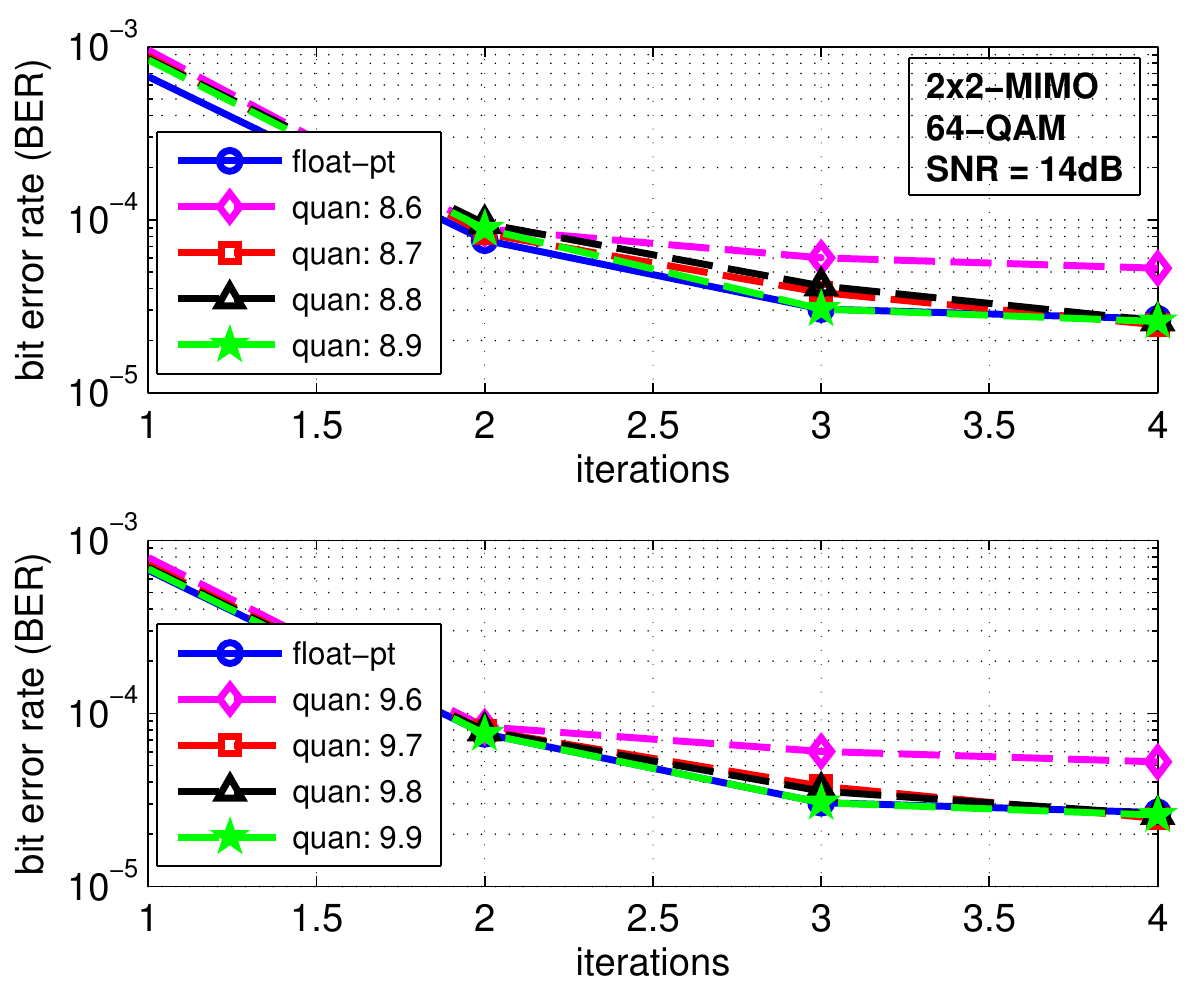}\vspace{-0.05in}
\caption{BER vs. outer detection-decoding iterations for various bit-precisions at $\text{SNR}\!=\!\unit[14]{dB}$ for a $2\times 2$ MIMO system with 64-QAM.}
\label{f:berplots_2x2_64QAM_QaunAnalysis}
\end{figure}

%
\vspace{-0.05in}
\subsection{Performance Results}\label{s:performance_sim}
\begin{figure}[t]
\centering
\includegraphics[scale=1]{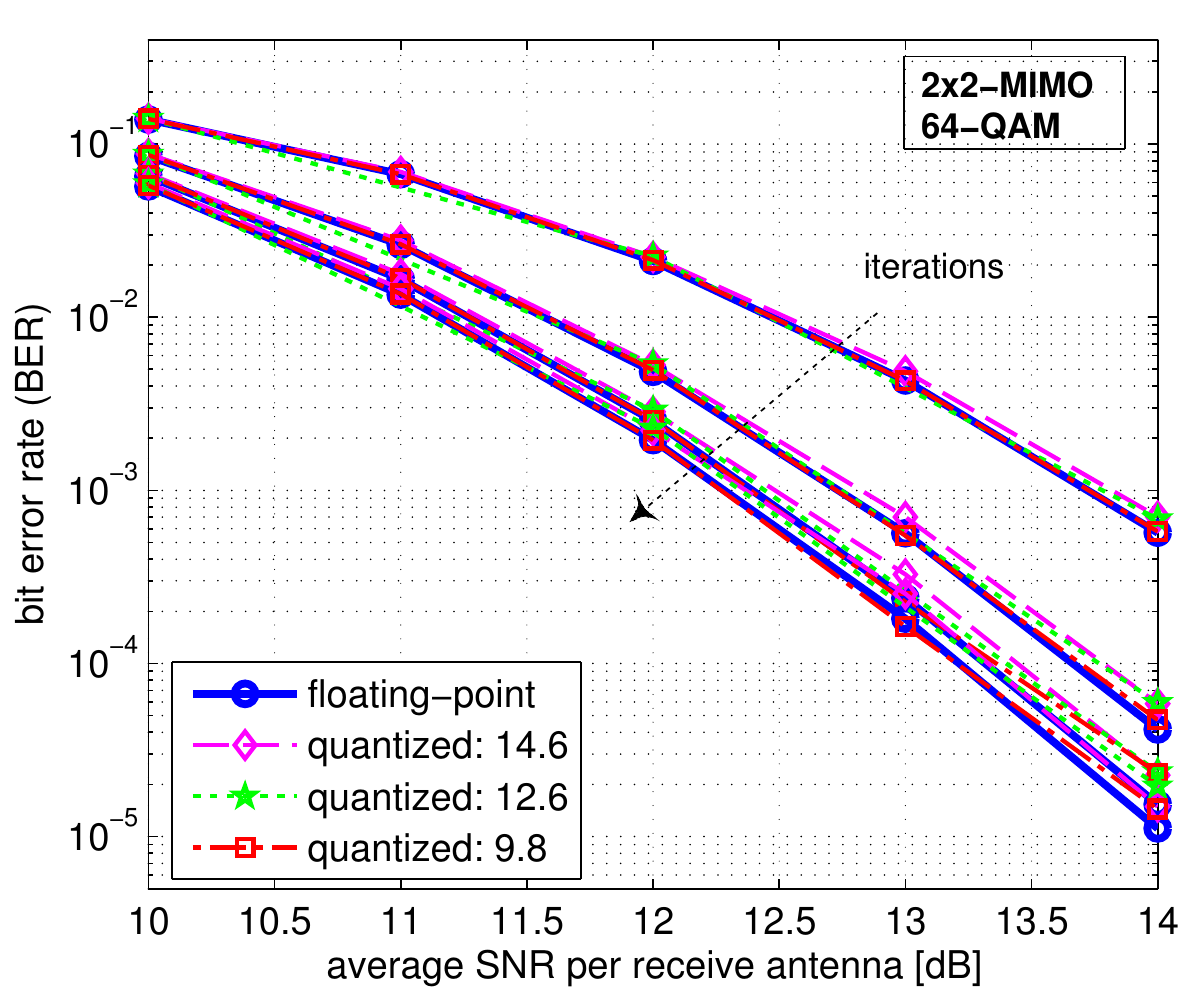}\vspace{-0.05in}
\caption{BER vs. SNR for various bit-precisions and up to 4 outer detection-decoding iterations, for a $2\times 2$ MIMO system with 64-QAM.}
\label{f:berplots_2x2_64QAM}
\end{figure}

\vspace{-0.05in}
The bit-precision of the detector architecture can be configured to enable tradeoff analysis between gate complexity and tolerable degradation in BER performance due to quantization noise. Figure~\ref{f:berplots_2x2_64QAM_QaunAnalysis} compares the BER performance of the detector core for 2 layers and 64-QAM under various integer and fractional bit-widths, versus floating-point performance, at $\text{SNR}\!=\!\unit[14]{dB}$. The x-axis denotes the number of joint detection and decoding iterations. The top figure corresponds to a fixed-point representation of $(I.F)\!=\!\{8.6,8.7,8.8,8.9\}$, where $I$ denotes integer bit-precision while $F$ denotes fractional bit-precision. The bottom figure corresponds to the representation of $(I.F)\!=\!\{9.6,9.7,9.8,9.9\}$. As can be seen, when $F$ starts to drop to 6, the BER starts to degrade. There is no significant improvement in BER in going beyond $I\!=\!9$ integer bits, as demonstrated also in Fig.~\ref{f:berplots_2x2_64QAM}.

\begin{figure}[hbtp]
\centering
\includegraphics[scale=1]{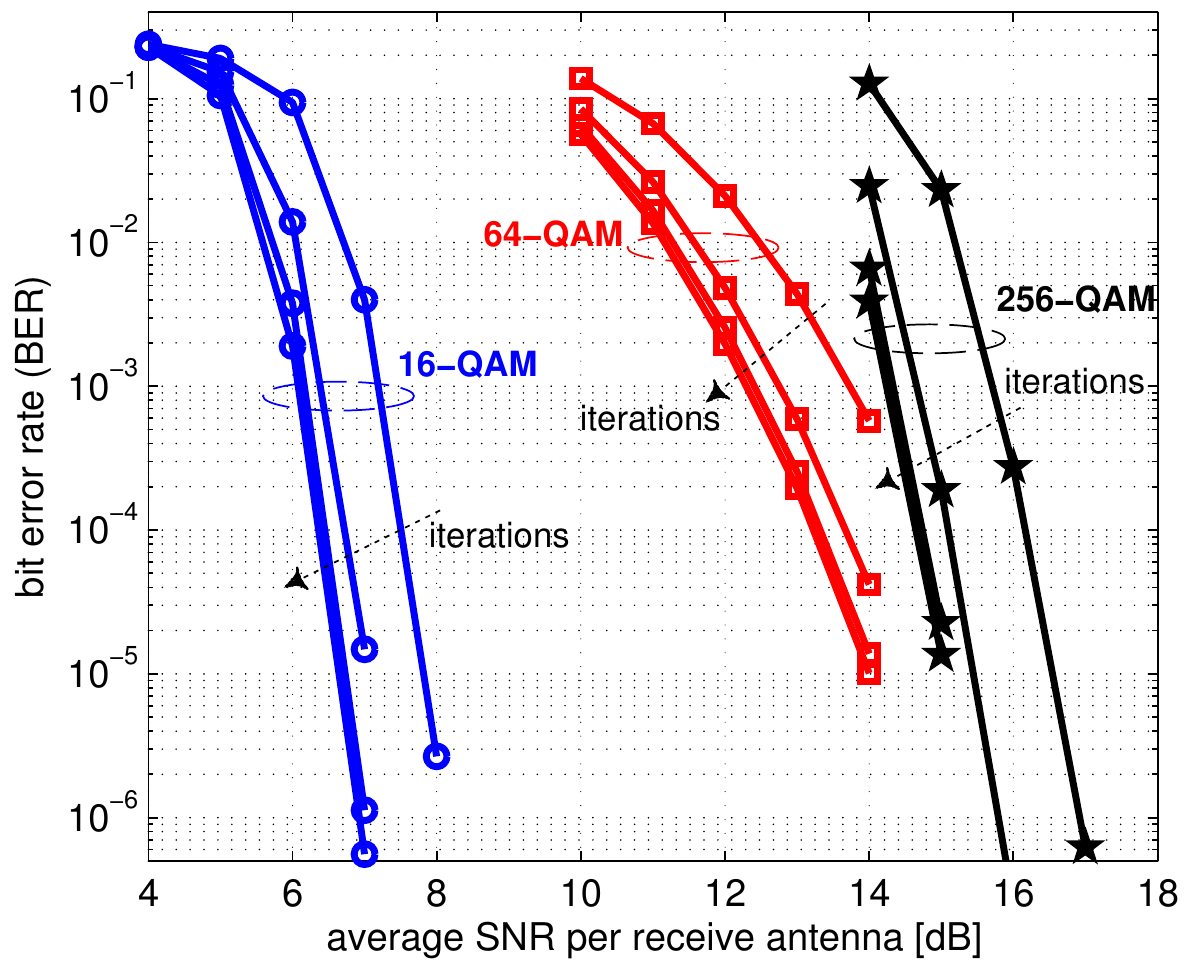}
\caption{BER vs. SNR for a $2\times 2$ MIMO system with 16-, 64-, and 256-QAM.}
\label{f:berplots_2x2_allQAM}
\end{figure}

Figure~\ref{f:berplots_2x2_allQAM} compares the BER performance of the core using 16-QAM, 64-QAM, and 256-QAM. The plots demonstrate that most of the coding gain is attained after 3 outer iterations, assuming the inner turbo decoder performs at most 4 full turbo decoding iterations.

\begin{figure}[hbtp]
\centering
\includegraphics[scale=1]{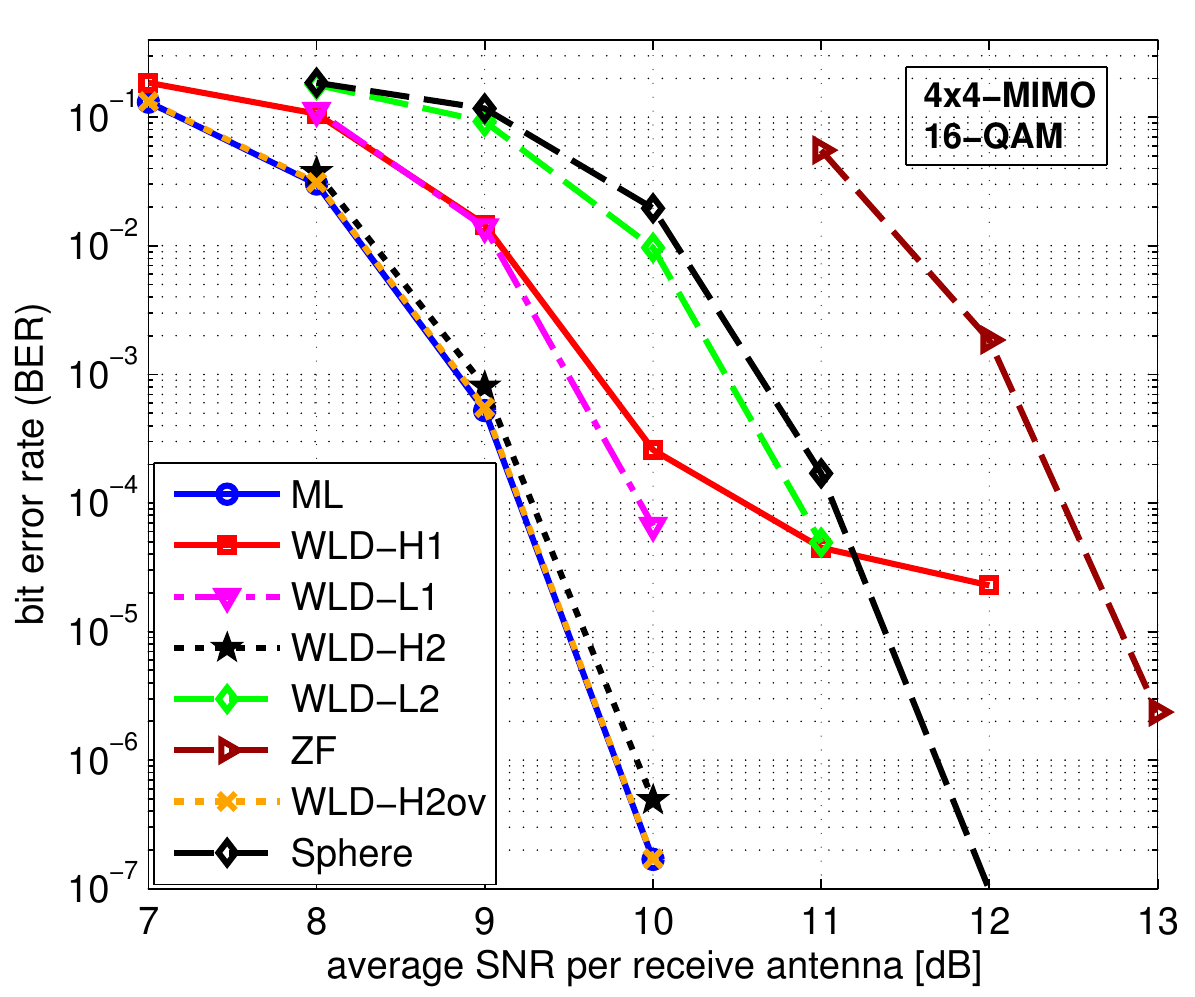}\vspace{-0.05in}
\caption{BER vs. SNR plots for a $4\!\times\! 4$ MIMO system with 16-QAM.}
\label{f:berplots_4x4_16QAM}
\end{figure}

\begin{figure}[hbtp]
\centering
\includegraphics[scale=1]{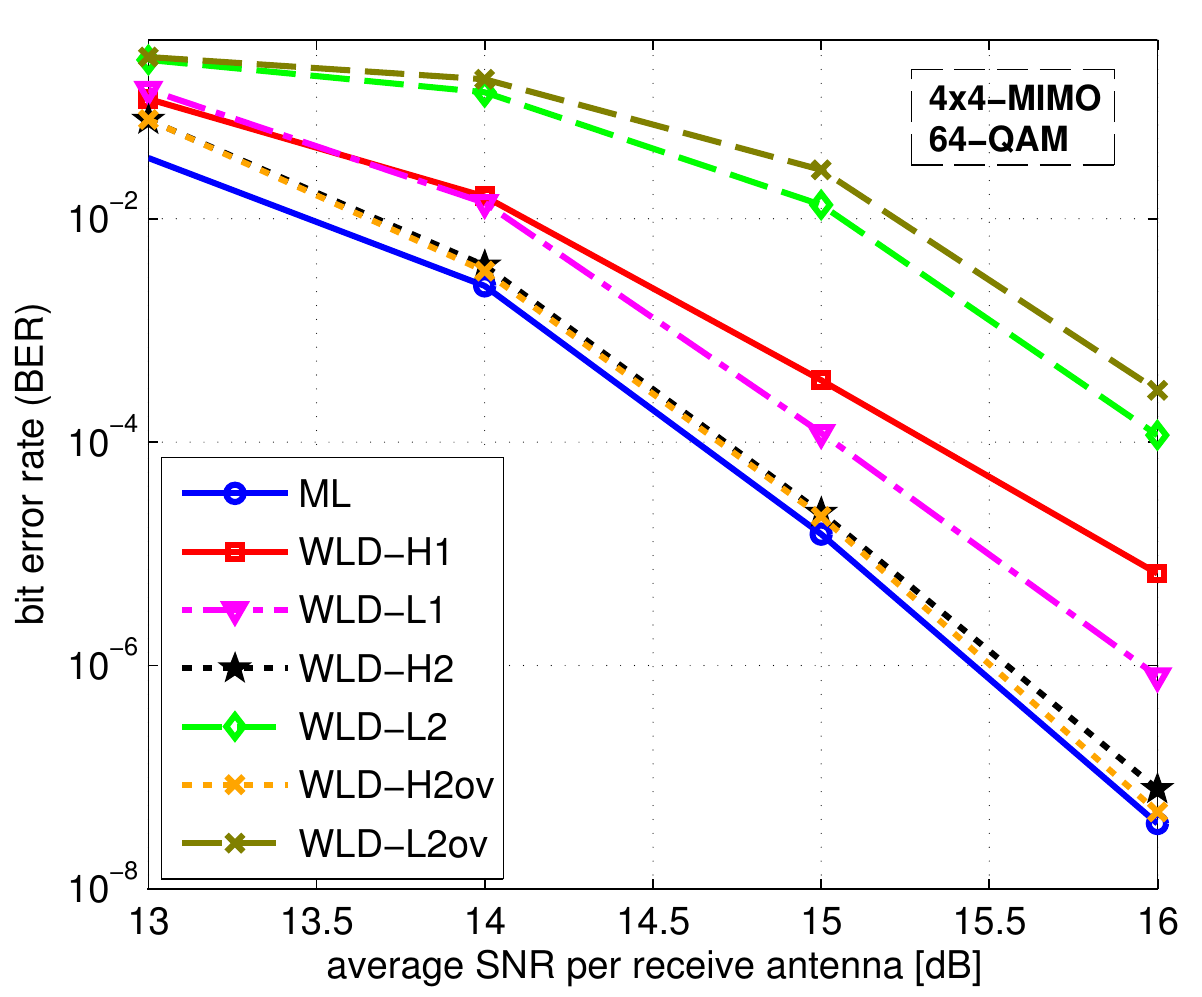}\vspace{-0.05in}
\caption{BER vs. SNR plots for a $4\!\times\! 4$ MIMO system with 64-QAM.}
\label{f:berplots_4x4_64QAM}
\end{figure}

In Figs.~\ref{f:berplots_4x4_16QAM} and~\ref{f:berplots_4x4_64QAM}, the BER  performance of a $4\times 4$ MIMO system using the proposed WLD scheme is simulated. In Fig.~\ref{f:berplots_4x4_16QAM}, the plots compare the BER versus SNR of the proposed WLD scheme with $E\!=\!1$ and 2 structures (Fig.~\ref{f:dataflow_matrices}a-\ref{f:dataflow_matrices}b), versus ML, zero-forcing (ZF), the approach of~\cite{2009_Ojard}, and the sphere decoder with radius clipping~\cite{2008_Studer}, for 16-QAM. Both overlapping and non-overlapping subsets are considered. Two scenarios for distance computations in~\eqref{eq:norms_inequality} are followed; one based on $\mbf{H}$ and one on $\mbf{L}$. The plots demonstrate that WLD with $E=2$ using $\mbf{H}$ distances with overlapping subsets performs virtually as ML, and is less than \unit[0.1]{dB} away from ML with no overlapping. Also, for single streams, $\mbf{L}$ distances perform better than $\mbf{H}$ distances. The plots correspond to one outer detection-decoding iteration, and 4 full internal turbo decoder iterations.

\begin{figure}[hbtp]
\centering
\includegraphics[scale=1]{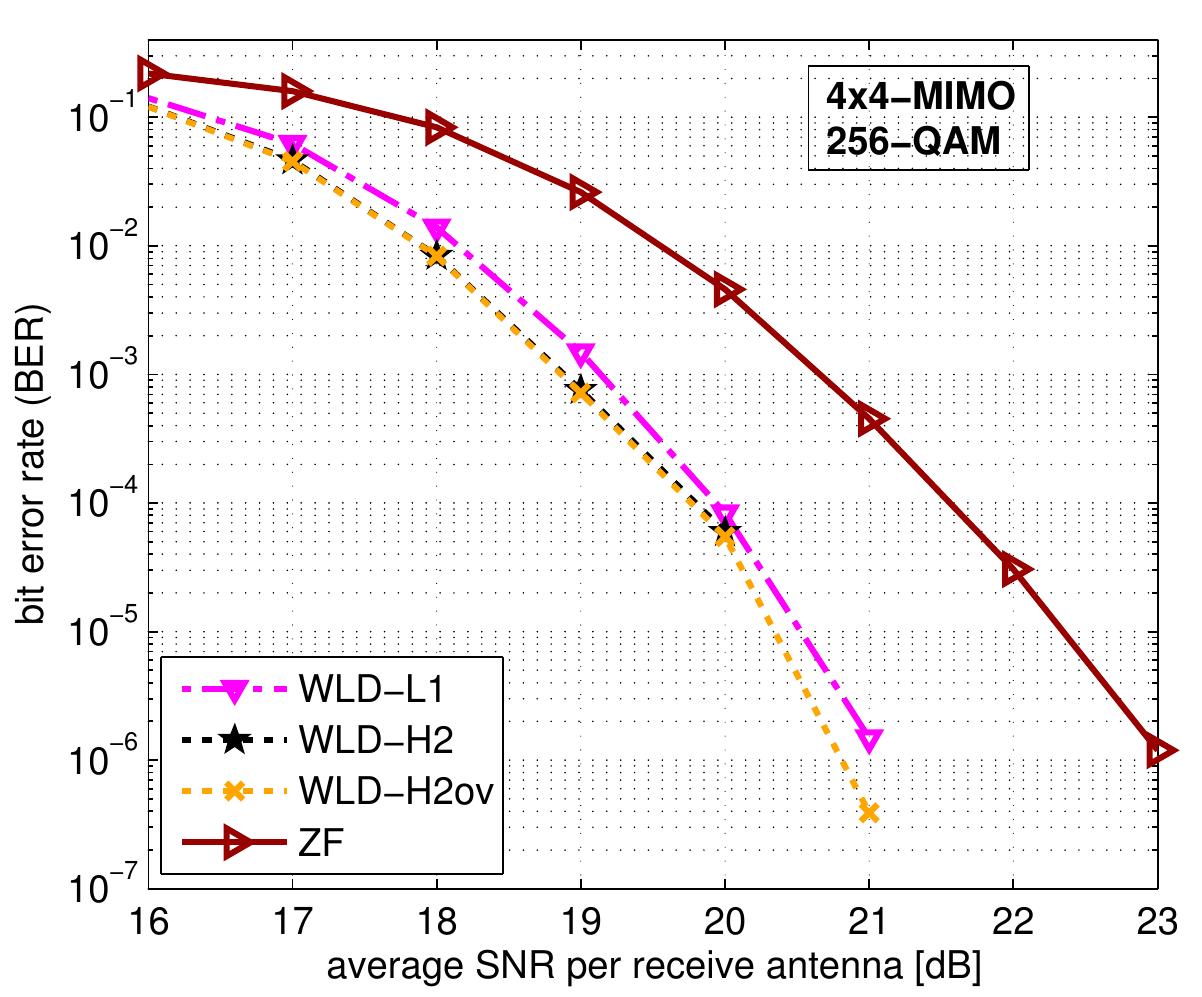}\vspace{-0.05in}
\caption{BER vs. SNR plots for a $4\!\times\! 4$ MIMO system with 256-QAM.}
\label{f:berplots_4x4_256QAM}
\end{figure}

Figure~\ref{f:berplots_4x4_64QAM} compares the BER performance for 64-QAM. The plots demonstrate again that the WLD scheme with $E\!=\!2$ using $\mbf{H}$ distances and overlapping subsets performs very close to ML. Figure~\ref{f:berplots_4x4_256QAM} shows the results for 256-QAM.

%
\subsection{Architecture Synthesis Results}\label{s:arch_synthesis_results}
Various architecture configurations for the $2\times 2$ core with different algorithmic features and architectural optimizations were synthesized, assuming 17-bit datapaths. The datapaths are pipelined with 6 stages and clocked at \unit[275]{MHz}. The input LLRs fed from the turbo decoder are 8 bits wide. The output LLRs from the detector are passed to a dynamic scaling block (not included in this work) that scales the bit-widths down to 8 bits before feeding them to the turbo decoder.

\begin{figure}[t]
\centering
\includegraphics[scale=1]{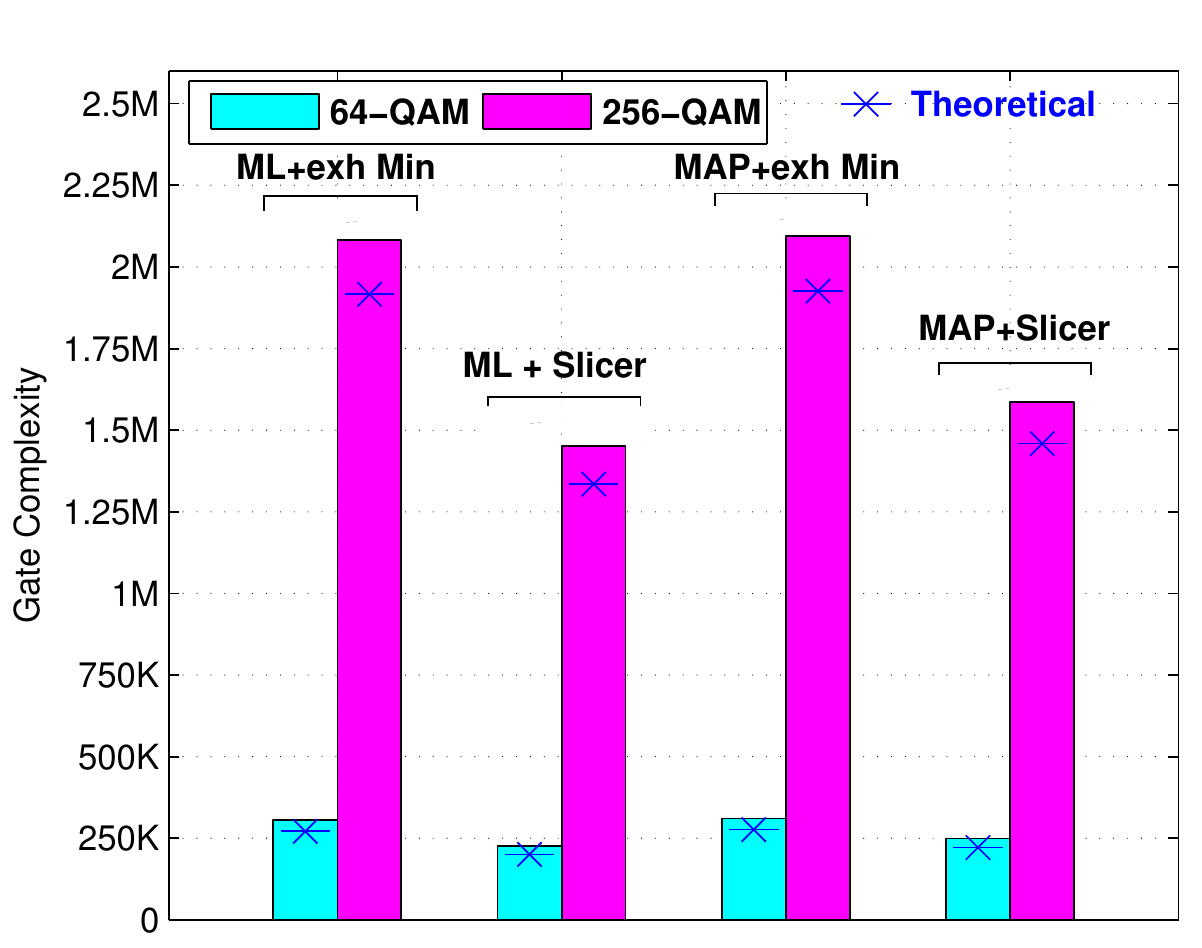}\vspace{-0.05in}
\caption{Hardware complexity of various synthesized detector cores.}
\label{f:hw_complexity}
\end{figure}

Figure~\ref{f:hw_complexity} shows the gate complexity of 8 different architectures. Four architectures support reconfigurable constellations up to 64-QAM, while the other four support up to 256-QAM. For the 64-QAM case, two architectures are designed to support soft-outputs only without soft-inputs (i.e. ML detection, see Section~\ref{s:2x2ML}): one based on distance minimizations using exhaustive search (Section~\ref{s:min_exh_search_arch}), and one based on minimization via slicing (Section~\ref{s:min_slicing_arch}). The other two 64-QAM architectures support both soft-outputs and soft-inputs (i.e. MAP detection, see Section~\ref{s:nonzero_priors}), one with minimization based on exhaustive search and one via slicing. The other four 256-QAM architectures are similar. All architectures have the same input/output interfaces, external buffers, and control logic. The reported gate counts in gate-equivalent (GE) are for the core logic only.

The plots demonstrate that there is a significant increase in complexity (between $6.35\mathsf{x}$-$6.82\mathsf{x}$) when supporting 256-QAM compared to 64-QAM. Furthermore, the slicer-based architectures using the proposed scheme in Section~\ref{s:min_slicing_arch} offer significant reduction in complexity compared to distance minimization by search (between $19.58\%$-$26.22\%$ for 64-QAM, and between $24.28\%$-$30.35\%$ for 256-QAM). Finally, for slicer-based architectures, supporting soft-inputs for MAP detection comes with an increase in gate count between $8.49\%$-$9.83\%$ compared to soft-output-only ML detection. For minimization-by-search architectures, the overhead of supporting soft-inputs is only between $0.51\%$-$1.71\%$. The gate counts predicted by the theoretical analyses in Section~\ref{s:complexity_par_arch} are also plotted in Fig.~\ref{f:hw_complexity}. The error ranges between $8\%$-$11\%$, which asserts the validity of the model used and the theoretical analysis performed.

\begin{figure}[t]
\centering
\includegraphics[scale=0.75]{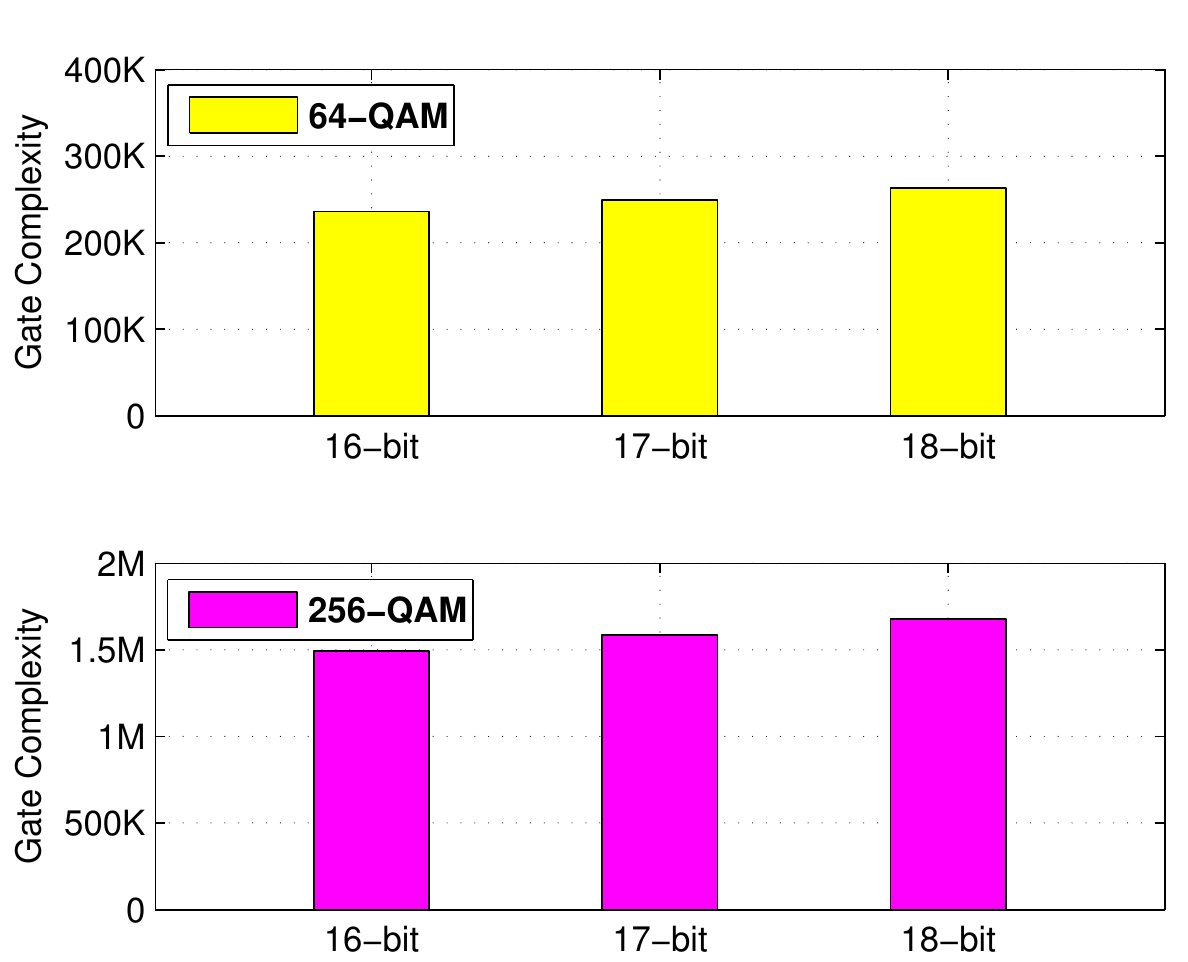}\vspace{-0.05in}
\caption{Hardware complexity as a function of bit-width.}
\label{f:hw_complexity_bitprecision}
\end{figure}

Figure~\ref{f:hw_complexity_bitprecision} plots the gate complexity of the slicer-based MAP cores as a function of bit-width. The complexity increases roughly between $5.2\%$-$5.9\%$ for every added bit. A similar trend was observed when synthesizing the 256-QAM core with soft-outputs only on a Virtex-6 FPGA. The area increases from 317937 LUTs ($33\%$) for 18 bits to 337210 LUTs ($35\%$) for 19 bits. The area jumps to 403498 LUTs $(42\%)$ when the integer bit-width is increased to 12 bits.

The core achieves an average SNR-independent throughput of \unit[2.2]{Gbps} for 2-layers with 256-QAM, when running in soft-input soft-output mode. In $4 \times 4$ mode, the core achieves a throughput of \unit[733]{Mbps} and consumes \unit[320.56]{mW} of power. This compares favorably with other detectors in the literature with throughput ranging from \unit[757]{Mbps} at \unit[410]{kGE}~\cite{2011_Studer_JSSC}; \unit[772]{Mbps} at \unit[212]{kGE}~\cite{2011_Borlenghi}; \unit[1.2]{Gbps} at \unit[1097]{kGE} for 16-QAM~\cite{2012_Sun_Cavallaro}; and \unit[2.2]{Gbps} at \unit[555]{kGE}~\cite{2013_Chen} for up to 64-QAM only. Table~\ref{t:implementation_summary} provides a comparative summary of our implemented detector and the detectors in\cite{2011_Studer_JSSC,2011_Borlenghi,2012_Sun_Cavallaro,2013_Chen}.

\begin{table*}[t]
\centering
\begin{threeparttable}[b]
\centering
\caption{Summary and comparison of implementation results}\vspace{-0.1in}
\label{t:implementation_summary}
\renewcommand{\arraystretch}{1.05}
\setlength{\tabcolsep}{0.25em}
\begin{tabular}{r|p{1.5cm}p{1.2cm}p{1.1cm}p{1.0cm}p{0.9cm}}
  \specialrule{.2em}{0em}{0em}
  Reference & This work & \cite{2011_Studer_JSSC} & \cite{2011_Borlenghi} & \cite{2012_Sun_Cavallaro} & \cite{2013_Chen}  \\\specialrule{.2em}{0em}{0em}
  Antennas & $\leq 4\!\times\!4 $  & $\leq 4\!\times\!4 $  & $\leq 4\!\times\!4 $  & $\leq 4\!\times\!4 $ & $4\!\times\!4 $  \\\hline
  Modulation [QAM] &$\leq 256 $  &  $\leq 64 $ &  $\leq 64 $ & 16 & 64 \\\hline
  \multirow{2}{*}{Algorithm} & \multirow{2}{*}{WLD}  & MMSE & \multirow{2}{*}{STS-SD} & Trellis  & \multirow{2}{*}{FSD}\\
                             &                       & -PIC &                         & search   & \\\hline
  Iterative & YES & YES & YES & YES & YES\\\specialrule{.2em}{0em}{0em}
  Technology [nm] & 90  & 90  & 90 & 65 & 90\\\hline
  Core Area [kGE]\tnote{a}~ & 1580 & 410\tnote{b} & 212  & 1097 & 555\\\hline
  Clock freq. [MHz] & 275  & 568 & 193  & 320  & 370 \\\hline
  Maximum & 2200 $(2\!\times\!2)$  & \multirow{2}{*}{757} & \multirow{2}{*}{772}  & \multirow{2}{*}{1200\tnote{c}} & \multirow{2}{*}{2200} \\
  Throughput [Mbps] & 733 $(4\!\times\!4)$  &  &  &  & \\\hline
  Normalized hardware & 0.72 $(2\!\times\!2)$ &  \multirow{2}{*}{0.54} & \multirow{2}{*}{0.28} & \multirow{2}{*}{0.91} & \multirow{2}{*}{0.25}\\
  efficiency [kGE/Mbps] &  2.16 $(4\!\times\!4)$ &  &  &  & \\\specialrule{.2em}{0em}{0em}
  Power consumption & 320.56   & 189.1  & 87.62 & \multirow{2}{*}{\textemdash} & 335.8\\
  in [mW] @ [Mbps]  & @ 733 & @ 757 & @ 772 &  & @ 2200\\\hline
  Energy efficiency & \multirow{2}{*}{0.44}  & \multirow{2}{*}{0.25}  & \multirow{2}{*}{0.11} & \multirow{2}{*}{\textemdash} & \multirow{2}{*}{0.15}\\ \specialrule{0em}{0em}{0em}
  in [nJ/bit] &  &  &  &  & \\\specialrule{.2em}{0em}{0em}
\end{tabular}
\begin{tablenotes}
\item [a] One gate-equivalent corresponds to a 2-input drive-1 NAND gate.\\
\item [b] Includes preprocessing circuitry.\\
\item [c] Technology scaling to \unit[90]{nm} CMOS technology according to $A\sim 1/s$, $t_{\text{pd}}\sim 1/s$, and $P_{\text{dyn}}\sim (1/s)(V_{\text{dd}}/V'_{\text{dd}})$~\cite{2011_Studer_JSSC}.
\end{tablenotes}

\end{threeparttable}
\end{table*}

%
\vspace{-0.05in}
\section{Conclusions}\label{s:conclusion}\vspace{-0.05in}
A configurable 2-layer soft-input soft-output MIMO detector core has been proposed as a basic building block for constructing detectors with more spatial streams. Optimizations targeting distance computations and slicing operations reduce the overall complexity when supporting constellations up to 256-QAM. By appropriately decomposing the MIMO channel, multi-layer detection is casted in terms of multiple parallel 2-layer detection problems, which can be mapped onto the 2-layer core. Various architectures have been developed to achieve a high target detection throughput. The proposed core has been applied as well to the design an optimal MU-MIMO detector for LTE. The core occupies an area of \unit[1.58]{MGE} and achieves a throughput of \unit [733]{Mbps} with \unit[320.56]{mW} of power for 256-QAM when synthesized in \unit[90]{nm} CMOS. Future work will target expanding the core to handle 1024-QAM.

\vspace{-0.1in}
\bibliographystyle{IEEEtran}
\vspace{-0.05in}
\bibliography{IEEEabrv,MIMOdetectorbib}

\begin{IEEEbiography}[{\includegraphics[width=1in,height=1.25in,clip,keepaspectratio]{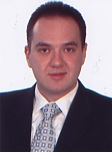}}]{Mohammad M. Mansour} (S'97-M'03-SM'08) received the B.E. (Hons.) and the M.E. degrees in computer and communications engineering from the American University of Beirut (AUB), Beirut, Lebanon, in 1996 and 1998, respectively, and the M.S. degree in mathematics and the Ph.D. degree in electrical engineering from the University of Illinois at Urbana–Champaign (UIUC), Champaign, IL, USA, in 2002 and 2003, respectively.

\noindent He was a Visiting Researcher at Broadcom, Sunnyvale, CA, USA, from 2012 to 2014, where he worked on the physical layer SoC architecture and algorithm development for LTE-Advanced. He was on research leave with Qualcomm Flarion Technologies in Bridgewater, NJ, USA, from 2006 to 2008, where he worked on modem design and implementation for 3GPP-LTE, 3GPP2-UMB, and peer-to-peer wireless networking physical layer SoC architecture and algorithm development. He was a Research Assistant at the Coordinated Science Laboratory (CSL), UIUC, from 1998 to 2003. He worked at National Semiconductor Corporation, San Francisco, CA, with the Wireless Research group in 2000. He was a Research Assistant with the Department of Electrical and Computer Engineering, AUB, in 1997, and a Teaching Assistant in 1996. He joined as a faculty member
with the Department of Electrical and Computer Engineering, AUB, in 2003, where he is currently an Associate Professor. His research interests are in the area of energy-efficient and high-performance VLSI circuits, architectures, algorithms, and systems for computing, communications, and signal processing.

\noindent Prof. Mansour is a member of the Design and Implementation of Signal Processing Systems (DISPS) Technical Committee Advisory Board of the IEEE Signal Processing Society. He served as a member of the DISPS Technical Committee from 2006 to 2013. He served as an Associate Editor for {\sc IEEE Transactions on Circuits and Systems II} (TCAS-II) from 2008 to 2013. He currently serves as an Associate Editor of the {\sc IEEE Transactions on VLSI Systems} since 2011, and an Associate Editor of the {\sc IEEE Signal Processing Letters} since 2012. He served as the Technical Co-Chair of the IEEE Workshop on Signal Processing Systems in 2011, and as a member of the Technical Program Committee of various international conferences and workshops. He was the recipient of the PHI Kappa PHI Honor Society Award twice in 2000 and 2001, and the recipient of the Hewlett Foundation Fellowship Award in 2006. He has six issued U.S. patents.
\end{IEEEbiography}
\vfill
\begin{IEEEbiography}[{\includegraphics[width=1in,height=1.25in,clip,keepaspectratio]{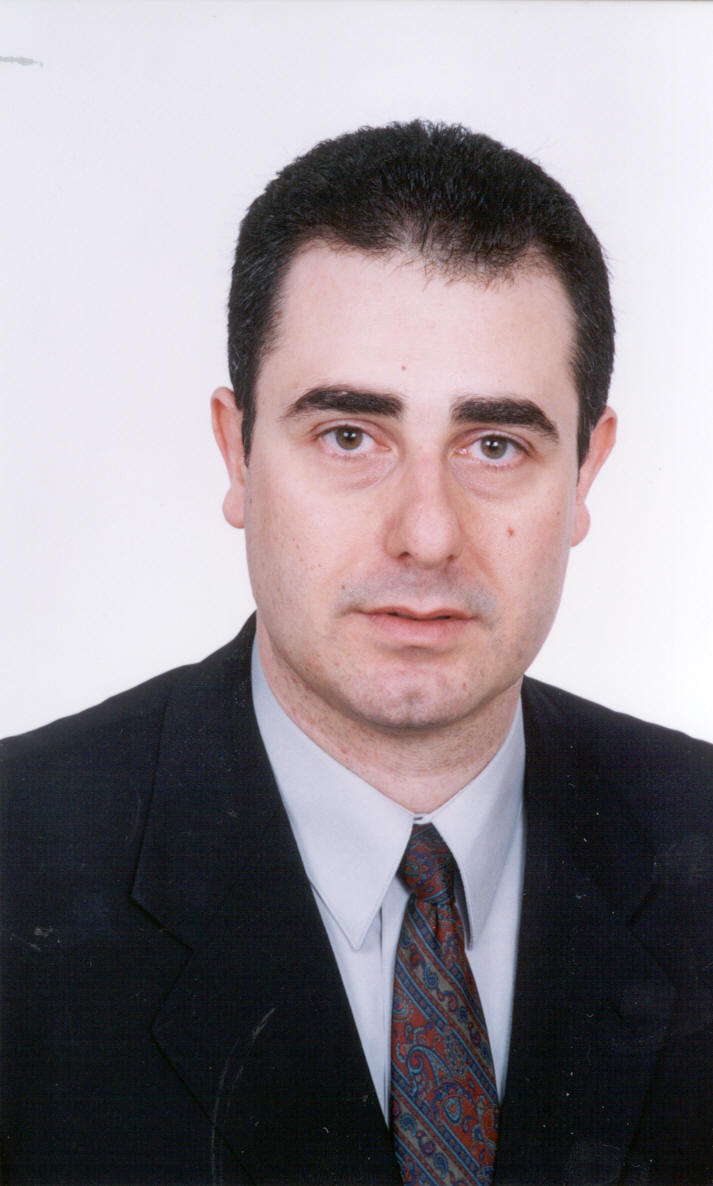}}]{Louay M.A. Jalloul} (M'91-SM'00) received the B.S. degree from the University of Oklahoma,
Norman, OK, USA, in 1985; the M.S. degree from
the Ohio State University, Columbus, OH, USA, in
1988; and the Ph.D. degree from Rutgers, The State
University of New Jersey, Piscataway, NJ, USA, in
1993, all in electrical engineering. He was a Research
Associate with the ElectroScience Laboratory,
Ohio State University; and the Wireless Information
Networks Laboratory (WINLAB), Rutgers.

He is currently a Technical Director with Broadcom
Corporation, Sunnyvale, CA, USA. Prior to that, he was a Senior Director
of Technology with Beceem Communications Inc. (a Silicon Valley startup
providing solutions for mobile broadband wireless communication systems).
From September 2004 to September 2005, he was an Associate Professor with
the Department of Electrical and Computer Engineering, American University
of Beirut, Beirut, Lebanon. In February 2001, he joined MorphICs Technology
Inc., Campbell, CA (acquired by Infineon Technologies AG in April 2003) as
the Director of Systems Architecture, where he led his team in the development
of the code-division multiple access (CDMA) cellular digital signal processor
for the third-generation wideband CDMA standard. From 1993 to 2001, he was
with Motorola Inc., taking on various functions in research and development.
He contributed to the early concepts of high-speed downlink packet access and IS-2000 evolution to voice and data
(1XEV-DV).

Dr. Jalloul has 57 issued U.S. patents and received numerous engineering awards for his innovations to Motorola products. He is a member of Eta Kappa Nu.
\end{IEEEbiography}

\end{document}

%% file: table_integer_multiples.tex
\begin{table}[t]
\centering
\caption{Constants that appear in $(E\!\abs{x_{1\re}}\!\pm\! F\!\abs{x_{1\im}})\!\abs{x_{2\re}}$ for 16-PAM}\vspace{-0.1in}
\label{t:integer_multiples}
\renewcommand{\arraystretch}{1.1}
\begin{tabular}{|@{\hspace{0.5mm}}c@{\hspace{0.5mm}}|@{\hspace{0.5mm}}c@{\hspace{0.5mm}}|@{\hspace{0.5mm}}c@{\hspace{0.5mm}}|@{\hspace{0.5mm}}c@{\hspace{0.5mm}}|}
  \hline
    $3\!=\!2\!+\!1$     &
    $5\!=\!4\!+\!1$     &
    $7\!=\!8\!-\!1$     &
    $9\!=\!8\!+\!1$     \\\hline
    $11\!=\!8\!+\!\mbf{3}$    &
    $13\!=\!16\!-\!\mbf{3}$   &
    $15\!=\!16\!-\!1$     &
    $21\!=\!16\!+\!\mbf{5}$ \\\hline
    $25\!=\!16\!+\!\mbf{9}$ &
    $27\!=\!32\!-\!\mbf{5}$ &
    $33\!=\!32\!+\!1$       &
    $35\!=\!32\!+\!\mbf{3}$ \\\hline
    $39\!=\!32\!+\!\mbf{7}$ &
    $45\!=\!32\!+\!\mbf{13}$ &
    $49\!=\!64\!-\!\mbf{15}$ &
    $55\!=\!64\!-\!\mbf{9}$  \\\hline
    $63\!=\!64\!-\!1$ &
    $65\!=\!64\!+\!1$ &
    $75\!=\!64\!+\!\mbf{11}$ &
    $77\!=\!64\!+\!\mbf{13}$ \\\hline
    $81\!=\!32\!+\!\mbf{49}$ &
    $91\!=\!64\!+\!\mbf{27}$ &
    $99\!=\!64\!+\!\mbf{35}$ &
    \red{$41\!=\!32\!+\!9$}   \\\hline
    $105\!=\!64\!+\!\red{\mbf{41}}$ &
    $117\!=\!128\!-\!\mbf{11}$ &
    $121\!=\!128\!-\!\mbf{7}$  &
    $135\!=\!128\!+\!\mbf{7}$  \\\hline
    $143\!=\!128\!+\!\mbf{15}$ &
    \blue{$37\!=\!33\!+\!4$} &
    $165\!=\!128\!+\!\blue{\mbf{37}}$ &
    $169\!=\!128\!+\!\red{\mbf{41}}$ \\\hline
    \orange{$61\!=\!64\!-\!3$} &
    $195\!=\!256\!-\!\orange{\mbf{61}}$  &
    \magenta{$31\!=\!32\!-\!1$} &
    $225\!=\!256\!-\!\magenta{\mbf{31}}$  \\\hline
\end{tabular}
\end{table}  

%% file: table_num_adders.tex
\begin{table}[t]
\scriptsize
\centering
\caption{Resources of detector core using exhaustive search}\vspace{-0.1in}
\label{t:num_adders_noslice_arch}
\renewcommand{\arraystretch}{1.25}
\begin{tabular}{|@{\hspace{0.5mm}}c@{\hspace{0.5mm}}|@{\hspace{0.5mm}}c@{\hspace{0.5mm}}|@{\hspace{0.5mm}}c@{\hspace{0.5mm}}|@{\hspace{0.5mm}}c@{\hspace{0.5mm}}|@{\hspace{0.5mm}}c@{\hspace{0.5mm}}|}
  \hline
  $\#$ adders (\& muxes)& 2-PAM & 4-PAM & 8-PAM & 16-PAM \\\hline\hline
  $Ax_{1\re}^2$ & 0 & 1 & 4 & 11 \\\hline
  $C\abs{x_{1\re}}$ & 0 & 1 & 3 & 7 \\\hline
  $\abs{\mbf{b}_{1\re}^T \bs{\lambda}_{1\re}^{}}$ & 0 & 2 & 6 & 12\\\hline
  $\bar{f}_{1\re}(x_{1\re})$ & 4 & 8 & 16 & 32\\\hline
  $D\abs{x_{1\re}}$ & 0 & 1 & 3 & 7 \\\hline
  $\abs{\mbf{b}_{1\im}^T \bs{\lambda}_{1\im}^{}}$ & 0 & 2 & 6 & 12\\\hline
  $\bar{f}_{1\im}(x_{1\im})$ & 4 & 8 & 16 & 32\\\hline
  $\bar{f}_1(x_1)\!=\!\bar{f}_{1\re}(x_{1\re})\!+\!\bar{f}_{1\im}(x_{1\im})$ & 4 & 16 & 64 & 256 \\ \hline
  $Bx_{2\re}^2$ & 0 & 1 & 4 & 11 \\\hline
  $G\abs{x_{2\re}}$ & 0 & 1 & 3 & 7 \\\hline
  $\abs{\mbf{b}_{2\re}^T \bs{\lambda}_{2\re}^{}}$ & 0 & 2 & 6 & 12\\\hline
  $B x_{2\re}^2 \!+\! Gx_{2\re} \!-\!\mbf{b}_{2\re}^T \bs{\lambda}_{2\re}^{}$ & 4 & 8 & 16 & 32\\\hline
  $H\abs{x_{2\im}}$ & 0 & 1 & 3 & 7 \\\hline
  $\abs{\mbf{b}_{2\im}^T \bs{\lambda}_{2\im}^{}}$ & 0 & 2 & 6 & 12\\\hline
  $B x_{2\im}^2 \!+\! Hx_{2\im} \!-\!\mbf{b}_{2\im}^T \bs{\lambda}_{2\im}^{}$ & 4 & 8 & 16 & 32\\\hline
  $(E\!\abs{x_{1\re}} \!\pm\! F\!\abs{x_{1\im}})\!\abs{x_{2\re}}$ & 2 & 16 & 122 & 936\\\hline

  $\bar{f}_{2\re}(x_{2\re}|x_1)$ & 8 & 64 & 512 & 4096 \\\hline
  $\overline{m}_{2\re}\!=\!\min\{\bar{f}_{2\re}(x_{2\re}|x_1)\}$ & 4 & 48 & 448 & 3840 \\
   \hspace{1.1in} muxes $\rightarrow$& 4 & 48 & 448 & 3840 \\\hline
  $\bar{f}_{2\im}(x_{2\im}|x_1)$ & 8 & 64 & 512 & 4096 \\\hline
  $\overline{m}_{2\im}\!=\!\min\{\bar{f}_{2\im}(x_{2\im}|x_1)\}$ & 4 & 48 & 448 & 3840 \\
  \hspace{1.1in} muxes $\rightarrow$& 4 & 48 & 448 & 3840 \\\hline
  $\bar{f}_1\!+\!\overline{m}_{2\re}\!+\!\overline{m}_{2\im}$ & 8 & 32 & 128 & 512 \\\hline\hline
  HD solution: $\min\{\bar{f}_1\!+\!\overline{m}_{2\re}\!+\!\overline{m}_{2\im}\}$ & 3 & 15 & 63  & 255 \\
  \hspace{1.1in} muxes $\rightarrow$ & 3 & 15 & 63 & 255 \\\hline\hline
  soft-output LLRs & 6 & 28  & 118  & 488 \\
  \hspace{1.1in} muxes $\rightarrow$ & 4 & 24 & 112 & 480 \\\hline\hline
  Total (soft-output) & 60 & 346 & 2460 & 18290 \\
                      & 12  & 120  & 1008  & 8160\\\hline
\end{tabular}
\end{table}

%% file: table_num_adders_slice.tex
\begin{table}[t]
\scriptsize
\centering
\caption{Resources of detector core using slicers}\vspace{-0.1in}
\label{t:num_adders_slicing}
\renewcommand{\arraystretch}{1.25}
\begin{tabular}{|@{\hspace{0.5mm}}c@{\hspace{0.5mm}}|@{\hspace{0.5mm}}c@{\hspace{0.5mm}}|@{\hspace{0.5mm}}c@{\hspace{0.5mm}}|@{\hspace{0.5mm}}c@{\hspace{0.5mm}}|@{\hspace{0.5mm}}c@{\hspace{0.5mm}}|}
  \hline
  $\#$ adders (\& muxes) & 2-PAM & 4-PAM & 8-PAM & 16-PAM \\\hline\hline
$\bar{f}_1(x_1)\!=\!\bar{f}_{1\re}(x_{1\re})\!+\!\bar{f}_{1\im}(x_{1\im})$ & 4 & 16 & 64 & 256 \\\hline\hline
  $B\!\abs{x_{2\re}\!+\!\bar{x}_{2\re}}$ & 0  & 0  & 2  & 6 \\\hline
  $\left(\mbf{b}_{2\re}\!-\!\bar{\mbf{b}}_{2\re}\right)^{\!T}\!\bs{\lambda}_{2\re}^{}$ &0 &2 &8 & 22 \\\hline
  $\frac{\left(\mbf{b}_{2\re}\!-\!\bar{\mbf{b}}_{2\re}\right)}{x_{2\re}\!-\!\bar{x}_{2\re}}^{\!T}\!\bs{\lambda}_{2\re}^{}$ &0 & 1&10 & 54\\\hline
  $R(x_{2\re},\bar{x}_{2\re})$ & 0& 4 & 24 & 112 \\\hline
  $\min/\max$ boundaries &0 &6 &42 & 210 \\
                   (MUXES)& 0&6 &42 & 210 \\\hline 
  $\min/\max~\text{boundaries}\!-\!G$ & 2 & 6 & 14 & 54 \\\hline
  $\abs{E\!x_{1\re} \!\pm\! F\!x_{1\im}}\!\cdot\!\abs{x_{2\re}}$ & 2 & 16 & 122 & 936\\\hline
  Compare $\abs{Ex_{1\re}\!\pm\!Fx_{1\im}}$ & & & &\\
  and $\min/\max~\text{boundaries}\!-\!G$ & 4 & 48 & 448 & 3840\\\hline
  $\bar{f}_{2\re}(\hat{x}_{2\re}|x_1)\!=\!\abs{E\!x_{1\re} \!\pm\! F\!x_{1\im}}\!\cdot\!\abs{\hat{x}_{2\re}}+$ & 4 & 16 & 64 & 256 \\
  $\left(B \hat{x}_{2\re}^2 \!+\! G\hat{x}_{2\re} \!-\!\mbf{b}^T\!(\hat{x}_{2\re}) \bs{\lambda}_{2\re}^{}\right)$ & 4 & 56 & 544 & 4736  \\\hline\hline

  $\left(\mbf{b}_{2\im}\!-\!\bar{\mbf{b}}_{2\im}\right)^{\!T}\!\bs{\lambda}_{2\im}^{}$ &0 &2 &8 & 22 \\\hline
  $\frac{\left(\mbf{b}_{2\im}\!-\!\bar{\mbf{b}}_{2\im}\right)}{x_{2\im}\!-\!\bar{x}_{2\im}}^{\!T}\!\bs{\lambda}_{2\im}^{}$ & 0&1 &10 & 54\\\hline
  $I(x_{2\im},\bar{x}_{2\im})$ & 0& 4 & 24 & 112 \\\hline
  $\min/\max$ boundaries & 0 & 6 & 42 & 210 \\
              (MUXES)    & 0 & 6 & 42 & 210 \\\hline 
  $\min/\max~\text{boundaries}\!-\!H$ & 2 & 6 & 14 & 54 \\\hline
  Compare $\abs{Ex_{1\im}\!\mp\!Fx_{1\re}}$ & & & &\\
  and $\min/\max~\text{boundaries}\!-\!H$ & 4 & 48 & 448 & 3840\\\hline $\bar{f}_{2\im}(\hat{x}_{2\im}|x_1)\!=\!\abs{Ex_{1\im}\!\mp\!Fx_{1\re}}\!\cdot\!\abs{\hat{x}_{2\im}}+$ & 4 & 16 & 64 & 256 \\
  $\left(B \hat{x}_{2\im}^2 \!+\! H\hat{x}_{2\im} \!-\!\mbf{b}^T\!(\hat{x}_{2\im}) \bs{\lambda}_{2\im}^{}\right)$ & 4 & 56 & 544 & 4736 \\\hline\hline
  $\bar{f}_1(x_1)\!+\!\bar{f}_{2\re}(\hat{x}_{2\re}|x_1)\!+\!\bar{f}_{2\im}(\hat{x}_{2\im}|x_1)$  & 4 &32 & 128 & 512 \\\hline\hline
  soft-output LLRs & 6 & 28  & 118  & 488 \\
  \hspace{1.1in} muxes $\rightarrow$ & 4 & 24 & 112 & 480 \\\hline\hline
  Total & 36 & 258 & 1654 & 11246\\
     (2:1)-MUXS   & 12 & 148 & 1284 & 10372\\\hline
\end{tabular}
\end{table}

%% file: Optimized_Configurable_Architectures.bbl
\begin{thebibliography}{10}
\providecommand{\url}[1]{#1}
\csname url@samestyle\endcsname
\providecommand{\newblock}{\relax}
\providecommand{\bibinfo}[2]{#2}
\providecommand{\BIBentrySTDinterwordspacing}{\spaceskip=0pt\relax}
\providecommand{\BIBentryALTinterwordstretchfactor}{4}
\providecommand{\BIBentryALTinterwordspacing}{\spaceskip=\fontdimen2\font plus
\BIBentryALTinterwordstretchfactor\fontdimen3\font minus
  \fontdimen4\font\relax}
\providecommand{\BIBforeignlanguage}[2]{{%
\expandafter\ifx\csname l@#1\endcsname\relax
\typeout{** WARNING: IEEEtran.bst: No hyphenation pattern has been}%
\typeout{** loaded for the language `#1'. Using the pattern for}%
\typeout{** the default language instead.}%
\else
\language=\csname l@#1\endcsname
\fi
#2}}
\providecommand{\BIBdecl}{\relax}
\BIBdecl

\bibitem{802.11ac}
\BIBentryALTinterwordspacing
\emph{IEEE Draft Standard - Part 11: Wireless LAN Medium Access Control and
  Physical Layer Specifications - Amendment 4: Enhancements for Very High
  Throughput for operation in bands below \unit[6]{GHz}}, IEEE Std.
  P802.11ac/D7.0, Dec 2013. [Online]. Available: \url{http://www.ieee.org}
\BIBentrySTDinterwordspacing

\bibitem{LTE_36.211}
\BIBentryALTinterwordspacing
\emph{Evolved Universal Terrestrial Radio Access {(E-UTRA)}; Physical Channels
  and Modulation}, 3GPP Std. TS 36.211. [Online]. Available:
  \url{http://www.3gpp.org}
\BIBentrySTDinterwordspacing

\bibitem{2003_Paulraj}
A.~Paulraj, R.~Nabar, and D.~Gore, \emph{Introduction to Space-Time Wireless
  Communications}.\hskip 1em plus 0.5em minus 0.4em\relax Cambridge, U.K.:
  Cambridge Univ. Press, 2003.

\bibitem{2006_Giannakis}
G.~B. Giannakis \emph{et~al.}, \emph{Space-Time Coding for Broadband Wireless
  Communications}.\hskip 1em plus 0.5em minus 0.4em\relax New York: John Wiley
  and Sons, 2006.

\bibitem{2007_Biglieri}
E.~Biglieri \emph{et~al.}, \emph{{MIMO} Wireless Communications}.\hskip 1em
  plus 0.5em minus 0.4em\relax Cambridge, U.K.: Cambridge Univ. Press, 2007.

\bibitem{2010_Oestges}
C.~Oestges and B.~Clerckx, \emph{{MIMO} Wireless Communications}.\hskip 1em
  plus 0.5em minus 0.4em\relax Oxford, U.K.: Elsevier Academic Press, 2007.

\bibitem{2014_Chockalingam}
A.~Chockalingam and B.~S. Rajan, \emph{Large MIMO Systems}.\hskip 1em plus
  0.5em minus 0.4em\relax Cambridge University Press, 2014.

\bibitem{2000_Hassibi}
B.~Hassibi, ``An efficient square-root algorithm for {BLAST},'' in \emph{Proc.
  IEEE Int. Conf. Acoustics, Speech, and Signal Process. (ICASSP)}, Istanbul,
  Turkey, Jun. 2000, pp. 5--9.

\bibitem{1999_Golden}
G.~D. Golden \emph{et~al.}, ``Detection algorithm and initial laboratory
  results using {V-BLAST} space-time communication architecture,'' \emph{IEE
  Electronics Letters}, vol.~35, no.~1, pp. 14--15, Jan. 1999.

\bibitem{2003_Wubben}
D.~W\"{u}bben, R.~B\"{o}hnke, V.~K\"{u}hn, and K.~Kammeyer, ``{MMSE} extension
  of {V-BLAST} based on sorted {QR} decomposition,'' in \emph{Proc. IEEE Vehic.
  Technol. Conf. (VTC)}, Orlando, Florida, Oct. 2003, pp. 508--512.

\bibitem{2011_Studer_JSSC}
C.~Studer, S.~Fateh, and D.~Seethaler, ``{ASIC} implementation of soft-input
  soft-output {MIMO} detection using {MMSE} parallel interference
  cancellation,'' \emph{{IEEE} Trans. Syst. Sci. Cybern.}, vol.~47, no.~7, pp.
  1754--1765, Jul. 2011.

\bibitem{1993_Viterbo_Biglieri}
E.~Viterbo and E.~Biglieri, ``A universal decoding algorithm for lattice
  codes,'' in \emph{14\`{e}me Colloque GRETSI}, Juan-Les-Pins, France, Sep.
  1993, pp. 611--614.

\bibitem{1999_Viterbo_Boutros}
E.~Viterbo and J.~Boutros, ``A universal lattice code decoder for fading
  channels,'' \emph{{IEEE} Trans. Inf. Theory}, vol.~45, no.~5, pp. 1639--1642,
  Jul. 1999.

\bibitem{2000_Damen}
O.~Damen, A.~Chkeif, and J.-C. Belfiore, ``Lattice code decoder for space-time
  codes,'' \emph{{IEEE} Commun. Lett.}, vol.~4, no.~5, pp. 161--163, May 2000.

\bibitem{2002_Agrell}
E.~Agrell \emph{et~al.}, ``Closest point search in lattices,'' \emph{{IEEE}
  Trans. Inf. Theory}, vol.~48, no.~8, pp. 2201--2214, Aug. 2002.

\bibitem{2003_Hochwald}
B.~Hochwald and S.~ten Brink, ``Achieving near-capacity on a multiple-antenna
  channel,'' \emph{{IEEE} Trans. Commun.}, vol.~51, no.~3, pp. 389--399, Mar.
  2003.

\bibitem{2005_Hassibi_Vikalo}
B.~Hassibi and H.~Vikalo, ``On sphere decoding algorithm. {I}. {Expected}
  complexity,'' \emph{{IEEE} Trans. Signal Process.}, vol.~53, no.~8, pp.
  2806--2818, Aug. 2005.

\bibitem{2005_Jalden}
J.~Jald\'{e}n and B.~Ottersten, ``On the complexity of sphere decoding in
  digital communications,'' \emph{{IEEE} Trans. Signal Process.}, vol.~53,
  no.~4, pp. 1474--1484, Apr. 2005.

\bibitem{2011_Seethaler}
D.~Seethaler, J.~Jald\'{e}n, C.~Studer, and H.~B\"{o}lcskei, ``On the
  complexity distribution of sphere decoding,'' \emph{{IEEE} Trans. Inf.
  Theory}, vol.~57, no.~9, pp. 5754--5768, Sep. 2011.

\bibitem{2002_Wong}
K.-W. Wong, C.-Y. Tsui, R.~S.-K. Cheng, and W.-H. Mow, ``A {VLSI} architecture
  of a {$K$}-best lattice decoding algorithm for {MIMO} channels,'' in
  \emph{Proc. IEEE Int. Symp. on Circuits and Systems (ISCAS)}, vol.~3,
  Scottsdale, Arizona, May 2002, pp. 273--276.

\bibitem{2006_Wenk_ISCAS}
M.~Wenk \emph{et~al.}, ``{K-best} {MIMO} detection {VLSI} architectures
  achieving up to \unit[424]{Mbps},'' in \emph{Proc. IEEE Int. Symp. on
  Circuits and Systems (ISCAS)}, Island of Kos, Greece, May 2006, pp.
  1151--1154.

\bibitem{2010_Mondal}
S.~Mondal, A.~Eltawil, C.-A. Shen, and K.~Salama, ``Design and implementation
  of a sort free {$K$}-best sphere decoder,'' \emph{{IEEE} Trans. {VLSI}
  Syst.}, vol.~18, no.~10, pp. 1497--1501, Oct. 2010.

\bibitem{2010_Liu}
L.~Liu, F.~Ye, X.~Ma, T.~Zhang, and J.~Ren, ``A {1.1-Gb/s} {115-pJ/bit}
  configurable {MIMO} detector using $\unit[0.13]{\mu m}$ {CMOS} technology,''
  \emph{{IEEE} Trans. Circuits Syst. {II}}, vol.~57, no.~9, pp. 701--705, Sep.
  2010.

\bibitem{2010_Shen_Eltawil}
C.-A. Shen, A.~Eltawil, and K.~Salama, ``Evaluation framework for {$K$}-best
  sphere decoders,'' \emph{J. of Circuits, Systems and Computers}, vol.~19,
  no.~5, pp. 975--995, Aug. 2010.

\bibitem{2012_Shabany}
M.~Shabany and P.~Gulak, ``A 675 {Mbps}, $4\times 4$ {64-QAM} {K-Best} {MIMO}
  detector in $\unit[0.13]{\mu m}$ {CMOS},'' \emph{{IEEE} Trans. {VLSI} Syst.},
  vol.~20, no.~1, pp. 135--147, Jan. 2012.

\bibitem{2013_Mahdavi}
M.~Mahdavi and M.~Shabany, ``Novel {MIMO} detection algorithm for high-order
  constellations in the complex domain,'' \emph{{IEEE} Trans. {VLSI} Syst.},
  vol.~21, no.~5, pp. 834--847, May 2013.

\bibitem{2004_Garret}
D.~Garrett \emph{et~al.}, ``Silicon complexity for maximum likelihood {MIMO}
  detection using spherical decoding,'' \emph{{IEEE} J. Solid-State Circuits},
  vol.~39, no.~9, pp. 1544--1552, Sep. 2004.

\bibitem{2004_Guo}
Z.~Guo and P.~Nilsson, ``A {VLSI} architecture of the {Schnorr-Euchner} decoder
  for {MIMO} systems,'' in \emph{Proc. IEEE CAS Symp. Emerging Technologies},
  vol.~1, Shanghai, China, May 2004, pp. 65--68.

\bibitem{2005_Burg}
A.~Burg \emph{et~al.}, ``{VLSI} implementation of {MIMO} detection using the
  sphere decoding algorithm,'' \emph{{IEEE} J. Solid-State Circuits}, vol.~40,
  no.~7, pp. 1566--1577, Jul. 2005.

\bibitem{2008_Studer}
C.~Studer, A.~Burg, and H.~B\"{o}lcskei, ``Soft-output sphere decoder:
  Algorithms and {VLSI} implementation,'' \emph{{IEEE} J. Sel. Areas Commun.},
  vol.~26, no.~2, pp. 290--300, Feb. 2008.

\bibitem{2009_Yang_Markovic_TCASI}
C.-H. Yang and D.~Markovic, ``A flexible {DSP} architecture for {MIMO} sphere
  decoding,'' \emph{{IEEE} Trans. Circuits Syst. {I}}, vol.~56, no.~10, pp.
  2301--2314, Oct. 2009.

\bibitem{2009_Yang_Markovic_ESSCIRC}
------, ``A \unit[2.89]{mW} \unit[50]{GOPS} $16 \times 16$ 16-core {MIMO}
  sphere decoder in \unit[90]{nm} {CMOS},'' in \emph{European Solid-State
  Circuits Conf. (ESSCIRC)}, Athens, Greece, Sep. 2009, pp. 344--347.

\bibitem{2011_Borlenghi}
F.~Borlenghi \emph{et~al.}, ``A \unit[772]{Mbit/s} \unit[8.81]{bit/nJ}
  \unit[90]{nm} {CMOS} soft-input soft-output sphere decoder,'' in \emph{IEEE
  Asian Solid State Circutis Conf. (A-SSCC)}, Jeju, Korea, Nov. 2011, pp.
  297--300.

\bibitem{2012_Liu_Lofgren}
L.~Liu, J.~Lofgren, and P.~Nilsson, ``Area-efficient configurable
  high-throughput signal detector supporting multiple {MIMO} modes,''
  \emph{{IEEE} Trans. Circuits Syst. {I}}, vol.~59, no.~9, pp. 2085--2096, Sep.
  2012.

\bibitem{2012_Sun_Cavallaro}
Y.~Sun and J.~R. Cavallaro, ``Trellis-search based soft-input soft-output
  {MIMO} detector: Algorithm and {VLSI} architecture,'' \emph{{IEEE} Trans.
  Signal Process.}, vol.~60, no.~5, pp. 2617--2627, May 2012.

\bibitem{2013_Chen}
X.~Chen, G.~He, and J.~Ma, ``{VLSI} implementation of a high-throughput
  iterative fixed-complexity sphere decoder,'' \emph{{IEEE} Trans. Circuits
  Syst. {II}}, vol.~60, no.~5, pp. 272--276, May 2013.

\bibitem{2014_sphereP1_mansour}
M.~M. Mansour, S.~Alex, and M.~Jalloul, ``Reduced complexity soft-output {MIMO}
  sphere detectors -- {Part I}: {Algorithmic} optimizations,'' \emph{{IEEE}
  Trans. Signal Process.}, vol.~62, no.~21, pp. 5505--5520, Nov. 2014.

\bibitem{2014_sphereP2_mansour}
------, ``Reduced complexity soft-output {MIMO} sphere detectors -- {Part II}:
  {Architectural} optimizations,'' \emph{{IEEE} Trans. Signal Process.},
  vol.~62, no.~21, pp. 5521--5535, Nov. 2014.

\bibitem{2014_Huang}
M.-Y. Huang and P.-Y. Tsai, ``Toward multi-gigabit wireless: Design of
  high-throughput {MIMO} detectors with hardware-efficient architecture,''
  \emph{{IEEE} Trans. Circuits Syst. {I}}, vol.~61, no.~2, pp. 613--624, Feb.
  2014.

\bibitem{2005a_Jiang}
Y.~Jiang, J.~Li, and W.~W. Hager, ``Joint transceiver design for {MIMO}
  communications using geometric mean decomposition,'' \emph{{IEEE} Trans.
  Signal Process.}, vol.~53, no.~10, pp. 3791--3803, Oct. 2005.

\bibitem{2005b_Jiang}
------, ``Uniform channel decomposition for {MIMO} communications,''
  \emph{{IEEE} Trans. Signal Process.}, vol.~53, no.~11, pp. 4283--4294, Nov.
  2005.

\bibitem{2008_Ariyavisitakul}
S.~Ariyavisitakul, J.~Zheng, E.~Ojard, and J.~Kim, ``Subspace beamforming for
  near-capacity {MIMO} performance,'' \emph{{IEEE} Trans. Signal Process.},
  vol.~56, no.~11, pp. 5729--5733, Nov. 2008.

\bibitem{2011_Chen}
Y.~Chen and S.~Brink, ``Near-capacity {MIMO} subspace detection,'' in
  \emph{Proc. IEEE Int. Symp. Personal Indoor and Mobile Radio Commun.
  (PIMRC)}, Toronto, Canada, Sep. 2011, pp. 1733--1737.

\bibitem{2006_Fitz}
M.~Siti and M.~P. Fitz, ``A novel soft-output layered orthogonal lattice
  detector for multiple antenna communications,'' in \emph{Proc. IEEE Int.
  Conf. Commun. (ICC)}, vol.~4, Istanbul, Turkey, Jun. 2006, pp. 1686--1691.

\bibitem{2007a_Siti}
------, ``On layer ordering techniques for near-optimal {MIMO} detectors,'' in
  \emph{Proc. IEEE Wireless Commun. and Netw. Conf. (WCNC)}, Hong Kong, Mar.
  2007, pp. 1199--1204.

\bibitem{2005_Yee_ICASSP}
M.~S. Yee, ``{Max-log-MAP} sphere decoder,'' in \emph{Proc. IEEE Int. Conf.
  Acoustics, Speech, and Signal Process. (ICASSP)}, vol.~3, Philadelphia, PA,
  Mar. 2005, pp. 1013--1016.

\bibitem{2009_Ojard}
\BIBentryALTinterwordspacing
E.~Ojard and S.~Ariyavisitakul, ``\BIBforeignlanguage{English}{Method and
  system for approximate maximum likelihood ({ML}) detection in a multiple
  input multiple output ({MIMO}) receiver},'' US Patent 12/207,721, Mar. 19,
  2009. [Online]. Available: \url{http://www.google.com/patents/US20090074114}
\BIBentrySTDinterwordspacing

\bibitem{2012_Zhang}
L.~A. C.~Zhang and T.~Meixia, ``{LTE-advanced} and {4G} wireless communications
  [{Guest Editorial}],'' \emph{{IEEE} Commun. Mag.}, vol.~50, no.~2, pp.
  102--103, Feb. 2012.

\bibitem{2014_b_Gomma}
A.~Gomma and L.~Jalloul, ``Efficient soft-input soft-output detection of
  dual-layer {MIMO} systems,'' \emph{{IEEE} Trans. Wireless Commun.}, vol.~3,
  no.~5, pp. 541--544, Oct. 2014.

\bibitem{2012_Yeung}
C.~K. Yeung, J.~Lee, and S.~Kim, ``A simple slicer for soft detection in
  gray-coded qam-modulated mimo ofdm systems,'' in \emph{IEEE 35th Sarnoff
  Sympos. (SARNOFF)}, Newark, NJ, May 2012, pp. 1--5.

\bibitem{2014_mansour_SPL_WLD}
M.~M. Mansour, ``A near-{ML} {MIMO} subspace detection algorithm,''
  \emph{{IEEE} Signal Process. Lett.}, vol.~22, no.~4, pp. 408--412, Apr. 2015.

\bibitem{1996_Golub}
G.~H. Golub and C.~F.~V. Loan, \emph{Matrix Computations}, 3rd~ed.\hskip 1em
  plus 0.5em minus 0.4em\relax Baltimore, MD: Johns Hopkins Univ. Press, 1996.

\bibitem{CEVA}
\BIBentryALTinterwordspacing
``{CEVA-XC4210 DSP} processors.'' [Online]. Available:
  \url{http://www.ceva-dsp.com/CEVA-XC4210}
\BIBentrySTDinterwordspacing

\bibitem{2009_Lee}
J.~Lee, J.-K. Huan, and J.~Zhang, ``{MIMO} technologies in {3GPP} {LTE} and
  {LTE-Advanced},'' \emph{EURASIP J. on Wireless Commun. and Netw.}, vol. 2009,
  no.~1, pp. 1--10, 2009.

\bibitem{2011_Duplicy}
J.~Duplicy \emph{et~al.}, ``{MU-MIMO} in {LTE} systems,'' \emph{EURASIP J. on
  Wireless Commun. and Netw.}, vol. 2011, no.~1, pp. 1--13, 2011.

\bibitem{2011_Bai}
Z.~Bai \emph{et~al.}, ``On the equivalence of {MMSE} and {IRC} receiver in
  {MU-MIMO} systems,'' \emph{{IEEE} Commun. Lett.}, vol.~15, no.~12, pp.
  1288--1290, Dec. 2011.

\bibitem{2011_Ghaffar_a}
R.~Ghaffar and R.~Knopp, ``Interference sensitivity for multiuser {MIMO} in
  {LTE},'' in \emph{IEEE Workshop on Sig. Proc. Advances in Wireless Commun.
  (SPAWC)}, Jun. 2011, pp. 506--510.

\bibitem{2011_Ghaffar_b}
------, ``Interference-aware receiver structure for multiuser {MIMO} and
  {LTE},'' \emph{EURASIP J. on Wireless Commun. and Netw.}, vol.~40, pp. 1--17,
  2011.

\bibitem{1998_Viterbi}
A.~Viterbi, ``An intuitive justifications and a simplified implementation of
  the {MAP} decoder for convolutional codes,'' \emph{{IEEE} J. Sel. Areas
  Commun.}, vol.~16, pp. 260--264, Feb. 1998.

\bibitem{2015_Gomaa}
\BIBentryALTinterwordspacing
A.~Gomaa \emph{et~al.}, ``Multi-user {MIMO} receivers with partial state
  information,'' \emph{{IEEE} Trans. Veh. Technol.}, Jan. 2015, (under review).
  [Online]. Available: \url{http://arxiv.org/abs/1502.00212}
\BIBentrySTDinterwordspacing

\bibitem{LTE_36.212}
\BIBentryALTinterwordspacing
\emph{Evolved Universal Terrestrial Radio Access {(E-UTRA)}; Multiplexing and
  channel coding}, 3GPP Std. TS 36.212. [Online]. Available:
  \url{http://www.3gpp.org}
\BIBentrySTDinterwordspacing

\bibitem{PedB_ChannelModel}
\BIBentryALTinterwordspacing
\emph{High Speed Downlink Packet Access: UE Radio Transmission and Reception
  FDD}, 3GPP Std. TR 25.890. [Online]. Available: \url{http://www.3gpp.org}
\BIBentrySTDinterwordspacing

\end{thebibliography}
